\documentclass[12pt,a4paper]{article}
\usepackage[a4paper,margin=2cm]{geometry}

\usepackage{graphicx}
\usepackage{amsmath}
\usepackage{mathtools}
\usepackage{amssymb}
\usepackage{grffile}
\usepackage{float}
\usepackage{xcolor}
\usepackage{gensymb}
\usepackage{hyperref}
\usepackage{dcolumn}
\usepackage{bm}
\usepackage{multirow}
\usepackage{makecell}
\usepackage{soul}
\usepackage{caption}
\usepackage{placeins}
\usepackage{bbm}

\newcommand{\twovector}[2]{\left(\begin{array}{c} #1 \\[.2cm] #2 \end{array}\right)}
\newcommand{\ket}[1]{\vert #1 \rangle}

\newcommand{\twoarray}[4]{\left(\begin{array}{cc} #1 & #2 \\[.2cm] #3 & #4 \end{array}\right)}

\vspace{-1.5em}
\title{Taming quantum interference: a route to high electrical conductance in carbon nanotube assemblies}

\vspace{-1.0em}
\author{
Teresa Kulka$^1$,
Agnieszka E. Lekawa-Raus$^{2,3}$,
John S. Bulmer$^{4,5,3}$,\\
Krzysztof Koziol$^{6}$,
Fedor F. Balakirev$^{7}$,
Irina V. Lebedeva$^{8,9,10}$,\\
Jacek A. Majewski$^{1,11}$, 
Magdalena Marganska$^{*,12,13}$,
Karolina Z. Milowska$^{*,8,14}$
}

\date{}

\begin{document}
\newgeometry{top=1.1cm,bottom=1.6cm,left=2cm,right=2cm}
\maketitle

\vspace{-2.0em}

\begin{center}
\footnotesize  
$^1$Faculty of Physics, University of Warsaw, Ludwika Pasteura 5, 02-093 Warsaw, Poland\\
$^2$Centre for Advanced Materials and Technologies (CEZAMAT), Warsaw University of Technology, Warsaw, Poland\\
$^3$ Department of Materials Science and Metallurgy, University of Cambridge, 27 Charles Babbage Rd, Cambridge, CB3 0FS, United Kingdom\\
$^4$ University of Cincinnati, 598 Rhodes Hall, Cincinnati, P.O. Box 210072, OH, USA \\
$^5$ Aerospace Systems Directorate, Air Force Research Laboratory, Wright-Patterson Air Force Base, 45433, OH, USA \\
$^6$ Centre for Materials, Cranfield University, Bld 56, Cranfield, Bedfordshire, MK43 0AL, United Kingdom\\
$^7$ National High Magnetic Field Laboratory, Los Alamos National Laboratory, Los Alamos, 87545, NM, USA \\
$^8$ CIC nanoGUNE, Tolosa Hiribidea 76, Donostia-San Sebastian, 20018, Spain\\
$^9$ Simune Atomistics, Donostia-San Sebastián, 20018, Spain\\
$^{10}$ Catalan Institute of Nanoscience and Nanotechnology (ICN2), CSIC and BIST, Campus UAB, Bellaterra, 08193 Barcelona, Spain\\
$^{11}$ Center for Terahertz Research and Applications - Centera 2, Centre for Advanced Materials and Technologies (CEZAMAT), Warsaw University of Technology, Warsaw, Poland\\
$^{12}$Institute of Theoretical Physics, Wroclaw University of Science and Technology, Wybrzeże Wyspiańskiego 27, 50-370, Wrocław, Poland\\
$^{13}$ Institute for Theoretical Physics, University of Regensburg, 93040, Regensburg, Germany\\
$^{14}$ Ikerbasque, Basque Foundation for Science, Plaza Euskadi, Bilbao, 48009, Spain\\
\end{center}
\begin{center}
\footnotesize
$^{*}$ E-mails: magdalena.marganska@pwr.edu.pl; kz.milowska@nanogune.eu 
\end{center}

\vspace{-2.0em}

\begin{abstract}
In nanostructured networks, transport is governed by junctions between neighbouring building blocks. Improving their alignment and removing defects is the intuitive route to better electron transport. At low temperatures, however, when transport becomes coherent, a junction cannot always be reduced to a single effective resistance, because electron-wave interference can strongly enhance or suppress transmission even in nominally ideal junctions. Using carbon nanotube (CNT) networks as a model system, we explore coherent transport through experimentally relevant junctions, from single and multiple single-walled CNT (SWCNT) contacts to double-walled CNT (DWCNT) and \color{black} multi-walled CNT (MWCNT) \color{black} junctions, with atomistic tight-binding non-equilibrium Green's-function calculations, also under a perpendicular magnetic field. We use analytically solvable minimal models to identify transport regimes expected for quasi-1D nanoscale junctions, and an electron-waveguide picture to interpret their CNT-specific manifestations. For single SWCNT--SWCNT junctions, high-transmission windows are set mainly by overlap length, doping and magnetic field. Gateway states can enhance conductance when some CNT subbands are gapped, and in some cases a magnetic field can restore transmission by lifting an interference blockade. In more complex architectures, added paths become selective: multi-junctions generate resonant filtering, while additional walls redistribute transmission instead of acting as independent channels. DWCNT junctions remain outer-wall dominated and SWCNT-like, whereas \color{black}  MWCNT junctions  redistribute transmission among coupled walls and show a more complex field response\color{black}. This \color{black}{ is consistent with} the lower, more field-sensitive conductance of multi-walled CNT (MWCNT) fibres, in accord with our ultrahigh-field measurements on SWCNT and MWCNT fibres. Ultimately, this work turns microscopic interference mechanisms into design principles for high-conductance, field-stable CNT conductors.
\end{abstract}

\clearpage
\restoregeometry

\section{Introduction}
Carbon nanotubes (CNTs) are fascinating one-dimensional materials~\cite{radushkevich1952, Iijima1991, Dresselhaus2001, SaitoDresselhaus1998}, with carbon atoms arranged in a hollow cylinder that can be obtained by rolling up a graphene sheet. Depending on the choice of the folding vector, their electronic band structure can be metallic or can contain a finite band gap. These properties have motivated numerous studies and made CNTs attractive for many applications, particularly in electrical devices~\cite{c10030069,hughes2024review,Popov2014}. While electron transport in isolated CNTs has been studied in depth~\cite{McEuen2002,laird2015}, transport in realistic CNT assemblies remains much less understood. In films and fibres, charge transfer proceeds not only along individual nanotubes but also across junctions between neighbouring tubes, where conduction requires intertube tunnelling. Because producing large-scale samples with uniform chirality remains extremely challenging~\cite{Zhang2005,Tian2011}, most CNT assemblies contain a heterogeneous mixture of nanotubes with different chiralities and electronic characters. In such assemblies, several modelling approaches build the network response from semiconducting and metallic nanotubes connected in \color{black} series \color{black} or parallel circuits~\cite{nakai2014,hayashi2019,hayashi2020}. These descriptions capture the role of nanotube electronic type and network connectivity, but they do not fully resolve the microscopic transmission of each CNT--CNT contact. Network-level studies further show that CNT contacts are key bottlenecks for conductance~\cite{wang2025,gabbett2024}, although they are often coarse-grained as effective junction resistances or simplified tunnelling contacts. Such a description is useful at the network scale, but it is not sufficient to explain why a given junction has a particular transport response, because junction behaviour depends on both nanotube and contact characteristics, as well as on external parameters such as temperature and magnetic field~\cite{manuscriptAGA}. Thus, theoretical modelling of well-defined junction architectures, which occur in real samples, brings us closer to understanding experimental transport data.

Real CNT assemblies are not limited to single-walled CNTs (SWCNTs). Both single- and multi-walled CNTs (MWCNTs) are widely used experimentally, with MWCNTs often preferred when larger quantities of material are required because they are less expensive and readily mass-produced~\cite{hughes2024review}. At the same time, experimental observations indicate that assemblies based on multi-walled systems can exhibit poorer transport performance than those based on SWCNTs, despite containing multiple concentric shells that can in principle carry current. 
This poorer performance does not contradict the established outer-shell-dominated picture of MWCNT transport, where current injection occurs mainly through the outer shells and intershell motion is tunnelling-limited rather than \color{black} providing \color{black} an independent parallel channel~\cite{Bourlon2004,Agrawal2007}. Instead, it raises a more specific question: if inner walls do not carry current as efficiently as outer walls, do they remain passive at junctions, or can their presence still reshape junction-level transmission? 

Magnetic fields provide a powerful probe of quantum transport in low-dimensional systems because they couple to orbital motion and modify phase coherence. In CNTs, axial magnetic fields have been widely used to study Aharonov--Bohm shifts of the nanotube subbands \cite{PhysRevLett.122.086802,PhysRevB.83.193407}. Perpendicular magnetic fields, however, remain much less explored for CNT junctions, even though they are directly relevant to high-field measurements on CNT films and fibres reaching 60\,T~\cite{BULMER2026121162,manuscriptAGA}. In this geometry the field does not simply probe an individual nanotube. Instead, it changes the relative phases of electronic paths across the intertube tunnelling region, making it a sensitive probe of junction-level interference and of the mechanisms controlling transport through CNT assemblies. 

To fully understand magnetotransport in CNT networks, we connect experimental observations with theoretical modelling. Building on our previous work linking CNT-assembly magnetotransport to junction-level transport limits~\cite{manuscriptAGA}, and motivated by experiments highlighting the reduced transport efficiency of multi-walled CNT systems, we carry out a rigorous theoretical study of charge transport through various CNT junction architectures relevant to real samples under a perpendicular magnetic field over a wide field range. Before turning to atomistic CNT junctions, we introduce a toy model of atomic chains to identify the interference motifs that can already arise in a coupled one-dimensional system and use them as a guide for interpreting the more complex CNT architectures. This minimal model picture then provides the basis for a coupled-waveguide interpretation of the CNT results, offering a physically intuitive description of coherent transport through CNT junction architectures. For simple SWCNT-SWCNT junctions, we derive and explain how the energy- and field-dependent transmission is governed by nanotube diameter, nanotube–nanotube overlap length, and nanotube chirality. We then extend the study to multiple SWCNT architectures, where additional finite-size and path-interference phenomena emerge beyond those of a single junction. Finally, we investigate double-walled CNT (DWCNT) \color{black}, triple-walled CNT (TWCNT) and four-walled CNT \color{black} junctions to determine how additional walls modify intertube coupling and, in turn, charge transmission. This allows us to rationalise why DWCNT junctions can remain close to the SWCNT case, whereas wider MWCNT junctions behave differently and provide poorer, more field-sensitive conductance despite containing more nominally conducting walls. In this way, the present study provides a physically grounded theoretical framework for identifying junction configurations that favour high, field-stable conductance in CNT-based networks, while turning microscopic interference mechanisms into design principles for macroscopic CNT conductors.

\section{Materials and methods}

\subsection{Experiment}

\subsubsection{CNT fibres spinning} 
CNT fibres used in this study were directly spun from a Floating Catalyst Chemical Vapour Deposition (FC-CVD) reactor as described previously~\cite{sundaram2011}. In brief, a carbon source and catalyst precursor (e.g., ferrocene) were continuously injected into a high-temperature reactor (temperatures above 1000\,$^o$C) together with a sulphur-containing promoter under a hydrogen carrier gas atmosphere. The catalyst decomposed in situ to form nanoscale iron particles, enabling CNT growth, while sulphur restricted the catalyst growth and facilitated CNT formation. The CNTs self-assembled into a continuous aerogel network within the gas phase, which was drawn from the reactor and spun into fibres. The as-spun fibres were densified in-line by an acetone spray and subsequently collected by winding onto a rotating collector. No further post-treatment was applied.

\subsubsection{CNT fibres \color{black} characterisation \color{black}} 
The surface morphology of the CNT fibres was examined using a scanning electron microscope (SEM) instrument (Zeiss, Gemini) with an accelerating voltage of 2.0\,kV . \color{black} The MWCNT fibre was additionally characterised using high-resolution transmission electron microscopy (HRTEM) with a Spectra 200 TEM.\color{black} The CNT fibres were \color{black} characterised \color{black} in terms of electrical resistance (digital multimeter 34461A, Keysight Technologies Inc., USA), mass (electronic microbalance UYA 2.4Y, Radwag, Radom, Poland), and length (Digital Microscope VX-S30B, Keyence International, Belgium). These parameters were used to determine the specific conductivity of the assemblies. Specific conductivity is a metric commonly applied to wet-spun assemblies \cite{LekawaRaus2014Carbon,LekawaRaus2014AFM}. Direct calculation of electrical conductivity based on \color{black} the assemblies' cross-sections \color{black} is usually unreliable due to the irregular and non-circular morphology. To avoid this limitation, conductivity was \color{black} normalised \color{black} using linear density, as reported in the literature.

Specific conductivity ($\sigma'$, in Siemens per \color{black} metre \color{black} over gram per cubic \color{black} centimetre \color{black}) was calculated according to:
\begin{equation}
	\sigma' = \frac{G \cdot L}{\mu_L} \cdot 10^{9}
\end{equation}
where $G$ is the electrical conductance (S), $L$ is the assembly length (m), and $\mu_L$ is the linear density (in tex=g/km). 

Magnetotransport measurements were performed in a four-probe configuration with a 3\,mm separation between voltage contacts, ensuring that the measured resistance reflects collective transport along the fibre. Individual CNT fibres were mounted on sapphire substrates with pre-patterned gold contacts and electrically connected using conductive silver paint. To minimise mechanical vibrations, the fibres were secured with an additional sapphire plate. Before \color{black} magnetoresistance \color{black} (MR) experiments, the samples' resistance was measured at room temperature using a Keithley 2000 multimeter.
Measurements were carried out at the National High Magnetic Field Laboratory (Los Alamos) using pulsed magnetic fields up to 60\,T applied perpendicular to the fibre axis. The sample was placed in a cryostat within a liquid-helium system, enabling temperature-dependent measurements down to 1.3\,K. Magnetoresistance was measured using an AC lock-in technique with excitation currents in the range of 30--310\,$\mu$A. Signal amplification and filtering were employed to extract the response under pulsed-field conditions. \color{black} Further \color{black} experimental details are given in Refs.~\cite{bulmer2017, BULMER2026121162, manuscriptAGA}.

The MR is defined as 100$\cdot$(R$_{XT}-$R$_{0T})$/R$_{0T}$\color{black}, where R$_{0T}$ is the resistance measured at temperature T and zero magnetic field \color{black} while R$_{XT}$ is the resistance measured at the same temperature under an applied magnetic field B.

\subsection{Modelling}

\begin{figure*}[t!]
	\centering
	\includegraphics[width=1.0\linewidth]{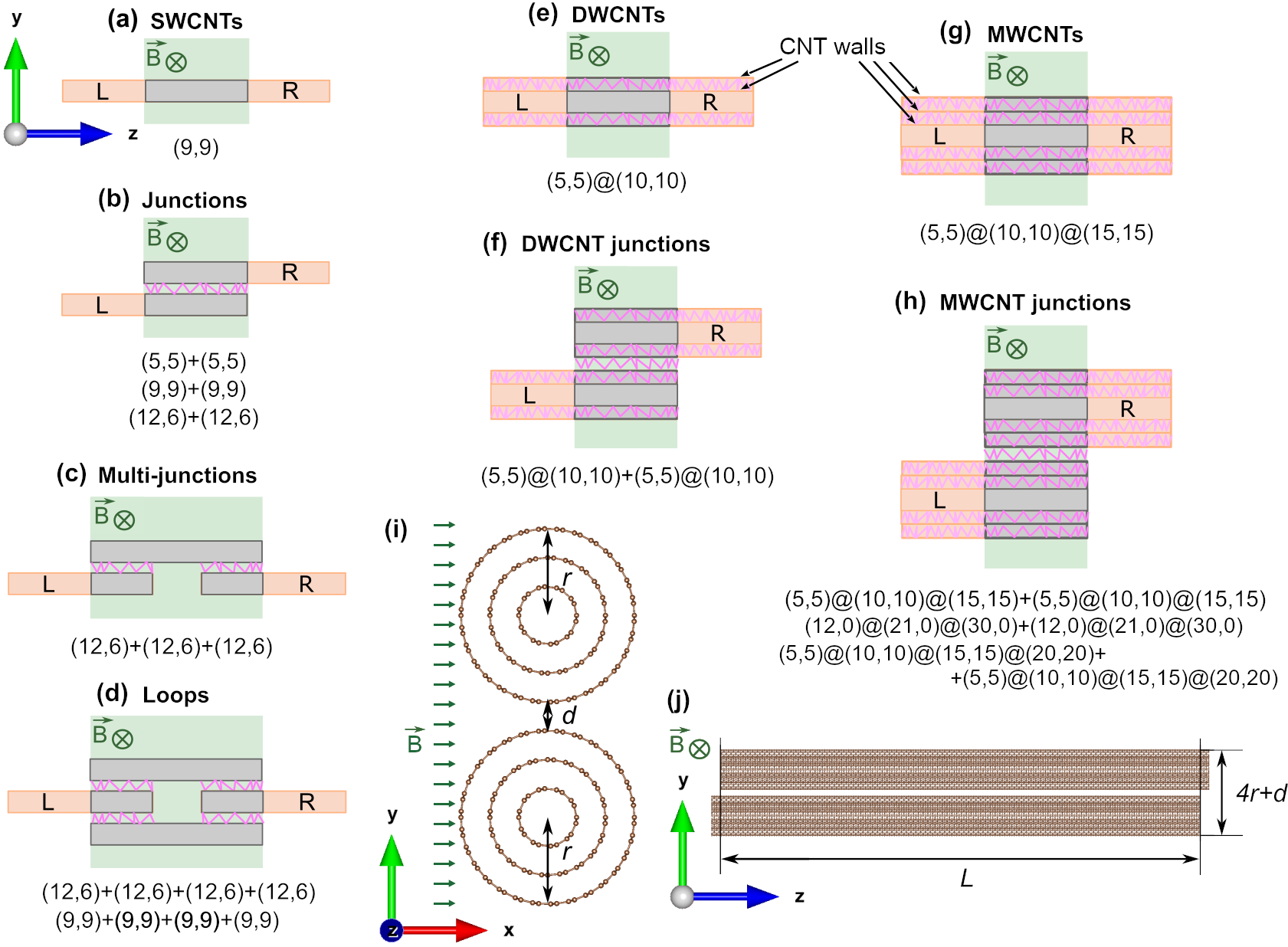}
	\caption{
		Models of  CNTs and CNT junction architectures considered in this work, with the chiralities of the constituent CNTs listed below.  Isolated CNT building blocks are shown for (a) SWCNT, (e) DWCNT and (g) \color{black} MWCNT\color{black}, with the corresponding junction architectures used in the calculations shown in (b--d), (f) and (h).  The pink zig-zag lines schematically indicate possible electron tunnelling paths between neighbouring SWCNTs, or between adjacent walls in MWCNTs. (i) Cross-sectional and (j) side views of the MWCNT junction with the radius $r$, intertube distance $d$, and overlap length $L$ marked.
	}
	\label{geometries}
\end{figure*}

We model representative metallic CNT junction architectures (Fig.~\ref{geometries}) relevant to experimentally investigated bundles and networks, without attempting to cover the full diversity of possible CNT contacts. Restricting the analysis to metallic nanotubes allows us to focus on the most conductive elements of CNT nanostructures and remains experimentally realistic, since metallic-enriched, chirality-enriched, and even single-chirality SWCNT samples can be obtained either by selective synthesis or by post-synthesis separation methods such as density-gradient ultracentrifugation, aqueous two-phase extraction, and chromatographic sorting, and assembled into films or networks~\cite{Haroz2010,Fagan2014,Yomogida2016,Gao2021}.
At cryogenic temperatures, the transport of semi-conducting CNTs freezes out\color{black}, \color{black} and all transmission is expected to go through networks of metallic CNTs~\cite{BULMER2026121162}. 

\begin{table}[h!]
	\small
	\caption{Basic structural parameters of SWCNTs used in this study}
	\begin{tabular*}{0.48\textwidth}{@{\extracolsep{\fill}}llll}
		\hline
		CNT type &(5,5) & (9,9) & (12,6) \\ 
		\hline
		Chiral angle $\theta$ & $30 \degree$ & $30 \degree$ & $19.1 \degree$ \\ 
		Diameter [\AA] & 6.78 & 12.21 & 12.44 \\
		Unit cell length [\AA] & 2.46 & 2.46 & 11.27 \\
		\hline
	\end{tabular*}
	\label{structure}
\end{table}

For SWCNT junctions, we consider CNTs with relatively small diameters (Table~\ref{structure}), such as $(5,5)$, $(9,9)$, and $(12,6)$, so that nanotube--nanotube overlap lengths up to 100\,nm remain computationally feasible. This choice is still experimentally relevant, since small-diameter and chirality-enriched SWCNT species can be isolated experimentally~\cite{Haroz2010,Fagan2014}. 
To isolate the role of additional shells, we consider DWCNTs using (5,5)@(10,10), as a distinct experimentally accessible CNT class bridging SWCNTs and MWCNTs~\cite{Moore2015}, and TWCNTs using (5,5)@(10,10)@(15,15) \color{black} and (12,0)@(21,0)@(30,0), as well as four-walled CNTs using (5,5)@(10,10) @(15,15)@(20,20) as progressively more complex \color{black} MWCNT \color{black} classes\color{black}~\cite{Fujisawa2016}. Multi-walled CNTs are denoted by the “@” symbol, with the innermost tube listed first and outer walls following in order.

All structures are treated as central scattering regions ($C$) connected to semi-infinite left ($L$) and right ($R$) electrodes (Fig.~\ref{geometries}), with each electrode built from the same type of CNTs as the corresponding junction. For armchair nanotubes\color{black}, \color{black} the electrodes are two unit cells long, whereas for chiral nanotubes they are one unit cell long. The carbon--carbon bond length is fixed at 1.42\,\AA, and the intertube separation in the junction is set to 3.356\,\AA, corresponding to the interlayer spacing in graphite~\cite{benedict1998}.

All junctions are subjected to an external perpendicular magnetic field (Fig.~\ref{geometries})  set in the $x$ direction, with the CNTs oriented along the $z$ direction. The junction geometry is oriented so that the intertube displacement lies along $y$, placing the magnetic field normal to the effective junction plane. \color{black} This corresponds to the maximum-flux orientation for a given junction geometry. \color{black}  In real fibres, a continuum of relative junction-to-field orientations is expected. Because the magnetic response scales with the effective flux through the junction region, the small diameters used in the SWCNT calculations require comparatively large fields to produce clear orbital effects~\cite{Nemec2006}. We therefore consider fields up to 200\,T, although non-destructive transport experiments are typically performed at lower fields up to about 12 T in most transport laboratories, and to about 60 to 100 T in \color{black} specialised \color{black} pulsed field facilities~\cite{Nanot2009,BULMER2026121162,manuscriptAGA}.
\color{black}{}

\subsubsection{Tight-binding Hamiltonian}

The formalism employed for \color{black} modelling \color{black} of electron transport through the CNT junctions is based on the empirical tight-binding (TB) approach. We consider one p-type orbital per carbon atom $i$ (indicated by $|i\rangle$), which corresponds to the dangling bond in the graphene sheet that forms the CNT.

We consider the basis \{$|i\rangle$\} to be non-orthogonal with the overlap matrix elements $s_{ij}$ being:
\begin{equation}
	s_{ij}=\langle i | j\rangle,
	\label{eq:S}
\end{equation}
and the Hamiltonian matrix elements:
\begin{equation}
	H_{ij} := \langle i|H | j\rangle = t_{ij},
	\label{eq:H}
\end{equation} 
where  $t_{ij}$ are hopping integrals between atomic sites $i$ and $j$ ($ i \ne j $), and are assumed to depend only on the distance $|\vec r_i - \vec r_j|$ between atoms $i$ and $j$ placed at atomic positions  $\vec r_i$ and $\vec r_j$, respectively. The diagonal matrix elements acquire the value that shifts the Fermi level to zero. We consider couplings between atomic orbitals up to the third \color{black} neighbours\color{black}. Both the hopping parameters and the overlap matrix elements for \color{black}a \color{black} non-orthogonal basis set are taken from Ref.~\cite{PhysRevB.81.245402} and depicted in \color{black} \color{black} Table~\ref{parameters}.

\begin{table}[h!tb]
	\small
	\caption{Tight-binding parameters used in this study. NN denotes nearest neighbour: 1st NN, 2nd NN and 3rd NN correspond to first-, second- and third-nearest-neighbour interactions, respectively}
	\begin{tabular*}{0.48\textwidth}{@{\extracolsep{\fill}}lll}
		\hline
		Hopping  &  Overlap matrix  &  Distance  \\ 
		parameters & elements & between atoms \\
		\hline		
		$t_1=-2.97$ eV  &    $s_1=0.073$ & $1^{\text{st}}$ NN, $d\leqslant1.5$ \AA \\
		$t_2=-0.073$ eV &    $s_2=0.018$ & $2^{\text{nd}}$ NN, $1.5$ \AA $<d\leqslant$ $2.5$ \AA\\
		$t_3=-0.33$ eV  &    $s_3=0.026$ & $3^{\text{rd}}$ NN, $2.5$ \AA $<d\leqslant$ $3.7$ \AA \\
		\hline
	\end{tabular*}
	\label{parameters}
\end{table}
The couplings between atoms separated by more than 3.7\,\AA\ are neglected. The inter-tube interactions are treated accordingly. 
This model describes \color{black} a \color{black} spin \color{black} unpolarised \color{black} system with the spin-orbit coupling neglected. Magnetic-field effects are included only through the orbital Peierls phase, and the Zeeman term is not considered. For electrons with a $g$-factor close to 2, the resulting spin splitting is not expected to change the orbital interference mechanisms that are the focus of this work.

The change of the Hamiltonian and overlap matrices ($H$ and $S$) under magnetic field is treated non-perturbatively and taken into account by Peierls substitution~\cite{Peierls1933, SaitoDresselhaus1998, PhysRevB.103.045402} that adds a phase factor to the hopping parameters and the overlap matrix elements, so that they transform in the following way:
\begin{equation}
	t_{ij}
	\;\rightarrow\;
	t_{ij}\exp\left(\frac{2\pi i}{\Phi_0} \int_{\mathbf r_i}^{\mathbf r_j} \mathbf A\cdot d\mathbf r\right),
	\label{eq:peierlsH}
\end{equation}
\begin{equation}
	s_{ij} \;\rightarrow\;
	s_{ij}\exp\left(\frac{2\pi i}{\Phi_0} \int_{\mathbf r_i}^{\mathbf r_j} \mathbf A\cdot d\mathbf r\right).
	\label{eq:peierlsS}
\end{equation}

The coordinates of the $i^{\text{th}}$ atom are $\mathbf r_i=(x_i,y_i,z_i)$, $\Phi_0=\frac{h}{e}$ is the flux quantum with $h$ being the Planck constant and $e$ the electron charge, and $\mathbf A$ is the magnetic vector potential. 
With this coordinate choice, we set $B_y=0$. First, for a single nanotube\color{black}, \color{black} this choice is not restrictive, because the tube is cylindrically symmetric with respect to rotations around the $z$ axis. Second, for a junction, the relative position of the two nanotubes breaks this symmetry. We choose the $y$ direction along the intertube displacement. A magnetic field parallel to the intertube displacement ($B_y\neq0$) would not cause an Aharonov--Bohm effect~\cite{PhysRev.115.485} between alternative paths and, therefore, would not change the transmission under magnetic field. The Peierls phase is evaluated using the actual atomic coordinates, so curvature and non-collinear intra- and intertube couplings are included. We choose the gauge leading to the vector potential $\mathbf A = (-y B_z,\, 0,\, y B_x)$ to preserve translational invariance along the $z$ direction. Then the integral in the phase factor takes the form:
\begin{equation}
	\int_{\mathbf r_i}^{\mathbf r_j} \mathbf A\cdot d\mathbf r
	=\frac{1}{2}(B_x\,\Delta z-B_z\,\Delta x)(y_i + y_j).
	\label{eq:phase}
\end{equation}
In these studies\color{black}, \color{black} we consider only \color{black}a \color{black} magnetic field perpendicular to the \color{black} nanotube \color{black} axis, {\it i.e.}\color{black}, \color{black} with $B_z = 0$. Then the path integral takes a simple form:
\begin{equation}
	\int_{\mathbf r_i}^{\mathbf r_j} \mathbf A\cdot d\mathbf r
	=\frac{1}{2}B_x\,\Delta z\,(y_i + y_j).
	\label{eq:phase-f}
\end{equation}

When considering the influence of the magnetic field on the electronic states of the CNT junctions, we neglect the Zeeman term. 
\color{black}{Its effect on transmission is straightforward -- it becomes a simple sum of the two spin channels at energies shifted by the Zeeman splitting. At our maximal fields of 200~T, for electrons with a $g$-factor close to 2, the energy splitting would be $\sim25$~meV, comparable to the distance between resonances in long junctions. The transmission would then appear more complicated, but it would still be only a sum of two independent spin contributions. The true complexity, caused by the ubiquitous interference, stems from the orbital effects.}

Solving the \color{black} generalised \color{black} eigenvalue problem with magnetic field dependent Hamiltonian and overlap matrices defined above, one finds the electronic states of the junction systems as a function of the external magnetic field. 

\subsubsection{Transport calculations}
\color{black} The device \color{black} consists of \color{black} a \color{black} left electrode $L$,\color{black} a \color{black}  scattering (central) region $C$, and \color{black} a \color{black} right electrode $R$. \color{black} The transmission \color{black} amplitude of an electron with energy $E$ between two electrodes $L$ and $R$ has been calculated employing non-equilibrium Green’s function (NEGF) formalism~\cite{Papior2017, PhysRevB.65.165401} and it reads (Landauer–Büttiker formula):
\begin{equation}
	T_{L\rightarrow R}(E)
	= \mathrm{Tr}\!\left[\mathbf \Gamma_R(E)\, \mathbf G_C(E)\, \mathbf \Gamma_L(E)\, \mathbf G_C^\dagger(E)\right],
	\label{eq:transmission}
\end{equation}
where $\mathbf \Gamma_{\alpha},\, \alpha=L,R$\color{black}, \color{black} is a matrix describing the coupling to the lead $L/R$:
\begin{equation}
	\mathbf \Gamma_{\alpha}(E)=i[\mathbf \Sigma_{\alpha}(E)-\mathbf \Sigma_{\alpha}^\dagger(E)],
	\label{eq:scattering}
\end{equation}
and $\mathbf G_C$ \color{black} is the \color{black} central region (dressed by self-energies) Green's function expressed in terms of its Hamiltonian $\mathbf H_C$, its overlap matrix $\mathbf S_C$, and \color{black} the \color{black} electrodes' self-energies $\mathbf \Sigma_{\alpha},\, \alpha=L,R$:
\begin{equation}
	\mathbf G_C^{-1}(E)= (E+i\eta) \mathbf S_C - \mathbf H_C - \mathbf \Sigma_L(E)-\mathbf \Sigma_R(E).
	\label{eq:G}
\end{equation}
Here $\eta\rightarrow0^+$. Self-energies account for the effect of the semi-infinite electrodes on the finite scattering region\color{black}, \color{black} and they arise from integrating out the leads' degrees of freedom. Each electrode $\alpha=L,R$ contributes one self-energy, which is calculated as:
\begin{equation}
	\mathbf \Sigma_{\alpha}(E)=(\mathbf H_{C\alpha}-E \mathbf S_{C\alpha}) \mathbf G_{\alpha}(E)(\mathbf H_{C\alpha}^{\dagger}-E \mathbf S_{C\alpha}^{\dagger}),
	\label{eq:sigma}
\end{equation}
with \color{black} the \color{black} lead Green's function:
\begin{equation}
	\mathbf G_{\alpha}^{-1}(E)= (E+i\eta) \mathbf S_{\alpha} - \mathbf H_{\alpha}
	\label{eq:Glead}
\end{equation}
computed via Sancho–Rubio recursion. \color{black} The total \color{black} Hamiltonian $\mathbf H$ and overlap \color{black} matrix \color{black} $\mathbf S$ matrices are:
\begin{equation}
	\mathbf H=
	\begin{pmatrix}
		\mathbf H_C & \mathbf H_{C\alpha} \\
		\mathbf H_{\alpha C} & \mathbf H_{\alpha}
	\end{pmatrix},\,
	\mathbf S=
	\begin{pmatrix}
		\mathbf S_C & \mathbf S_{C\alpha} \\
		\mathbf S_{\alpha C} & \mathbf S_{\alpha}
	\end{pmatrix},\,
	\alpha=L,R,
\end{equation}
where $\mathbf H_C$, $\mathbf S_C$ \color{black} are the \color{black} scattering region Hamiltonian and overlap \color{black} matrices\color{black}, $\mathbf H_{\alpha}$, $\mathbf S_{\alpha}$ \color{black} are the \color{black} semi-infinite electrode \color{black} Hamiltonian and overlap matrices\color{black}, \color{black} and \color{black} $\mathbf H_{C\alpha}$, $\mathbf S_{C\alpha}$ \color{black} describe the \color{black} coupling between \color{black} the \color{black} central region and \color{black} the \color{black} lead. $\mathbf H$ and $\mathbf S$ are Hermitian\color{black}; \color{black} thus\color{black}, \color{black} $\mathbf H_{\alpha C}=\mathbf H_{C\alpha}^{\dagger}$ and $\mathbf S_{\alpha C}=\mathbf S_{C\alpha}^{\dagger}$, which remains true when we modify these matrices by including the magnetic field via Peierls substitution. This can be done by including \color{black}the  changes in  \color{black} the Hamiltonian $\delta \mathbf H$, overlap $\delta \mathbf S$, and self-energies $\delta \mathbf \Sigma$ under \color{black} a \color{black} magnetic field in the Green's function:
\begin{equation}
	\mathbf G_C^{-1}(E)= (E+i\eta) (\mathbf S_C +\delta \mathbf S) - \mathbf H_C - \delta \mathbf H - \mathbf \Sigma_L(E) - \mathbf \Sigma_R(E) - \delta \mathbf \Sigma.
	\label{eq:dG}
\end{equation}
In the zero-temperature, low-bias limit, the conductance is obtained from the transmission evaluated at the electrochemical potential, $E=\mu$,  which can be shifted by doping in the structure.

In addition to the total transmission, we analyse its local decomposition into interorbital bond contributions, henceforth referred to as bond-resolved transmission contributions. These quantities describe how the coherent transmission is distributed over the orbital couplings $\nu$--$\mu$ for states injected from the left electrode, and are calculated as:
\begin{equation}
	\tau_{L,\nu\mu} = \frac{e}{h}\,\mathrm{Im}\!\left[ \mathbf{A}_{L,\nu\mu}\bigl(\mathbf{H}_{\mu\nu}-E\mathbf{S}_{\mu\nu}\bigr) - \mathbf{A}_{L,\mu\nu}\bigl(\mathbf{H}_{\nu\mu}-E\mathbf{S}_{\nu\mu}\bigr)
	\right]\color{black}, \color{black}
	\label{eq:spectral}
\end{equation}
where $A_{L,\nu\mu}$ is the spectral function associated with the left electrode. To quantify how the transmission is distributed among the constituent nanotubes of a junction, we define the tube-resolved transmission contribution, denoted by
$\tau_{\mathrm{tube}}$. For a selected nanotube, this quantity is obtained by summing $\tau_{\nu\mu}$ over all orbitals $\nu$ belonging to that nanotube and over all orbitals $\mu$ coupled to them:
\begin{equation}
	\tau_{\mathrm{tube}} = \sum_{\nu \in \mathrm{tube}} \sum_{\mu} \tau_{\nu\mu}.
	\label{eq:tau}
\end{equation}
In this way, $\tau_{\mathrm{tube}}$ measures the contribution of an individual nanotube to the total coherent transmission through the multi-walled nanotube or nanotube junction. As in the rest of the present TB--NEGF treatment,
inelastic processes such as electron--phonon scattering are not
included.

\subsubsection{Numerical procedure}
The calculation process performed in the present studies consists of the two main parts.  In the first part, performed using the \texttt{sisl} Python library version 0.15.1~\cite{Papior2023sisl}, the structures and their spin-\color{black} unpolarised \color{black} TB Hamiltonian and overlap matrices with magnetic field included via the Peierls substitution are generated. In the second part, dealing with the calculations of junctions' transport properties, the code incorporating NEGF formalism (\texttt{TBtrans} version 4.5.1~\cite{Papior2017}) is employed.

A long-standing issue in quantum-transport calculations under \color{black} a \color{black} magnetic field is the treatment of the field in the semi-infinite leads~\cite{datta1995}. In general\color{black}, \color{black} there is a non-zero magnetic field present in the lead\color{black}\color{black} which complicates the computations. However, it turns out that in many cases it is sufficient to assume that any magnetic field is present only in the scattering region.  The Peierls substitution makes \color{black} the  Hamiltonian \color{black} $\mathbf H$ and overlap \color{black} matrix \color{black} $\mathbf S$ complex\color{black}; \color{black} however, the open-source \texttt{sisl} and \texttt{TBtrans} codes support only \color{black} a \color{black} complex $\mathbf H$ matrix in the scattering (central) region. Therefore, we modified \texttt{sisl} v0.15.1 and \texttt{TBtrans} (based on \texttt{SIESTA} 5.2.0-alpha~\cite{Soler2002, Garcia2020}) to allow complex $\mathbf H$ and $\mathbf S$ \color{black} matrices \color{black} additionally in both electrodes and \color{black} a \color{black} complex $\mathbf S$ \color{black} matrix \color{black} in the scattering region. While for \color{black} a \color{black} parallel magnetic field that would be a problem due to the strong Landau level splitting in the electrodes, it turns out that the results for \color{black} a \color{black} perpendicular magnetic field are fairly independent \color{black} of  \color{black} whether one uses \color{black} a  \color{black} modified or non-modified version of \texttt{sisl} and \texttt{TBtrans} codes (see the benchmarking presented in Figure~\ref{bench} in Supplementary Material). However, the calculations with \color{black} the  \color{black} non-modified code version (\textit{i.e.}, without the magnetic field in \color{black} the \color{black} leads taken into account) are around 2-3 times computationally less expensive. Therefore, we use non-modified versions of \texttt{sisl} and \texttt{TBtrans} for the majority of performed computations. Only band structures in magnetic field are obtained using modified versions of the codes. In transport calculations with the \texttt{TBtrans} code, the energy spacing of 0.001 eV, the $1\times1\times1$ k-grid, and the electronic temperature of 300 K, \color{black} used as the Fermi--Dirac occupation parameter for numerical integration, \color{black} \color{black} were \color{black} chosen.

\subsubsection{Structural relaxation of selected CNT architectures}

\color{black}{The impact of structural relaxation on electronic transmission through selected DWCNT and TWCNT junctions was investigated via geometry optimisations performed with the density-functional theory SIESTA code \cite{Garcia2020, Soler2002} using the Atomistic Simulation Advanced Platform (ASAP) package \cite{SIMUNE2025, Marchesin2022}. The unit cells and atomic positions were fully optimised for both the isolated nanotubes (serving as electrodes) and the nanotube junctions (constituting the scattering region). Exchange-correlation effects were treated using the generalised gradient approximation of Perdew, Burke, and Ernzerhof (PBE) \cite{Perdew1996}, supplemented with the Grimme D3 correction \cite{Grimme2010} featuring rational Becke--Johnson (BJ) damping \cite{Grimme2011} to account for van der Waals interactions. Norm-conserving ONCVPSP pseudopotentials \cite{Hamann2013} in the PSML format \cite{Garcia2018} were obtained from the PseudoDojo library \cite{vanSetten2018}. A triple-zeta polarised (TZP) basis set was employed alongside a real-space mesh cutoff of 500 Ry. The Brillouin zone was sampled using 24 $k$-points along the nanotube axis \cite{Monkhorst1976}. Geometry optimisations were driven by the Broyden--Fletcher--Goldfarb--Shanno (BFGS) algorithm as implemented in the Atomic Simulation Environment (ASE) \cite{Larsen2017} until the maximum residual force fell below 0.01 eV/\AA{}.}

\color{black}{To obtain a qualitative overview of how structural relaxation affects the electronic transport, two-probe device structures were constructed by combining the optimised unit cells of the isolated lead nanotubes and the junction regions. Structural continuity across the device interface was maintained by matching the centres of mass of the lead nanotubes with those in the junctions and rotating the isolated nanotubes around their longitudinal axes to align their spatial orientations. The maximum deviations in atomic cross-sectional positions between the isolated nanotubes and those in the junctions remain within 0.25~\AA, with mean atomic displacements below 0.08~\AA. As a first-order approximation, omitting a transition region between the isolated leads and the junction geometries is reasonable due to the small magnitude of these displacements. Furthermore, local edge relaxations at the free nanotube ends within the scattering region were neglected under the assumption that transport is dominated by interwall tunnelling across the broad contact area. A fully quantitative treatment accounting for explicit lead-junction transition regions and reconstructed nanotube terminations remains a topic for future detailed investigation.}

\color{black}

\section{Results and discussion}

\subsection{Experimental comparison of SWCNT and MWCNT assembly conductivity}

\begin{figure*}[h!tb]
	\centering
	\includegraphics[width=0.98\linewidth]{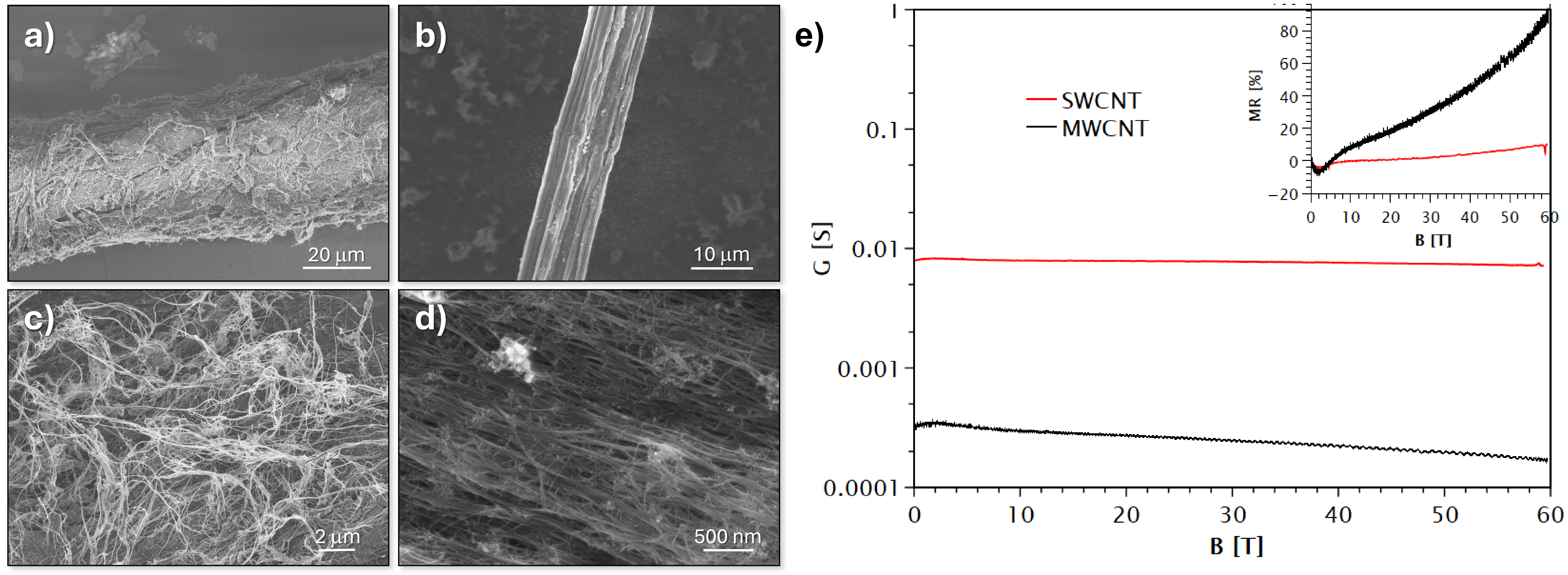}
	\caption{Morphology and magnetotransport of SWCNT and MWCNT fibres. (a--d) SEM images of CNT fibres directly spun from the FC-CVD reactor.  (a) The SWCNT-based fibre and (b) a fibre made of a mixture of MWCNTs of different numbers of walls. Corresponding higher-magnification images highlight the alignment of individual (c) SWCNTs and (d) MWCNTs within the fibres. (e) Change in conductance of a 3\,mm long MWCNT fibre sample against a magnetic field measured at 1.495\,K. Previously reported data~\cite{manuscriptAGA} for SWCNT fibre (3\,mm-long) measured at 1.3\,K are included for comparison. The inset shows magnetoresistance MR [\%] as a function of magnetic field for both samples.}
	\label{sem}
\end{figure*}

Single-walled (SWCNTs) and multi-walled carbon nanotubes (MWCNTs) are widely used as building blocks for macroscopic CNT assemblies, including films, fibres, printed conductive paths, and composite materials~\cite{lepak2024multifiller, LekawaRaus2014AFM, chen2020}. However, MWCNT-based structures often exhibit inferior electrical performance compared to those fabricated from SWCNTs~\cite{lepak2024multifiller, LekawaRaus2014AFM}. To examine this behaviour, two types of CNT fibres were compared: a predominantly SWCNT fibre synthesised using methane and carbon disulfide as feedstocks, and an MWCNT fibre produced from methane and thiophene. The morphology of both fibres is shown in Fig.~\ref{sem}. \color{black}Additional characterisation of the SWCNT sample is also presented in Ref.~\cite{sundaram2011}, while that of the MWCNT sample is provided in the SI (Fig.~\ref{HRTEM}). \color{black}  The SWCNT fibre exhibited a specific conductivity of (1.2$\pm$0.2)$\cdot$10$^6$\,S\,m$^{-1}$ (g\,cm$^{-3}$)$^{-1}$, whereas the MWCNT fibre showed a significantly lower value of (0.2$\pm$0.1)$\cdot$10$^6$\,S\,m$^{-1}$ (g\,cm$^{-3}$)$^{-1}$. Although the SWCNT fibre had a larger diameter (Fig.~\ref{sem}a) compared to the MWCNT fibre (Fig.~\ref{sem}b), both displayed a similar linear density of 0.04$\pm$0.01\,tex. The larger diameter of the SWCNT fibre is attributed to its lower degree of densification and poor axial alignment of CNT bundles, as evident in Figs.~\ref{sem}a and c. This supports \color{black} the \color{black} conclusion that SWCNT-based fibres possess superior intrinsic conductivity, which could be enhanced even more with improved densification.

Magnetic-field measurements provide an additional probe of this transport contrast (Fig.~\ref{sem}e). The conductance of the SWCNT fibre changes only weakly with magnetic field, showing only a small decrease even at 60\,T, whereas the MWCNT fibre exhibits a much stronger field-induced conductance reduction. This contrast is particularly evident in the MR curves shown in the inset, where a small low-field conductance enhancement is followed by a much faster high-field conductance suppression in the MWCNT fibre.

\subsection{Physics of the junction transport}
\label{analogies}

\subsubsection{Minimal model -- an atomic chain junction}

The oscillatory transmission behaviour that will appear later in CNT junction architectures as a function of electron energy, overlap length and perpendicular magnetic field is already present in a simpler toy model of two partly overlapping atomic chains connected to leads. 
In this minimal setting, the origin of the different transport behaviours can be identified more clearly.
\begin{figure*}[h]
	\includegraphics[width=\textwidth]{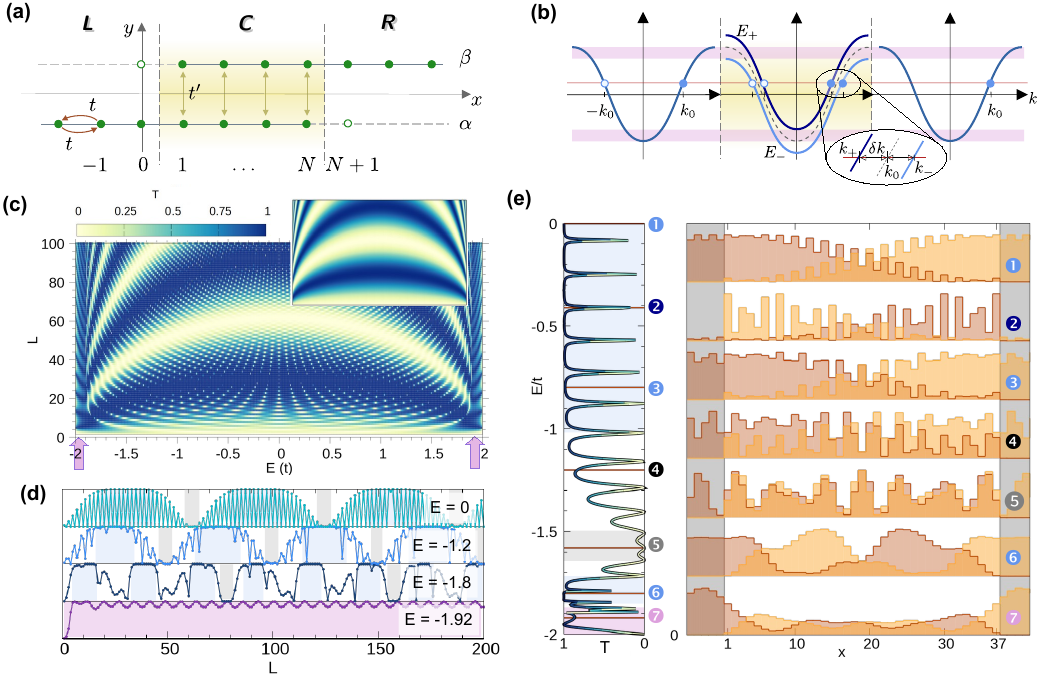}
	\caption{\label{fig:chain-junction-basic}
		Transmission and transport regimes in overlapping atomic chains. (a) Sketch of the system, with $t$ the hopping integral along each chain and $t'$ the hopping between chains $\alpha$ and $\beta$. (b) The energy bands of the individual chains and the two \color{black} hybridised \color{black} bands $E_\pm$ in the central region. The circles mark the incoming ($k_0$), reflected ($-k_0$) and transmitted ($k_0$ in $R$) momenta at a given energy, as well as the $\pm k_\pm$ at this energy in the junction. The light violet stripes mark the range of the {\em gateway states} (see text for description of the transport regimes). (c) Numerically calculated $T(E,L)$ -- transmission through the junction, as a function of energy $E$ and junction length $L$. The fine structure with fast oscillation reflects the variation of $k_0(E)$, while the broad pattern of crescents with enhanced and suppressed $T(E,L)$ is governed by $\sin^2(\delta k\,L)$ plotted in the inset. The violet arrows mark the range of the gateway state transport. (d) The line cuts of $T(E,L)$ at four values of $E$. The fast and slow oscillations are present in the two-band region only. (e) Transmission and local density of states (LDOS) along the junction with 37 cells, at seven values of $E$, rescaled by the maximum LDOS value. Five side nodes for each chain are plotted on a grey background. Violet background marks the range of {\em gateway state} transport, light blue marks the region of {\em forward current} transport\color{black}; \color{black} grey\color{black}, \color{black} the {\em standing waves} range. Dark blue marks the electron {\em trapped} in the junction.}
\end{figure*}
The tight-binding Hamiltonian of this system contains the hopping $t$ along the individual chains, denoted as $\alpha$ and $\beta$, and the coupling $t'$ between the chains in the region of the junction (see Fig.~\ref{fig:chain-junction-basic}(a)). Each of the left ($L$), central ($C$) and right ($R$) systems has translational symmetry in the bulk, so they can be described using the wavevector $k$. The energy bands of $L$ and $R$ chains are the same, and when the two chains overlap in $C$, their double degeneracy is removed by the inter-chain coupling $t'$, forming the fully \color{black} hybridised \color{black} bands $E_\pm$, which correspond to electronic states distributed equally over both chains in the central region, as shown in Fig.~\ref{fig:chain-junction-basic}(b).

The quantum mechanical transmission problem can be set up in the usual way, by building piecewise the wave function of an electron at energy $E$ and matching it at the two interfaces ($x=0$ and $x=L=N+1$). The boundary conditions in this case are those of continuity of the wave function on each chain at both interfaces and the continuity of the derivative along the unbroken chain at each interface. The wave function on the left ($L$) is a linear combination of incoming ($k_0$) and reflected ($-k_0$) states\color{black}; \color{black} on the right ($R$)\color{black}, it \color{black} corresponds only to the transmitted state $k_0$\color{black}; \color{black} and in the junction ($C$)\color{black}, it \color{black} contains the four momenta $\pm k_\pm$.  
Further details of the calculation are given in Sec.~\ref{supp:chain-analytics} of the Supplementary Material\color{black}; \color{black} here we show only a simplified formula for the transmission,
\begin{equation}
	\label{eq:chain-transmission-formula}
	T(E,L) = 64\,k_0^2\frac{\left[k_0\,\cos(k_0L)\,\sin(\delta k\,L)-\delta k\,\cos(\delta k\,L)\,\sin(k_0 L)\right]^2}{\left|\left(k_0+\delta k +e^{-2i\delta k L}(k_0-\delta k)\right)^2 -4 e^{-i2(k_0-\delta k)L} k_0^2\right|^2}.
\end{equation}
This result has been obtained with the assumption that $k_\pm = k_0 \pm \delta k$, as sketched in Fig.~\ref{fig:chain-junction-basic}(b). It does not hold near the band edges, but is a good approximation near the \color{black} centre \color{black} of the bands. The transmission depends on energy through $k_0(E)$ and $\delta k(E)$. Note that for small $t'$, similar to the intertube couplings in the later sections, $\delta k \ll k_0$ and the dominant term in the numerator of Eq.~\ref{eq:chain-transmission-formula} is the first one, proportional to $k_0^2$. The analytical results have been tested against numerical calculations with \color{black} our \color{black} own Green's function code written using the Armadillo library~\cite{sanderson2025}. The numerical results for $T(E,L)$ shown in Fig.~\ref{fig:chain-junction-basic}(c) agree well with Eq.~\ref{eq:chain-transmission-formula} (for comparison see Supplementary Material).
Its most prominent \color{black} features\color{black}, the large scale crescents of enhanced and suppressed transmission, are due to the $\sin^2(\delta k\,L)$ factor in the numerator, plotted for illustration in the inset. \\ 
Depending on the junction length and the energy at which we calculate the transmission, we can distinguish several different transport regimes. 

{\em Forward current}. Transmission is strongly enhanced when the boundary conditions can be satisfied by a linear combination of $+k_+$ and $+k_-$ only -- both terms carry only the forward momentum\color{black}; \color{black} the current flows through the junction almost uninterrupted, and transmission is \color{black} almost \color{black} equal to 1, as in Fig.~\ref{fig:chain-junction-basic}(c). In Fig.~\ref{fig:chain-junction-basic}(d) we see it for $E/t=0,-1.2,-1.8$ as $T(L)\approx 1$ plateaux. In Fig.~\ref{fig:chain-junction-basic}(e) cases 1, 3 and 6 belong to this regime.

{\em Standing waves}. In contrast, the large scale suppression of transmission occurs when the phase difference $2\delta kL$ between the forward propagating states with $k_0+\delta k$ and $k_0-\delta k$, gathered upon the transit through the junction, is a multiple of $2\pi$. Then, since the boundary condition forces the wave function on the chain $\beta$ to vanish at its left end, the wave function \color{black} also vanishes \color{black} at its right end -- at the exit from the junction. Transport is then suppressed by the destructive interference of the $k_+$ and $k_-$ waves. This feature can be seen in the LDOS for case 5 shown in Fig.~\ref{fig:chain-junction-basic}(e), where at $E/t=-1.62$ a standing wave pattern forms on both chains, with minima at both ends, and transmission is accordingly suppressed. 

{\em Trapping states}. The underlying fast oscillation of the transmission is due both to $\cos^2(k_0L)$ in the numerator and to the divergences in the denominator. When $k_0 L\approx \pi/2$, the standing wave pattern forms again, this time due to trapping the electron inside the junction, as can be seen at $E=-0.4t$ (case 2) in Fig.~\ref{fig:chain-junction-basic}(e). 

{\em Gateway states}. Especially interesting is the range of energies where the \color{black} travelling \color{black} states belong to only one band, marked in Fig.~\ref{fig:chain-junction-basic} in light violet. If this was a single band, the mismatch of wave vectors between the central and the side regions would lead to a perfect reflection. However, the presence of the second band provides the system with a crucial ingredient: evanescent states with imaginary wave vectors, whose presence at the interfaces facilitates the entry of an electron into (and its exit from) the junction. Such gateway states are known from molecular electronics, where they form at the groups linking the central molecule to the leads~\cite{joachim2005,sangtarash2018}. 
They facilitate the transport in the energy windows $|E|\in(|2t|-|t'|,|2t|)$, as can be seen in Fig.~\ref{fig:chain-junction-basic}(e) at $E=-1.92\,t$.

\begin{figure*}[h]
	\includegraphics[width=\textwidth]{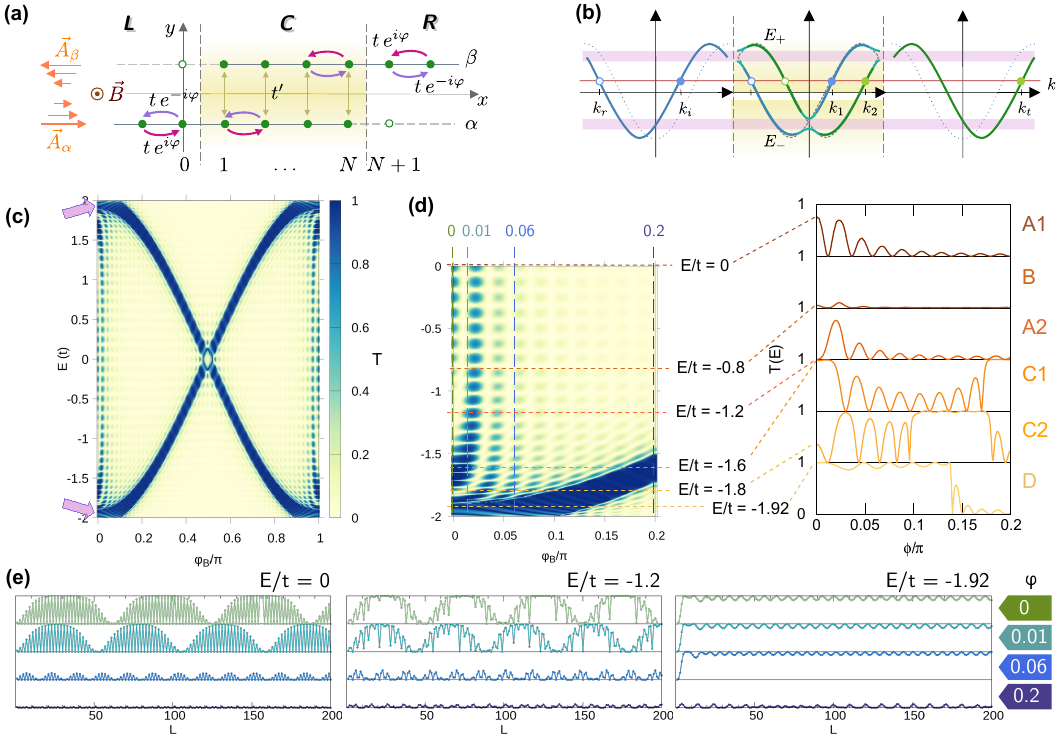}
	\caption{\label{fig:chain-junction-magnetic}
		Transport across an atomic chain junction in \color{black} a \color{black} perpendicular magnetic field. (a) Sketch of the system with the orientation of $\vec{B}$ and the vector potential $\vec{A}(y)$. Hopping integrals along $x$ are modified by the Peierls phase, opposite on chains $\alpha$ and $\beta$. (b) The electronic bands are shifted by the Peierls phase, $+\varphi$ on chain $\alpha$ and $-\varphi$ on chain $\beta$. In the junction\color{black}, \color{black} the shifted bands cross only at $k=0,\pi$, and only in the \color{black} neighbourhood \color{black} of these points \color{black} can \color{black} the electronic states of the two chains can \color{black} hybridise\color{black}. These energy ranges are also those where transport is facilitated by gateway states, and with increasing $\vec{B}$ they shift towards $E=0$. (c) Numerically calculated $T(E,\varphi)$ for a junction with 49 unit cells. Light violet arrows mark the regions of transport facilitated by the gateway states, flowing through the band. (d) Zoom into the low $\varphi$ and $E$ ranges, with line cuts at chosen values of $E$. (e) $T(L)$ for three values of energy and four values of the Peierls phase, marked in (d).	}
\end{figure*}

The orbital effect of the magnetic field perpendicular to the junction is described by the Peierls phase modifying the hopping. An analytical calculation and further details can be found in Sec.~\ref{supp:chain-magnetic} of the Supplementary Material, here we \color{black} summarise \color{black} our main results. In the gauge we chose to keep the translational invariance, hoppings along the chain are modified while the interchain hoppings are not, as shown in Fig.~\ref{fig:chain-junction-magnetic}(a). The Peierls phases on the two chains lying on both sides of the $x$ axis are opposite, and the dispersions in regions $L$ and $R$ are phase-shifted in opposite directions, as depicted in Fig.~\ref{fig:chain-junction-magnetic}(b). In consequence, there are only small energy and momentum ranges in the junction where the electronic states of both chains can hybridize. For mismatched $L$ and $R$ dispersions \color{black} hybridisation \color{black} opens a band gap, in which also only one band is present while the second provides the gateway states. For small interchain coupling the band mismatch reduces the \color{black} hybridisation \color{black} very quickly, resulting in strong suppression of the transmission everywhere except for the {\em gateway states} range, as shown in Fig.~\ref{fig:chain-junction-magnetic}(d)\color{black} and \color{black}(e). The resulting $T(\varphi)$ curves can take various shapes, depending on the examined energy range. Type A behaves like $|\sin(\varphi)/\varphi|^2$ (rounded maxima and minima) with initial suppression (A1) or enhancement (A2) of transmission by the magnetic field. Type B corresponds to a trapping state throughout the whole range of $\varphi$. Type C crosses a range of gateway states where the dimensionless transmission is nearly unity and almost constant; when entering or leaving this range\color{black}, \color{black} $T(\varphi)$ behaves like $|\sin(\varphi)/\varphi|$ with rounded peaks but sharp minima. Type D lies initially within the gateway states range and displays clear oscillations at high $T$.  The most interesting observation here is that at small field values its application can initially enhance the transmission, as we see in the range of $E\in(-1.4t,-t)$ in Fig.~\ref{fig:chain-junction-magnetic}(d), or even  bring the dimensionless transmission close to unity within a wide range of applied fields when the junction is driven into the gateway states regime. 

\subsubsection{Nanotubes as electron waveguides}

The minimal model discussed above provides a natural bridge to the real systems considered in this work, namely CNT junctions (Fig.~\ref{geometries}). 
Because \color{black} neighbouring \color{black} nanotubes in the overlap region can exchange the amplitude of the electronic wavefunction through intertube coupling, such junctions can be viewed as electronic analogues of coupled waveguides~\cite{yariv1973}. 
Within this picture, the left electrode corresponds to the input waveguide in which the electronic wave is launched, and the right electrode to the output waveguide in which the transmitted wave is collected. The overlap region forms the coupling section, with its finite length determining the relative phase of the interfering partial waves.

This viewpoint applies not only to SWCNT--SWCNT junctions, but also to DWCNT--DWCNT and MWCNT--MWCNT junctions, where transmission likewise involves coupling between neighbouring tubes or neighbouring walls. Intertube coupling then causes an oscillatory exchange of electronic-wavefunction amplitude between neighbouring nanotubes, analogous to the evanescent coupling of light between two nearby optical waveguides.

Although the analogy is not exact, with electrons propagating in a crystalline lattice with a specific band structure and light propagating in a continuum dielectric medium, the coupled-wave picture captures the main features of coherent transmission through a CNT junction. In particular, multiple reflections within a finite overlap region give rise to Fabry-P\'erot-like interference, while in some junction geometries the magnetic-field dependence of transmission will be interpreted in terms analogous to Fraunhofer diffraction.

\subsection{Magneto-transport through single SWCNT--SWCNT junctions}
\label{SWCNT-junctions}

With the chain toy model and the coupled-wave picture in mind, we first decode the physics of single SWCNT--SWCNT junctions (type A systems), which constitute the basic transport element of CNT assemblies~\cite{manuscriptAGA}. In what follows, we derive the relations governing the energy- and magnetic-field-dependent transmission in terms of nanotube diameter, chirality, and nanotube--nanotube overlap length. We therefore consider three metallic--metallic SWCNT--SWCNT junctions, (5,5)+(5,5), (9,9)+(9,9), and (12,6)+(12,6). Here, (5,5) and (9,9) CNTs have the same chirality but different diameters, whereas (9,9) and (12,6) CNTs have similar diameters but different chiralities.

\subsubsection{Oscillations of transmission with electron energy}\label{sec:T(E)}

\begin{figure*}[h!]
	\centering
	\includegraphics[width=0.95\linewidth]{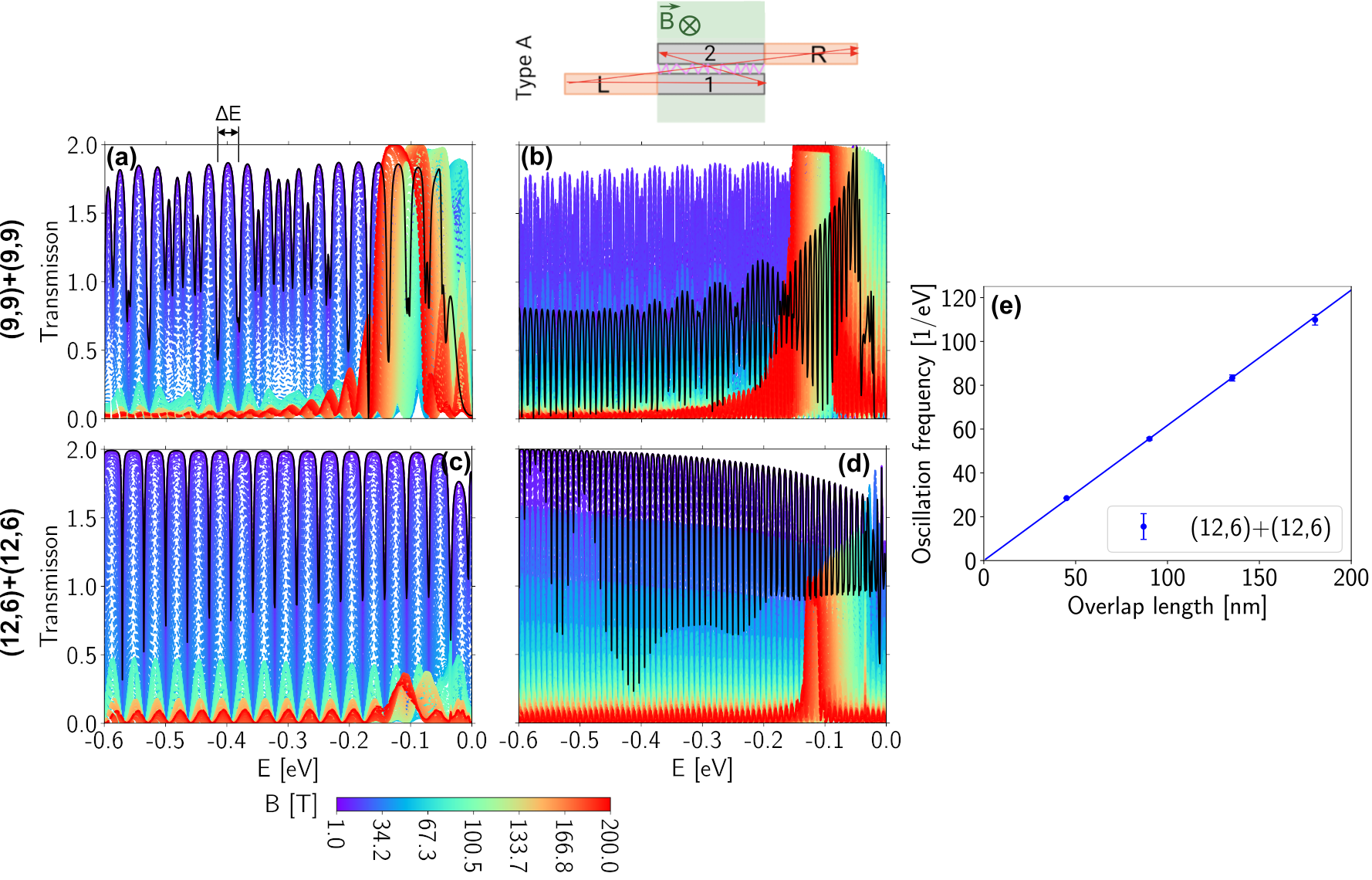}
	\caption{ Electron transport through simple junctions of single-walled CNTs (type A systems) under an external perpendicular magnetic field. The junction schematics above the panels show some of the possible electron paths (red arrows). (a--d) Oscillations of the computed zero-bias transmission as a function of electron energy $E$ for (9,9)+(9,9) SWCNT junctions with overlap lengths of (a) 49.19\,nm (200 unit cells) and (b) 196.76\,nm (800 unit cells), and for (12,6)+(12,6) SWCNT junctions with overlap lengths of (c) 45.08\,nm (40 unit cells) and (d) 180.33\,nm (160 unit cells), shown for magnetic fields $B$ up to 200\,T in the energy window [-0.6,0]\,eV. (e) Transmission oscillation frequency, defined as $f_h =1/\Delta E$, where $\Delta E$ is the energy spacing between adjacent transmission minima, as a function of the junction overlap length $L$ for \color{black} the \color{black} (12,6)+(12,6) junction. The blue solid line represents $f_h=L/\beta$, with $\beta=\pi\hbar v_F=1.62\,\mathrm{eV}\cdot\mathrm{nm}$ calculated using $v_F=0.7834\cdot10^{6}\,\mathrm{m\,s^{-1}}$ for the (12,6) CNT. 	}
	\label{transmission_zoom}
\end{figure*}

To understand transport through the junctions, we analyse the transmission function given by Eq.~\ref{eq:transmission}, which at zero temperature (at $T=0$) determines the linear conductance. Within the coherent Landauer–Büttiker formalism, the transmission is an energy-dependent quantity, $T(E)$, obtained as the trace over all transmitting channels. At zero temperature, the zero-bias conductance is given by $G=G_{0}T(E_F)$, where $G_{0}=2e^{2}/h$, \color{black} and \color{black} $E_{F}$ is the Fermi level. Varying the Fermi level therefore mimics electrostatic gating or chemical doping and makes the transport response explicitly dependent on electron energy. We therefore begin by analysing the transmission spectra of simple SWCNT--SWCNT junctions as functions of $E$. 

Figure~\ref{transmission_zoom}(a--d) shows the computed transmission spectra for (9,9)+(9,9) and (12,6)+(12,6) metallic--metallic SWCNT junctions in the experimentally relevant $p$-doped range from $E=0$\,eV (undoped system) to $E=-0.6$\,eV (nitric-acid-treated CNTs~\cite{Hayashi2016}), and for magnetic fields up to 200\,T. For each junction, we show results for two overlap lengths, denoted as \textit{short} and \textit{long}, with the \textit{long} junction having an overlap region four times longer than the \textit{short} one. In Fig.~\ref{transmission} we show these results for a wider energy range from -1.5~eV to 1.5~eV. 
In a junction, charge transport is no longer governed solely by coherent propagation along a continuous nanotube. Electrons must tunnel between neighbouring nanotubes across the overlap region, and the intertube coupling strongly modulates the transmission compared with that of an isolated nanotube~\cite{manuscriptAGA}. Instead of the step-like transmission of a continuous nanotube (cf. Fig.~\ref{99_single}(a)), the transmission spectra contain sharp resonances near the van Hove singularities (cf. Fig.~\ref{transmission}(a,b)) and broad regions with pronounced oscillations visible in Fig.~\ref{transmission_zoom}. 

The transmission spectra in Figure~\ref{transmission_zoom}(a--d) show fairly regular oscillatory behaviour as a function of electron energy, with several features resembling those of the minimal-chain model. Transmission maxima occur when the electronic wave is transferred efficiently through the overlap region, analogous to the forward-current regime. The frequent local minima can instead be associated with trapping states in the finite junction region. In the \textit{short} (9,9)+(9,9) junction, Figure~\ref{transmission_zoom}(a), the high-field curves close to the Fermi level reach the largest transmission values, indicating the dominance of gateway-state transport. This gateway-state contribution becomes more pronounced with increasing overlap length, as seen in Figure~\ref{transmission_zoom}(b,d).

To quantify these oscillations, we focus on the dominant high-frequency component, $f_h$. In analogy with optical spectra, where an oscillatory signal can be characterised by a spectral period and its inverse, frequency, we define the energy period $\Delta E$ as the spacing between adjacent transmission maxima, as indicated in Figure~\ref{transmission_zoom}(a), and use $f_h=1/\Delta E$. The spectra
also show a lower-frequency modulation of the dominant oscillation, which we denote qualitatively as $f_l$. 
For the \textit{short} (9,9)+(9,9) junction (Figure~\ref{transmission_zoom}(a)), this modulation is clearly visible and makes the extraction of a unique $\Delta E$ less reliable. For the \textit{short} (12,6)+(12,6) junction (Figure~\ref{transmission_zoom}(c)), the modulation appears mainly as a gradual decrease of the $f_h$ maxima when approaching the Fermi level, and is easier to see over the wider energy range shown in Figure~\ref{transmission} (c). For the \textit{long} junctions (Figure~\ref{transmission_zoom}(b,d)), the envelope modulation becomes more pronounced, while the $f_h$ oscillations become denser. The extracted values of $f_h$, shown in Figure~\ref{transmission_zoom}(e), therefore reveal a clear dependence on the overlap length $L$. This trend can be understood from the following simple phase-accumulation argument.

Near the Fermi level, metallic CNTs exhibit a Dirac-cone-like linear dispersion, $E-E_F=\hbar v_F(k-k_F)$~\cite{Malysheva2008,Kuchment2007}, where $v_F$ is the Fermi velocity and $k_F$ is the wave vector at the band-crossing point. Taking $E_F=0$ and defining $\tilde{k}=k-k_F$, this becomes $E=\hbar v_F\tilde{k}$.
Within the junction, electrons can bounce backward and forward from the edges of the junction, as schematically depicted above the panels in Figure~\ref{transmission_zoom}, and form standing-wave-like confined states that correspond to the local energy levels of electrons confined in the junction region of length $L$, which are the same as the trapping states in the minimal model. The wave vector of the confined electrons can only take discrete values $\tilde{k}_n=n\pi/L$, with $n$ being an integer. The transmission of incident electrons from the left to the right lead through the junction is the least efficient when their wave vectors, and therefore their energies, are close to those of the confined states. Therefore, minima of the transmission should be expected for energies $E_n=\hbar v_F\tilde{k}_n=\hbar v_F n\pi/L$.  The energy distance between two adjacent maxima, \textit{e.g.}, for $n+1$ and $n$, is then $\Delta E=\hbar v_F\pi/L$. This determines the high-frequency component of the transmission oscillations with energy as $f_h=L/\beta$, with $\beta=\hbar v_F\pi$, so that $f_h$ exhibits a linear dependence on the overlap length $L$. This linear dependence is confirmed in Figure~\ref{transmission_zoom}(e), where the extracted $f_h$ values for the (12,6)+(12,6) junctions follow $f_h=L/\beta$. As to the low frequency modulation $f_l$, it can have two origins:  one is a mismatch between the momenta from the \color{black} hybridised \color{black} branches of the dispersion $\delta k$, analogous to those in the minimal model (cf. Fig.~\ref{fig:chain-junction-basic}(b)). Another possible origin is the mismatch between the velocities of two forward-propagating branches of the electronic dispersion in the CNTs, so-called Sagnac interference observed in single-CNT devices.~\cite{dirnaichner2016,lotfizadeh2021} 

These results challenge the common design intuition that the highest conductance should be obtained simply by using highly aligned, defect-free metallic nanotubes. Complete de-doping, even in nominally monochiral metallic samples, is unlikely to provide the most robust route to high conductance. In the simple junctions studied here, the undoped Fermi level does not generally coincide with the highest-transmission region of a spectrum (Figure~\ref{transmission_zoom}). 
Light or moderate $p$-doping is therefore expected to be more favourable, particularly when the material must remain conductive under \color{black} a \color{black} high perpendicular magnetic field. This conclusion is consistent with conductivity enhancement reported for chemically doped, electronic-type-separated SWCNT films~\cite{Puchades2015}, but should be understood statistically rather than as a precise tuning rule. In macroscopic assemblies, the local doping level and the overlap length are both distributed, so different junctions sample different parts of the oscillatory transmission spectrum. Since $\Delta E=\pi\hbar v_F/L$, longer overlaps make the transmission maxima and minima more closely spaced in energy. Highly aligned samples with extended parallel contacts may therefore become more sensitive to small variations in local doping, which can reduce the averaged transmission through the assembly.

\subsubsection{Oscillations of transmission with overlap length}

\begin{figure*}[ht!b]
	\centering
	\includegraphics[width=1.0\linewidth]{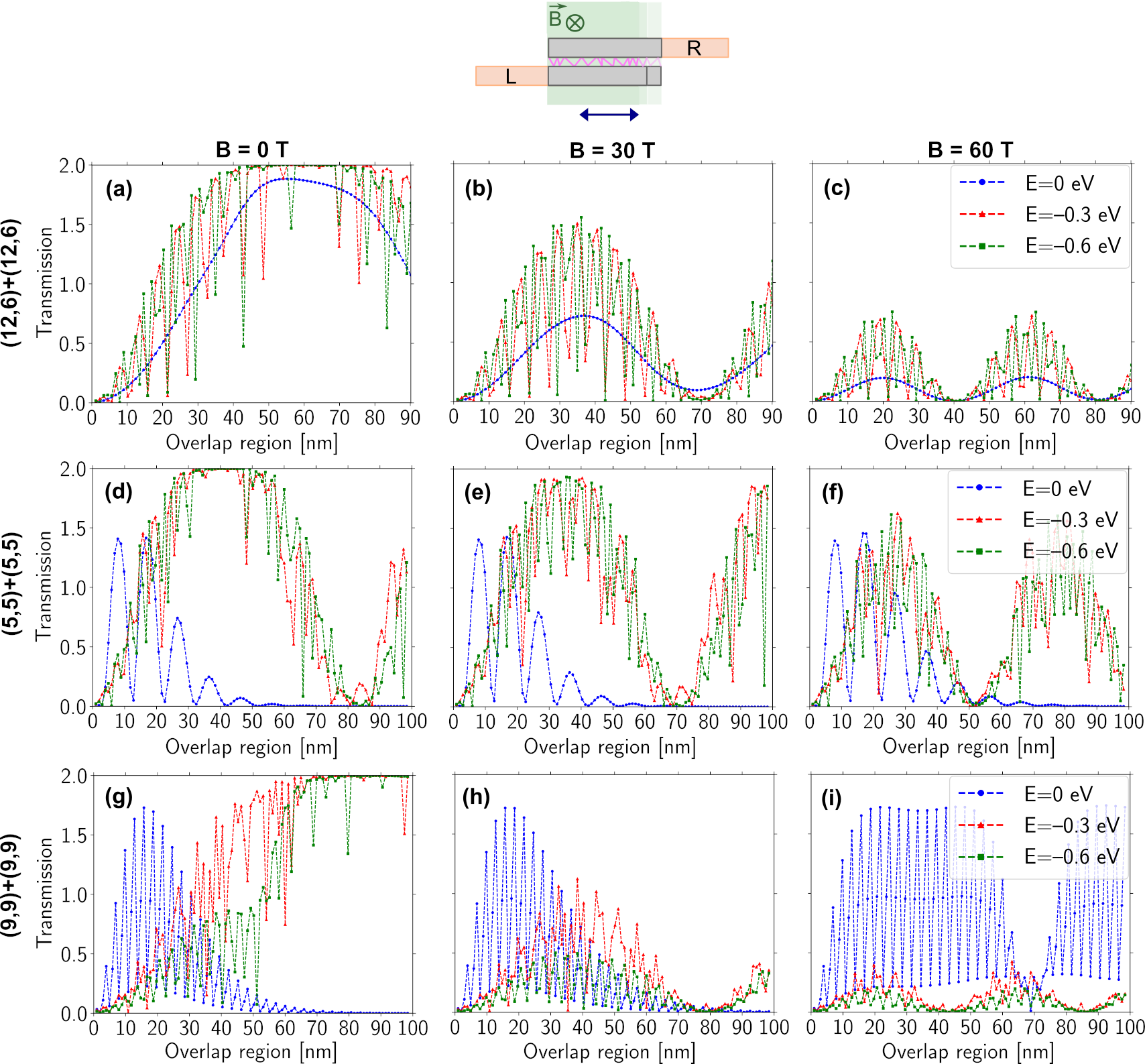}
	\caption{
		Electron transport through simple SWCNT--SWCNT junctions under an external perpendicular magnetic field. Computed zero-bias transmission as a function of overlap length in the range 0--100\,nm for (a--c) (12,6)+(12,6), (d--f) (5,5)+(5,5), and (g--i) (9,9)+(9,9) junctions, shown for $B$=0\,T, 30\,T, and 60\,T, and for electron energies $E=0$, -0.3\,eV, and -0.6\,eV.
	}
	\label{osc_swcnt}
\end{figure*}

We next quantify the role of the overlap length $L$ in magnetotransport through type A SWCNT--SWCNT junctions. Figure~\ref{osc_swcnt} presents the computed zero-bias transmission as a function of $L$ in the range 0--100\,nm for the
(12,6)+(12,6), (5,5)+(5,5), and (9,9)+(9,9) junctions, shown for $B=0$, 30, and 60\,T, and for electrochemical potentials $\mu=0$, $-0.3$, and $-0.6$\,eV.

The transmission shows a strong oscillatory dependence on the overlap length $L$ for all three SWCNT--SWCNT junctions. This behaviour closely follows the interference picture introduced in the chain toy model (Figure~\ref{fig:chain-junction-basic}). Fast local oscillations are superimposed on a broader envelope: the local minima can be
associated with trapping-state-like suppression inside the finite junction region, while the large envelope minima correspond to the standing-wave regime, where destructive interference suppresses transmission through the overlap region
(Figure~\ref{fig:chain-junction-basic}(e)). The envelope maxima, in turn, correspond to forward-current-like transport, where the electronic wave is transferred efficiently through the junction. This interpretation is consistent with previous zero-field studies, where overlap-dependent conductance oscillations in parallel CNT contacts were attributed to intertube coupling and quantum interference of incident, reflected and transmitted waves~\cite{Tripathy2016}. 
A closely related standing-wave picture was also reported for telescopic DWCNTs, where changing the overlap length switches interlayer transport between nearly perfect, partial and blocked transmission, with the blocked cases associated with \color{black} localised \color{black} states in the overlap region -- our trapping states~\cite{wittermeier2022}.
At $B=0$, the armchair junctions studied here also show strong suppression of transmission for sufficiently long overlaps, consistent with the zero-field conductance quenching predicted for long parallel CNT contacts~\cite{Xu2013}, which arises from intertube-coupling-induced gap opening and destructive interference. In practice, the appearance of this quenching depends on the specific junction Hamiltonian and on the initial relative orientation between the two nanotubes, as discussed previously for CNT assemblies~\cite{manuscriptAGA}.

\begin{figure*}[h!tb]
	\centering
	\includegraphics[width=0.9\linewidth]{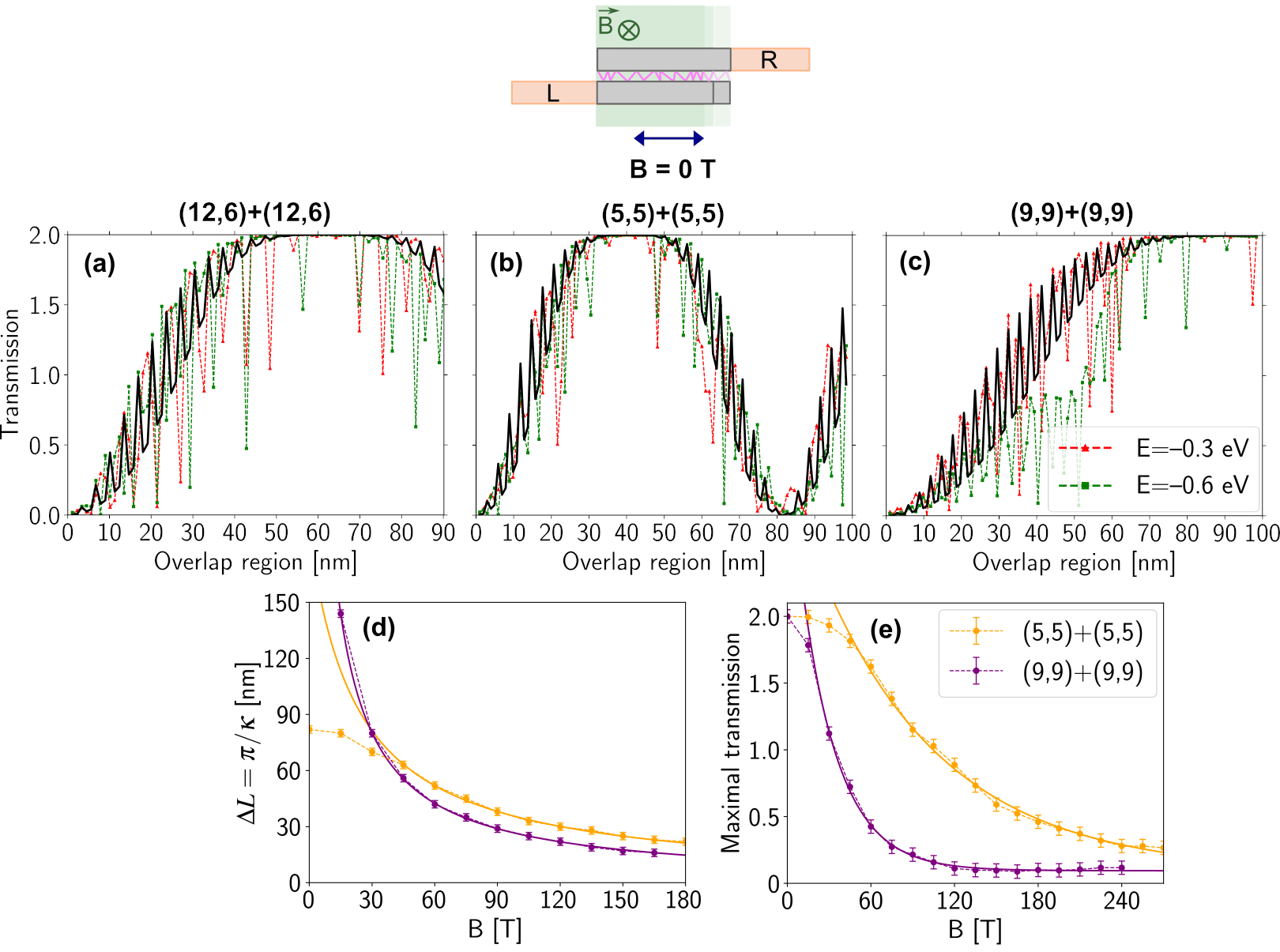}
	\caption{Electron transport through simple SWCNT--SWCNT junctions under an external perpendicular magnetic field. Comparison between the computed zero-bias transmission and the fitted expression given in equation~\eqref{eq:yakobson} (solid black line) as a function of overlap length at $B=0$\,T and $E=-0.3$\,eV for (a) (12,6)+(12,6), (b) (5,5)+(5,5), and (c) (9,9)+(9,9) junctions, with $\varkappa=(40\,\mathrm{nm})^{-1}$, $(26\,\mathrm{nm})^{-1}$, and $(60\,\mathrm{nm})^{-1}$, respectively.  (d) Oscillation period $\Delta L$ as a function of magnetic field $B$ in the range 0--180 T at $E=-0.3$ eV for (5,5)+(5,5) (orange dots) and (9,9)+(9,9) (purple dots) junctions. The solid lines are fits to $\pi/(a\cdot B+b)$, with $a=0.00073$ and $b=0.01675$ for the (5,5)+(5,5) junction ($\chi^2=0.78$), and $a=0.00116$ and $b=0.00443$ for the (9,9)+(9,9) junction ($\chi^2=0.39$). The fit for the (5,5)+(5,5) junction was performed for $B\geq45$ T. (e) Exponential decay of the maximal value of transmission with magnetic field at $E=-0.3$ eV for the same junctions, fitted to $a\cdot e^{-b\cdot B}+c$, with $a=2.9306$, $b=0.0112$, $c=0.0874$ for the (5,5)+(5,5) junction ($\chi^2=2.45$), and $a=3.1785$, $b=0.0372$, $c=0.0943$ for the (9,9)+(9,9) junction ($\chi^2=1.48$). The fit for the (5,5)+(5,5) junction was performed for $B\geq45$ T, while for (9,9)+(9,9) $B\geq30$ T. }
	\label{osc_swcnt_0}
\end{figure*}

At $B=0$, away from this quenched undoped regime, the doped traces $\mu=-0.3$ and $-0.6$\,eV show the two-scale overlap-length oscillations reported for CNT  contacts~\cite{Xu2013,Tripathy2016}, with a short-period component superimposed on a broad envelope. To rationalise these overlap-dependent oscillations, we model the junction as two coupled electron waveguides, with the intertube coupling described by an effective parameter $\varkappa$. This gives an analytical expression for the zero-field transmission as a function of overlap length, equivalent to the formula introduced in
Ref.~\cite{Xu2013}, and derived here using a coupled-wave picture in the Supplementary Material (Section~\ref{sec:kappa_derivation}). We use this expression below to fit the computed spectra:
\begin{equation}
	T(L)=\frac{2\sin^2(\varkappa L)[1+\cos(2kL)]}{1+2\cos(2kL)\sin^2(\varkappa L)+\sin^4(\varkappa L)}.
	\label{eq:yakobson}
\end{equation}
Equation~\eqref{eq:yakobson} contains two characteristic length scales. The term $\sin^2(\varkappa L)$ gives the long-period oscillation, with period $\pi/\varkappa$, whereas the terms containing $\cos(2kL)$ give the short-period oscillation, with period $\pi/k$. For metallic CNTs, $k=2\pi/(3d)$, where $d$ is the translational unit
cell length, equal to $d_{(12,6)}=1.127$\,nm for the (12,6) CNT and $d_{(n,n)}=0.246$\,nm for the armchair (5,5) and (9,9) CNTs. We fitted the $B=0$, $\mu=-0.3$\,eV transmission curves obtained from the numerical calculations for all SWCNT--SWCNT junctions considered here using Eq.~\eqref{eq:yakobson}, with $\varkappa$ as the fitting parameter. For this comparison, Eq.~\eqref{eq:yakobson} was multiplied by a factor of two, because the analytical expression describes a single channel, whereas the presence of two Dirac cones in the CNTs provides them with two transmission channels. The resulting fits are shown in Figure~\ref{osc_swcnt_0}(a--c), with the corresponding fitted values of $\varkappa$ given in the caption. Note the similarity between Eq.~\eqref{eq:yakobson} and the minimal model's Eq.~\eqref{eq:chain-transmission-formula}, with $\varkappa$ equivalent to $\delta k$. 

Although two interacting SWCNTs forming a junction have been discussed previously at zero field~\cite{Xu2013,Tripathy2016}, the influence of a perpendicular magnetic field on the junction transmission has not yet been investigated. We now use this coupled-wave picture to quantify how the perpendicular magnetic field modifies the overlap-length oscillations. Since $k$ is fixed by the nanotube geometry, the short-period component with period $\pi/k$ is expected to remain essentially unchanged. We describe the field response through a field-dependent effective coupling $\varkappa(B)$, which controls the long-period envelope. This is consistent with Figure~\ref{osc_swcnt}, where increasing $B$ leaves the short-period oscillations almost unchanged, but shortens the envelope period and suppresses the transmission maxima. This trend is visible in Figure~\ref{osc_swcnt} for the doped junctions, $\mu=-0.3$ and $-0.6$\,eV, for all three SWCNT--SWCNT systems, including the (5,5)+(5,5) junction. The response at the undoped Fermi level is less uniform. For example, the (9,9)+(9,9) junction shows strong suppression at long overlaps for $B=0$ and 30\,T, whereas at $B=60$\,T pronounced oscillations remain.
The oscillation period $\Delta L$ decreases systematically with magnetic field, as shown for $\mu=-0.3$\,eV in Figure~\ref{osc_swcnt_0}(d). This dependence is well captured by $\Delta L=\pi/(aB+b)$, with the fitted parameters and $\chi^2$ values given in the caption. The fit quality is good over most of the field range, with the main deviations occurring for the (5,5)+(5,5) junction below $B=45$\,T. Assuming that the long-period oscillation remains governed by $\Delta L=\pi/\varkappa(B)$, the fit implies an approximately linear increase of the effective intertube coupling parameter $\varkappa$ with magnetic field. The field-induced quenching of the transmission maxima is well described by an empirical exponential decay. Figure~\ref{osc_swcnt_0}(e) shows this behaviour at $\mu=-0.3$\,eV, fitted with $a\exp(-bB)+c$.

These results quantify how the overlap length $L$, previously identified as a key junction parameter in CNT assembly magnetotransport~\cite{manuscriptAGA}, controls the transmission of individual SWCNT--SWCNT junctions. Even for a fixed SWCNT--SWCNT junction type, small changes in $L$ can shift the system between high-transmission and strongly suppressed regimes. 
The practical consequence is that, when the overlap-length distribution cannot be controlled precisely, the most favourable junctions are those for which high-transmission windows occur most frequently and remain robust under magnetic field.
Within the armchair series studied here, this favours the smaller (5,5)+(5,5) junction over the larger (9,9)+(9,9) junction: Figure~\ref{osc_swcnt_0}(d) shows that its envelope period $\Delta L$ is shorter for $B<45$\,T, increasing the probability of sampling a high-transmission overlap, while Figure~\ref{osc_swcnt_0}(e) shows that its maximum transmission is quenched more slowly with magnetic field. Thus, for assemblies in which junction overlap lengths are broadly distributed, smaller metallic SWCNTs are expected to provide a more statistically robust route to high conductance, especially when transport must remain efficient under \color{black} a \color{black} high perpendicular magnetic field.

\subsubsection{Oscillations of transmission with magnetic field}

\begin{figure*}[ht!b]
	\centering
	\includegraphics[width=1.0\linewidth]{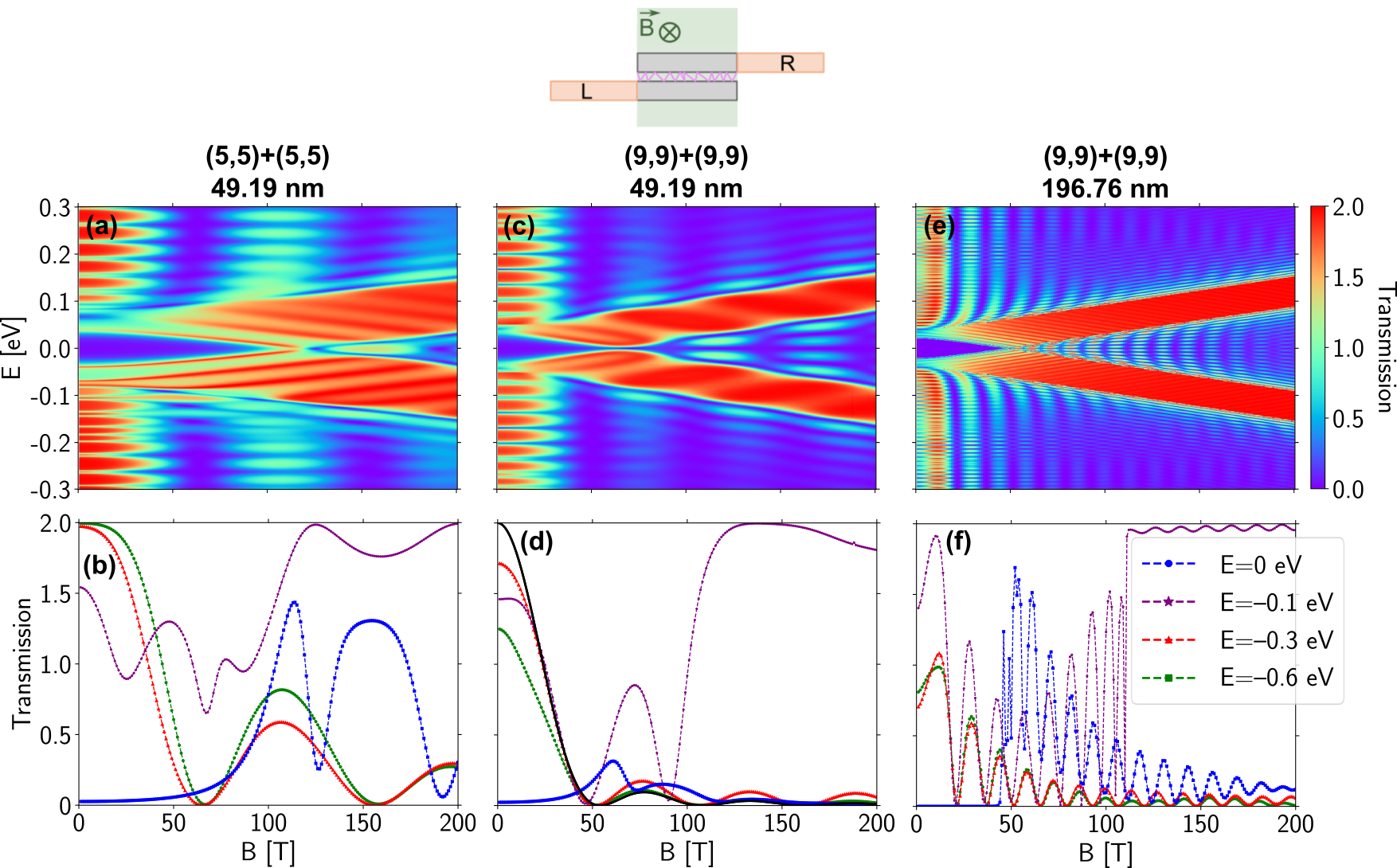}
	\caption{ Computed zero-bias transmission maps for single CNT junctions as a function of electron energy $E$ and external perpendicular magnetic field $B$: (a) (5,5)+(5,5) with an overlap length of 49.19\,nm (200 unit cells), (c) (9,9)+(9,9) with an overlap length of 49.19\,nm (200 unit cells), and (e) (9,9)+(9,9) with an overlap length of 196.76\,nm (800 unit cells). The transmission is roughly symmetric in energy with respect to the Fermi level, $E=0$. (b,d,f) Cross-sections of the corresponding transmission maps, plotted at $E=0$, $-0.3$~eV, and $-0.6$~eV, showing transmission oscillations as a function of magnetic field $B$. In panel (d), the solid black line is the fitted Fraunhofer diffraction formula (Eq.~\ref{eq:Fraunhofer}). }
	\label{55_map}
\end{figure*}

We now turn to transmission oscillations as a function of magnetic field. Before discussing junction transport, we briefly recall the behaviour of an isolated nanotube in a perpendicular magnetic field. As shown previously for a single SWCNT~\cite{manuscriptAGA}, the magnetic-field-induced change in electron transport is very small. Consistent with that result, the transmission of an isolated (9,9) SWCNT changes by only about $10^{-5}$ between $B=0$ and $B=200$\,T, as shown in Figure~\ref{99_single}(a,b). 

In contrast, SWCNT--SWCNT junctions show a pronounced magnetic-field response. Figures~\ref{55_map} and~\ref{55_map_SI} show transmission maps as functions of electron energy, $E=-0.3$ to $0.3$\,eV, and magnetic field, $B=0$ to 200\,T, for the
type A SWCNT--SWCNT junctions. The accompanying line cuts show $T(B)$ at $E=0$, $-0.1$, $-0.3$, and $-0.6$\,eV.

A characteristic feature of all transmission maps is the presence of two high-transmission stripes, visible as red V-shaped ridges. These stripes are the CNT-junction analogue of the high-transmission gateway-state regions found in the chain toy model under \color{black} a \color{black} perpendicular magnetic field (Figure~\ref{fig:chain-junction-magnetic}(c,d)).  In the toy model, the magnetic field shifts the dispersions of the two chains
relative to each other, so that efficient transport survives only in restricted energy windows where the incoming state can couple into the junction through gateway states. The same mechanism is reflected here: with increasing magnetic field, the high-transmission windows move approximately linearly in energy.

The positions of the high-transmission stripes can be described by an approximately linear relation, $E\simeq \pm\alpha B$, where the two signs correspond to the electron-like and hole-like branches of the V-shaped pattern. The slope $\alpha$ is controlled mainly by the effective magnetic flux through the junction and therefore by the
nanotube diameter. This is seen by comparing the (5,5)+(5,5) and (9,9)+(9,9) junctions: the smaller-diameter (5,5)+(5,5) junction has a smaller $\alpha$, whereas the wider (9,9)+(9,9) junction shows a stronger field-induced energy shift (Figure~\ref{55_map}(a,c)). By contrast, $\alpha$ is only weakly affected by the overlap length (Figure~\ref{55_map}(c,e)) and by chirality, as seen by comparing the (9,9)+(9,9) and (12,6)+(12,6) junctions of similar diameter (Figure~\ref{55_map}(e) and Figure~\ref{55_map_SI}(a,c)). Chirality mainly modifies the width and detailed shape of the high-transmission stripes rather than their overall slope.

\begin{figure*}[ht!b]
	\centering
	\includegraphics[width=1.0\linewidth]{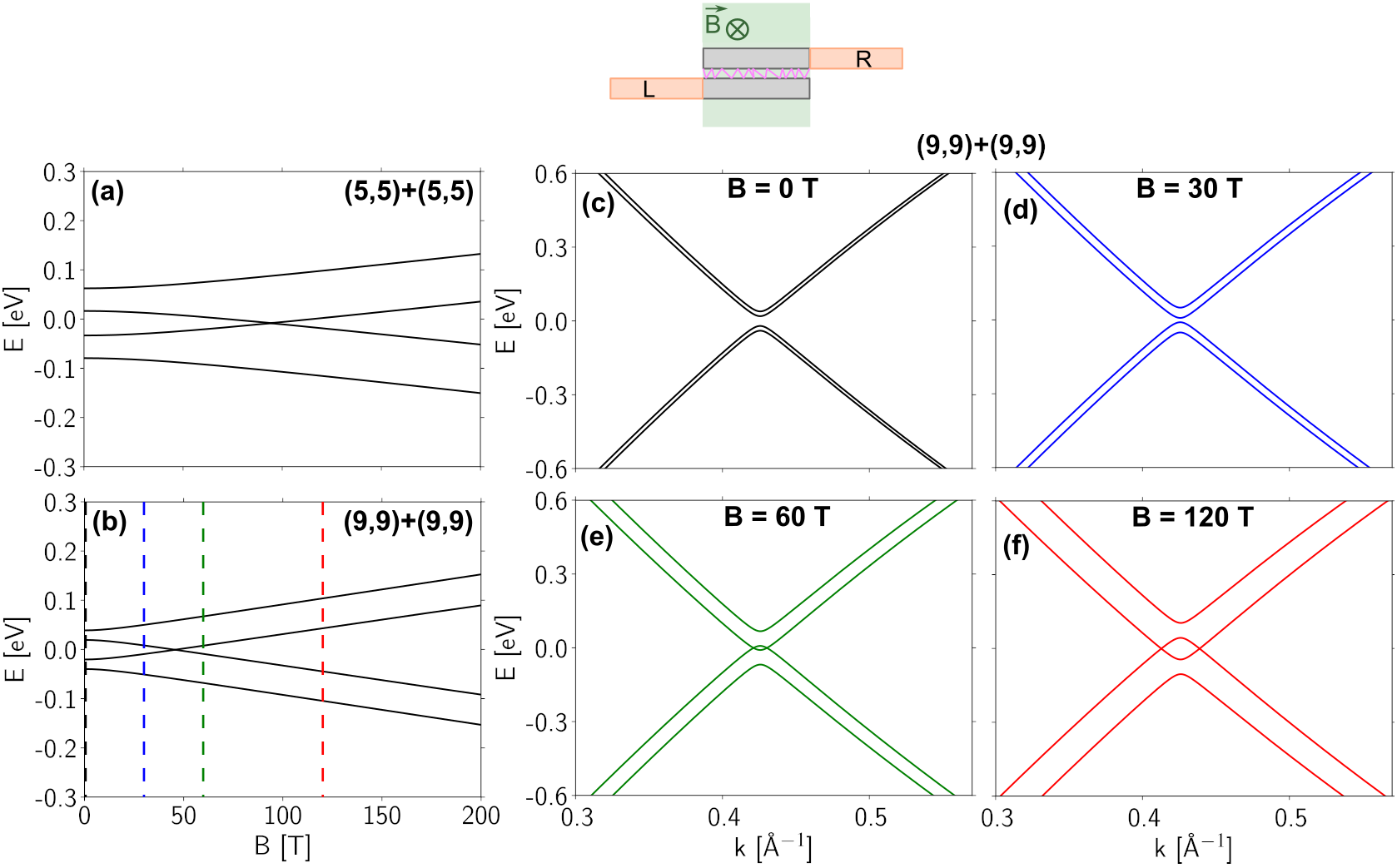}
	\caption{ Electronic structure of short CNT junctions under an external perpendicular magnetic field. (a,b) Energy spectra near the Fermi level, $E=0$, for (5,5)+(5,5) and (9,9)+(9,9) CNT junctions, respectively, with an overlap length of 0.49\,nm (2 unit cells), shown as a function of magnetic field $B$. (c--f) Band structures of the (9,9)+(9,9) CNT junction with the same overlap length as a function of wave vector $k$ near $k_F$ for $B=0\,T$, 30\,T, 60\,T, and 120\,T, respectively. }
	\label{99_bands}
\end{figure*}

To understand the electronic-structure origin of these high-transmission stripes, we compare the transmission maps with the field-dependent band structures shown in Figure~\ref{99_bands}. The key behaviour is already captured by the short-overlap (5,5)+(5,5) and (9,9)+(9,9) junctions (Figure~\ref{99_bands}(a,b)). At $B=0$, intertube coupling opens a gap around the Fermi level in both systems. This gap narrows with increasing magnetic field and closes at about 100\,T for the (5,5)+(5,5) junction and about 50\,T for the (9,9)+(9,9) junction. 
The lower closing field of the (9,9)+(9,9) junction is consistent with its larger diameter, which gives a larger magnetic phase shift at the same field.
The transmission maps in Figure~\ref{55_map}(a,c,e) mirror this band-structure evolution. The purple triangle near $E=0$ corresponds to energies inside the main gap, where transmission is suppressed. The red high-transmission stripes follow the band edges closest to the Fermi level at $k=k_F$ and the corresponding \color{black} energy range with a single band and gateway states\color{black}. The approximate symmetry with respect to $E=0$ reflects the electron-like and hole-like branches: the stripe at $E>0$ corresponds to electron transmission, whereas the stripe at $E<0$ corresponds to hole transmission.

This map--band correspondence can be illustrated using the (9,9)+(9,9) junction, whose band structure is shown at selected magnetic fields in Figure~\ref{99_bands}(c--f). At $B=0$, the bands near the Fermi level are nearly degenerate and separated by a gap of about 0.04\,eV (Figure~\ref{99_bands}(c)). Applying a perpendicular field lifts the degeneracy further and narrows the gap, as seen already at $B=30$\,T (Figure~\ref{99_bands}(d)). The gap closes at around 50\,T and, by $B=60$\,T, has reopened for two subbands (Figure~\ref{99_bands}(e)); at $B=120$\,T the reopened gap increases to about 0.09\,eV (Figure~\ref{99_bands}(f)). Because the parent (9,9) nanotube is metallic, the relevant low-energy spectrum consists of two cone-like subband pairs around the Fermi level. Each pair gives one branch of the V-shaped high-transmission pattern in Figure~\ref{55_map}(c,e). As $B$ increases, the separation between the tips of these cone-like bands grows, so progressively higher electron-like ($E>0$) or hole-like ($E<0$) energies are required to access the gateway-state transport window.

Beyond the V-shaped gateway-state stripes discussed above, the line cuts of the maps (Figure~\ref{55_map}(b,d,f) and Figure~\ref{55_map_SI}(b,d)) reveal field-driven oscillations of the transmission. 
As anticipated by the chain minimal model in \color{black} a \color{black} perpendicular field (Figure~\ref{fig:chain-junction-magnetic}(d)), the character of these cuts depends on whether the fixed energy crosses a gateway-state window. For energies that do not intersect the high-transmission stripes, in particular $E=-0.3$ and $-0.6$\,eV, the junction remains outside the gateway-state regime. These cuts are analogous to the Type A behaviour of the toy model: constructive phase conditions give forward-current-like transmission maxima, whereas destructive phase conditions suppress the transmission. 
The corresponding $T(B)$ curves therefore oscillate with a decaying amplitude while remaining visible within the experimentally accessible field range, $B\leq60$\,T. 
Here, we describe these oscillations analytically using a Fraunhofer-like interference picture, which is fitted to the selected off-stripe line cuts in Figure~\ref{55_map}(d) and Figure~\ref{55_map_SI}(b). In optical Fraunhofer diffraction, coherent light passing through a slit produces an intensity pattern because waves emitted from different points of the aperture arrive with different phases~\cite{Feynman1965,Beenakker1991,Ando2005}. The finite CNT--CNT overlap plays an analogous role: an electron can tunnel between the two nanotubes at different positions along the junction. Since these paths enclose different magnetic fluxes, a perpendicular magnetic field imposes different Aharonov--Bohm phases~\cite{PhysRev.115.485} on the corresponding electronic waves. Their coherent sum produces alternating constructive and destructive interference in the transmission.

In this picture, a contribution to the transmission associated with tunnelling at position $z$ along the overlap region acquires an Aharonov--Bohm phase $2\pi\Phi(z)/\Phi_0$, where $\Phi(z)=B(2r+d)z$ and $\Phi_0=h/e$ is the flux quantum. The transmission amplitude is obtained by coherently summing these contributions over the overlap length $L$, and the transmission is the squared modulus of this amplitude. Assuming a uniform tunnelling amplitude along the junction gives
\begin{equation}
	T(B)=T_{\mathrm{max}}
	\frac{ \sin^2\left(\pi\Phi/\Phi_0\right)}
	{\left(\pi\Phi/\Phi_0\right)^2},
	\qquad
	\Phi=B(2r+d)L .
	\label{eq:Fraunhofer}
\end{equation}
This expression, derived in the Supplementary Material (Section~\ref{sec:fraunhofer}), is fitted directly to the computed $T(B)$ curves for the short-overlap junctions
(Figure~\ref{55_map}(d) and Figure~\ref{55_map_SI}(b)). Here, $r$ is the nanotube radius, $d$ is the intertube distance, and $T_{\mathrm{max}}$ sets the transmission scale. Because $\Phi\propto L$, the oscillation period in magnetic field decreases with increasing overlap length.

For longer junctions, with overlaps of about 180--200\,nm, the Fraunhofer-like transmission probability in Eq.~\eqref{eq:Fraunhofer} no longer captures the computed line cuts (Figure~\ref{55_map}(f) and Figure~\ref{55_map_SI}(d)). 
The deviations for the longer junctions are expected because the Fraunhofer-like expression in Eq.~\eqref{eq:Fraunhofer} assumes a single coherent summation over tunnelling positions along the overlap. This is analogous to optical Fraunhofer diffraction, where waves propagate from the aperture to a distant screen and no return path from the screen back to the aperture is included.
In a CNT junction, however, the finite overlap region is not only an aperture-like tunnelling region. For longer overlaps, the probability of repeated intertube hopping and internal reflections increases. The junction then no longer behaves as a simple single-pass aperture, and additional phase accumulation inside the overlap region modifies the transmission. For this reason, the long-overlap line cuts are compared with the phenomenological envelope $\left|\sin(\pi\Phi/\Phi_0)/(\pi\Phi/\Phi_0)\right|$ (Figure~\ref{fraunhofer}), rather than with the squared Fraunhofer transmission probability. Similar diffraction-like envelope functions are commonly used to describe magnetic-field interference in extended Josephson junction arrays~\cite{Lucci2016}.

The line cuts at $E=-0.1$~eV show a different behaviour, because they intersect the V-shaped gateway-state stripes. This corresponds to the Type C/D behaviour of the chain toy model, where a magnetic field can drive the system into a gateway-state transport window instead of only suppressing transmission.   As the magnetic field shifts the gateway-state transport windows through this energy, the transmission increases rather than following only the decaying Fraunhofer-like envelope. For the armchair junctions in Figure~\ref{55_map}, this field-driven access to the gateway-state regime can restore the transmission close to its maximum value, $T\simeq2$. For the (12,6)+(12,6) junctions in Figure~\ref{55_map_SI}, the same mechanism produces a clear transmission enhancement at the fields where the line cut crosses the high-transmission stripe, although the transmission does not recover to $T=2$. 

Taken together, these results identify geometry as the main control of the field-driven response in type A SWCNT--SWCNT junctions. The slope $\alpha$ of the V-shaped high-transmission stripes is set mainly by nanotube diameter, whereas the period of the Fraunhofer-like oscillations is governed by the magnetic flux through the overlap region, and therefore by diameter and overlap length. Chirality mainly modifies the width and detailed shape of the high-transmission regions. Thus, for an individual junction, geometry fixes the location of high transmission stripes in the $(E,B)$ plane, while the perpendicular magnetic field moves the junction into or out of these windows. In macroscopic CNT films and fibres, these field-accessible high-transmission windows are sampled by an ensemble of junctions with distributed overlap lengths, local doping levels and relative atomic registries. 
A moderate spread of junction phases may be beneficial, because different percolating paths can sample different parts of the gateway-state windows and the network becomes less sensitive to a single unfavourable interference condition. This should not be confused with structural disorder, which would create \color{black} localised \color{black} state bottlenecks and suppress transmission. High conductance is therefore favoured by a robust distribution of metallic junctions whose high-transmission windows remain accessible within the relevant doping and magnetic field range, rather than by forcing all junctions to share the same phase condition.

\subsection{Magneto-transport in multiple junctions of SWCNTs}

Beyond the two-tube type A junctions, we next consider type B SWCNT architectures: a three-tube multi-junction and a four-tube loop, shown in Figure~\ref{geometries}(c,d) and in the schemes above the panels in Figure~\ref{loop}. In both systems\color{black}, \color{black} the nanotubes are identical and collinear. The multi-junction contains two equivalent overlap regions, whereas the loop contains four. The overlap regions, taken to have the same length within each architecture, define the sites where electrons can tunnel between neighbouring nanotubes.

\subsubsection{Multi-junctions}

\begin{figure*}[t!]
	\centering
	\includegraphics[width=1.0\linewidth]{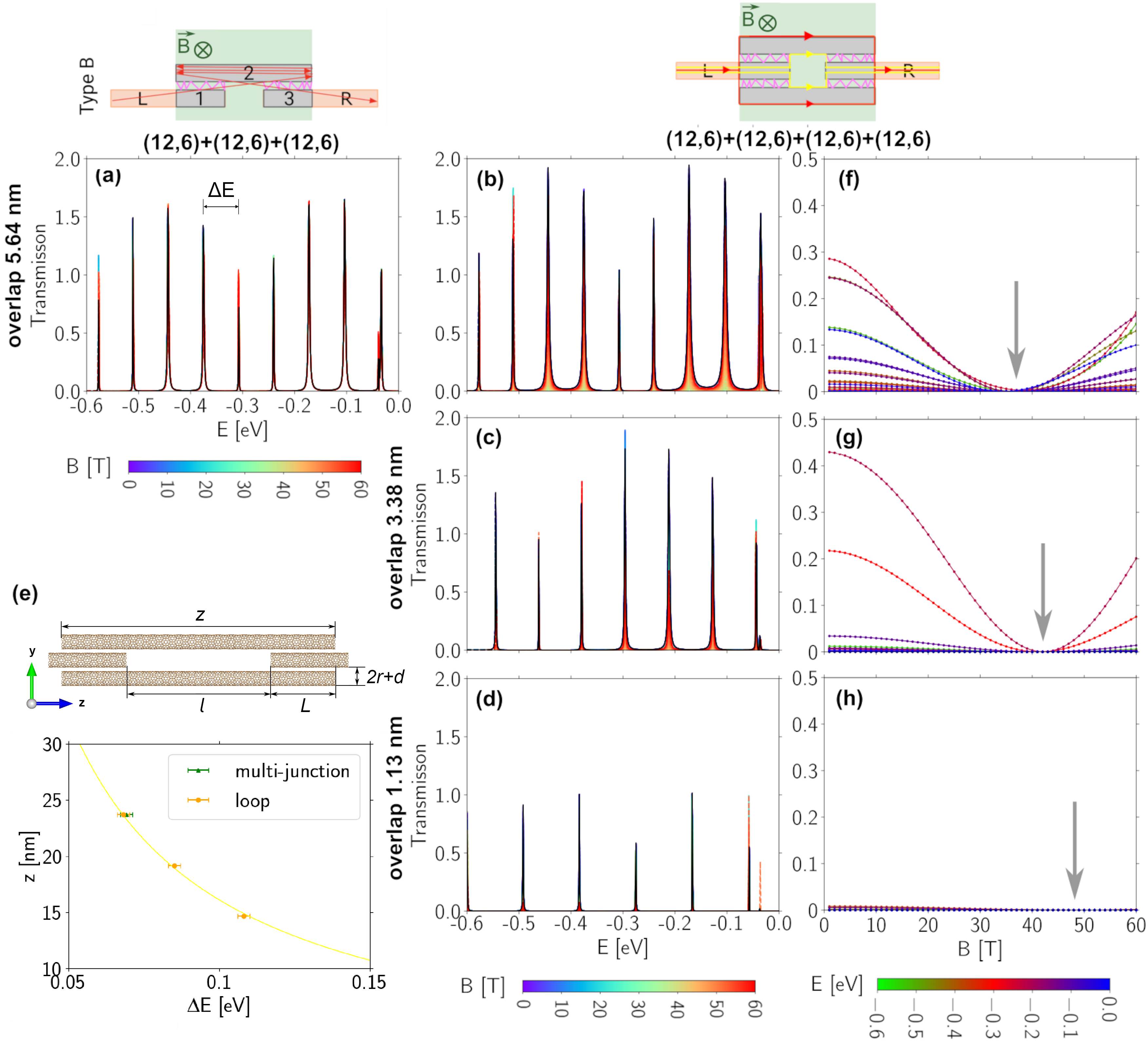}
	\caption{Electron transport through complex architectures (type B systems) made of metallic (12,6) CNTs under an external perpendicular magnetic field. The schematics above the panels illustrate the corresponding multi-junction and loop architectures and some possible electron paths (red and yellow arrows). (a) Computed zero-bias transmission spectrum of the multi-junction with both overlap regions 5.64\,nm long, shown in the energy window [-0.6,0]\,eV for magnetic fields up to 60\,T. (b--d) Corresponding zero-bias transmission spectra of loops with equal overlap lengths of 5.64\,nm (five unit cells), 3.38\,nm (three unit cells), and 1.13\,nm (one unit cell), respectively.  (e) The side-view atomistic representation of the loop architecture with the marked structural parameters, and the dependence of the nanotube length $z$ on the energy spacing $\Delta E$ between the transmission peaks visible in panels (a--d). The yellow line is a fit to $\Delta$E=$\beta$/z with $\beta$=1.62\,eV$\cdot$nm. (f--h) Transmission through loops with overlap lengths matching those in panels (b--d) as a function of perpendicular magnetic field for different doping levels. Arrows mark the field at which complete extinction of transmission occurs.}
	\label{loop}
\end{figure*}

The transmission of the three-tube multi-junction is shown in Figure~\ref{loop}(a) in the energy window $[-0.6,0]$\,eV for magnetic fields up to 60\,T, with the corresponding extended energy range shown in Figure~\ref{loops2}(a). In contrast to the type A system, the transmission spectrum of system B is dominated by a series of very narrow, nearly regularly spaced resonances, whose positions depend only weakly on the perpendicular magnetic field. 
Between these resonances, the transmission is almost completely suppressed. This reflects the fact that the overlap regions in the multi-junction are very short: even a single (12,6)+(12,6) junction with such a short overlap would transmit poorly over most of the energy range (Figure~\ref{osc_swcnt}(a)), and the additional junction further restricts transmission to resonant energies.
This behaviour is consistent with the double-junction chain toy model shown in Figure~\ref{fig:chain-multijunction} and discussed in Section~\ref{supp:chain-multijunction} of the Supplementary Material. When reflection at each junction is weak, the transmission through two junctions is close to the product of the individual transmissions. However, in the standing-wave and intermediate regimes, where reflection is stronger, the second junction splits the transmission into narrow sub-peaks. The narrow transmission maxima that survive this filtering in the CNT multi-junction therefore form a Fabry--Pérot-like resonance series with a weak amplitude modulation. We attribute this resonant transmission to quantum confinement in the finite nanotube segment that couples the two semi-infinite current-carrying nanotubes. Electrons are transmitted efficiently only when their energy matches one of the quasi-bound states of this finite segment. Equivalently, these quasi-bound states can be viewed as standing waves formed by reflection at the two ends of the finite nanotube, as indicated schematically above Figure~\ref{loop}(a). 

\subsubsection{Loops}

The loop architecture is formed by four identical collinear (12,6) SWCNTs, as shown in Figure~\ref{loop}(e). In this geometry, the four equivalent overlap regions have length $L$, while $z$ denotes the length of the finite upper and lower nanotube arms that form the resonant part of the loop. We consider three loop configurations with very short overlap lengths $L$. The separation $\ell$ between the two overlap regions is kept fixed, so changing $L$ also changes the finite nanotube length $z$. The computed transmission curves for these structures are shown in Figure~\ref{loop}(b--d), with the corresponding extended energy range given in Figure~\ref{loops2}(b--d). As in the three-tube multi-junction, the loop transmission is dominated by narrow Fabry--Pérot-like resonances. Their spacing, width and amplitude depend strongly on $L$: as $L$ increases, the resonances become more closely spaced in energy and their amplitude and width increase. This behaviour is consistent with quantum confinement in the finite upper and lower arms of the loop, where the allowed quasi-bound states set the resonant transmission energies.

To test this interpretation, we extracted the energy spacing $\Delta E$ between neighbouring resonant peaks and plotted it as a function of the finite nanotube-arm length $z$ in Figure~\ref{loop}(e). In an optical Fabry--Pérot interferometer, the
spacing between resonances is inversely proportional to the distance between the mirrors. Here, the analogous cavity length is the length $z$ of the finite upper and lower nanotube arms. We therefore fit the data with $ \Delta E=\frac{\beta}{z}$.
The fitted value, $\beta=1.62$\,eV$\cdot$nm, coincides with $\beta=\pi\hbar v_F$ obtained in Section~\ref{sec:T(E)} for the (12,6) CNT. The value of $\Delta E$ extracted for the three-tube multi-junction is also included in Figure~\ref{loop}(e)\color{black} and \color{black} and lies on the same line, confirming that both type B architectures behave as Fabry--Pérot-like interferometers whose resonant spacing is controlled by the length of the finite nanotube segment.

The resonant peaks also show a weak but systematic dependence on the perpendicular magnetic field. This is clearer in the fixed-energy line cuts in Figure~\ref{loop}(f--h), where the transmission is fully suppressed at a characteristic field $B^*$. For the (12,6) loop geometries shown in Figure~\ref{loop}(f--h), the extracted values of $B^*$ decrease with increasing overlap length $L$. The same trend is found for the corresponding (9,9) loop series, as \color{black} summarised \color{black} in Table~\ref{loopB}. This behaviour is consistent with Aharonov--Bohm interference between electronic paths propagating through the upper and lower arms of the loop. The two paths enclose magnetic flux and therefore acquire different magnetic-field-dependent phases. Destructive interference occurs when the phase difference reaches $\pi$, giving the first transmission node. 
Comparing the numerical $B^*$ values extracted from the line cuts with the analytical estimates in Table~\ref{loopB} shows that they are closest to $B^*_{\mathrm{av}}$, calculated from the average enclosed area $S_{\mathrm{av}}$. This indicates that the dominant electronic path encircles a magnetic flux between the minimum- and maximum-flux paths, marked by the yellow and red loops in the schemes above panels (b) and (f) of Figure~\ref{loop}.

The analysis of type B architectures therefore shows that increasing the number of possible pathways does not necessarily increase conductance. Additional CNT--CNT contacts can still transmit well if each contact lies in a favourable interference condition, but they also add further phase constraints, making high transmission more selective in energy and local geometry. Loop-like motifs add an additional limitation. These loops introduce Aharonov--Bohm phase sensitivity: at certain magnetic fields, destructive interference between the two loop arms can suppress transmission almost completely. Thus, high conductance in macroscopic CNT conductors is favoured by longer nanotubes, controlled bundling and sufficient alignment to reduce unnecessary crossings, branching points and loop-like bottlenecks, rather than by a highly connected network of many competing paths.

\subsection{Magneto-transport in MWCNT--MWCNT junctions}

\begin{figure*}[h!tb]
	\centering
	\includegraphics[width=1.0\linewidth]{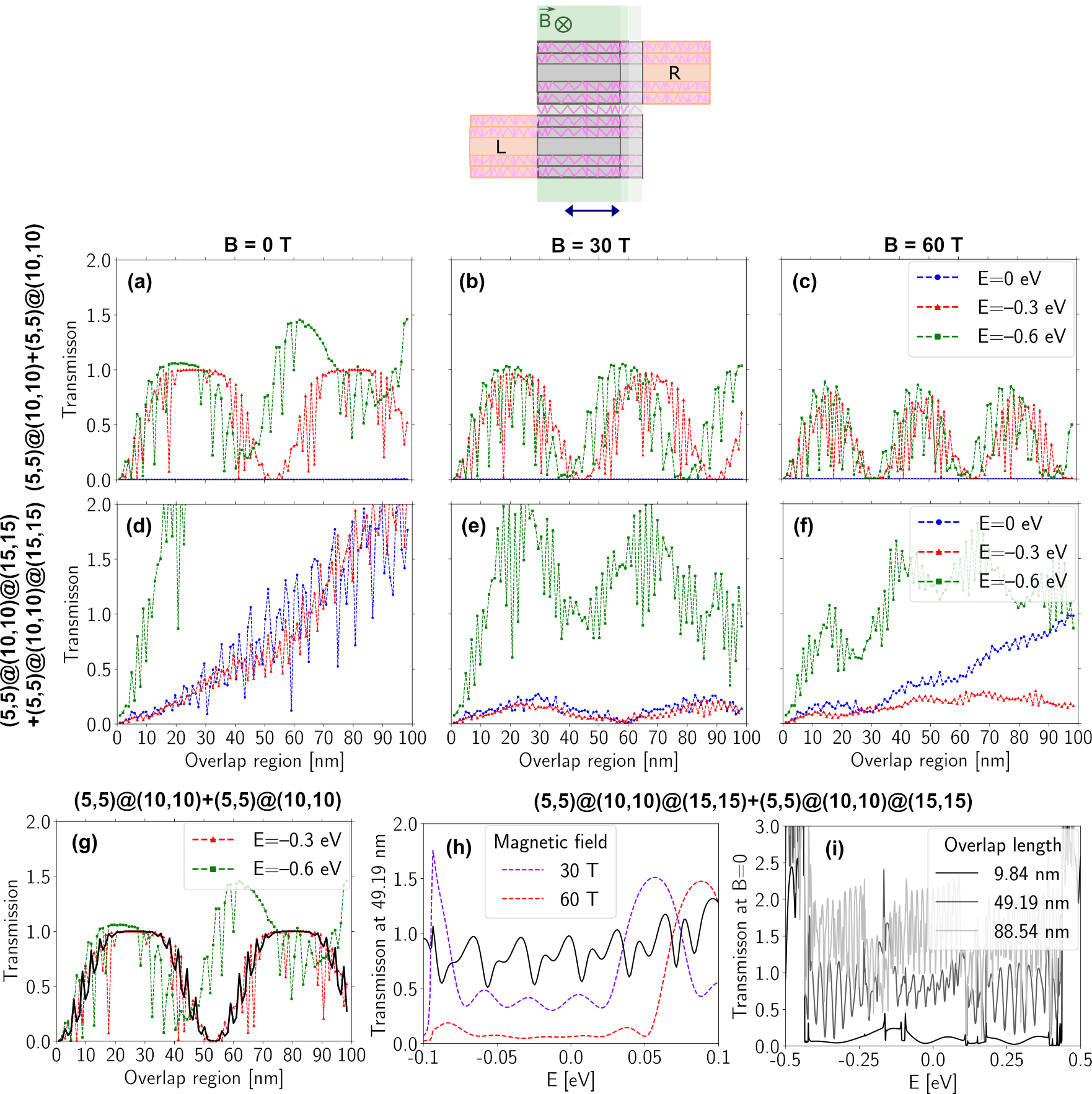}
	\caption{Electron transport through DWCNT and MWCNT junctions under an external perpendicular magnetic field. Computed zero-bias transmission spectra of (a--c) (5,5)@(10,10)+(5,5)@(10,10) DWCNT and (d--f) (5,5)@(10,10)@(15,15)+(5,5)@(10,10)@(15,15) MWCNT junctions as functions of overlap length, shown for different magnetic-field strengths and for electron energies $E=0$, -0.3\,eV, and -0.6\,eV. For the DWCNT--DWCNT junction, the transmission at the Fermi level is zero because of band-gap opening. (g) Comparison of the computed zero-bias transmission of the DWCNT--DWCNT junction at $B$=0\,T as a function of overlap length with the transmission calculated from equation~\eqref{eq:yakobson} (solid black line), with $\varkappa=(17\,\mathrm{nm})^{-1}$. (h) Computed zero-bias transmission spectra of the MWCNT--MWCNT junction with an overlap length of 49.19\,nm (200 unit cells) as a function of electron energy $E$ for different values of magnetic field $B$\color{black}, with the black line corresponding to $B=0$\,T. \color{black} (i) Transmission of the same junction at $B$=0\,T for overlap lengths of 9.84\,nm (40 unit cells), 49.19\,nm (200 unit cells), and 88.57\,nm (360 unit cells). }
	\label{osc_mwcnt}
\end{figure*}

We now ask whether the interference mechanisms identified for SWCNT--SWCNT junctions persist when each conducting element is multi-walled, and whether they can rationalise the lower conductance observed experimentally for MWCNT-based fibres. We therefore compare simple parallel DWCNT--DWCNT and TWCNT--TWCNT junctions, built from the (5,5)@(10,10) DWCNT and the (5,5)@(10,10)@(15,15) TWCNT (Figure~\ref{geometries}(e--h)).

Before turning to junctions, we first inspect the corresponding isolated DWCNT (Figure~\ref{geometries}(e)) and TWCNT systems (Figure~\ref{geometries}(g)). Their transmission (Figure~\ref{99_single}(c,e)) is not simply a two- or threefold multiple of the SWCNT transmission spectrum (Figure~\ref{99_single}(a)). Although the isolated DWCNT and TWCNT can reach the nominal transmissions expected from two and three metallic walls, interwall coupling produces dips in $T(E)$, so a given doping level can sample fewer effective channels. Similar interwall-interference effects, including antiresonance dips, \color{black} localised \color{black} states and overlap-dependent conductance oscillations, have been reported for telescopic DWCNT systems~\cite{kim2002,wittermeier2022}.
The tube-resolved decomposition in Table~\ref{tau1} shows a small but consistent imbalance between walls. In the DWCNT, the inner wall contributes slightly less than one half of the total transmission, indicating that transmission is preferentially carried by the outer wall. The same tendency appears in the TWCNT: the inner wall gives the smallest contribution, below one third of the total transmission. Notably, the outer wall is not the dominant one. The largest contribution comes from the middle wall.  This mirrors the trend found previously in concentric three-layer CNT bundles, where the innermost nanotube was the least conducting and nanotubes in the second layer carried larger per-tube transmission contributions than those in the outer layer~\cite{BULMER2026121162}.
The perpendicular magnetic-field response of the isolated DWCNT and TWCNT (Figure~\ref{99_single}(d,f)) remains as weak as in the SWCNT case (Figure~\ref{99_single}(b)), except in narrow energy windows where the field-induced change in transmission reaches the order of $10^{-1}$.

Figure~\ref{osc_mwcnt}(a--f) presents the computed transmission as a function of overlap length $L$ up to 100\,nm for the DWCNT--DWCNT junction (a--c) and the TWCNT--TWCNT junction (d--f), shown for $B=0$, 30, and 60\,T and for $E=0$, $-0.3$, and $-0.6$\,eV.
We first analyse the DWCNT--DWCNT junction. Despite the additional wall, its transmission retains the broad overlap-length oscillations characteristic of the SWCNT--SWCNT junctions.
In particular, the DWCNT--DWCNT transmission is strongly quenched near the undoped Fermi level, although it does not become strictly zero (Table~\ref{tau2}). This suppression occurs despite all constituent walls being metallic and reflects a coupling-induced low-transmission window, associated with destructive interference between the coupled wall channels at $E=0$.  Away from this gap, at $E=-0.3$ and $-0.6$\,eV, the transmission remains finite and exhibits broad oscillations with overlap length. As in the SWCNT--SWCNT junctions, increasing $B$ shortens the oscillation period and suppresses the transmission maxima. The DWCNT--DWCNT oscillations can still be described qualitatively by the coupled-wave expression in Eq.~\eqref{eq:yakobson}. 
Figure~\ref{osc_mwcnt}(g) compares the $B=0$ numerical data at $E=-0.3$\,eV with this analytical expression. Unlike \color{black} in \color{black} the SWCNT--SWCNT fits, no additional factor of two was applied here. At this energy, the DWCNT--DWCNT junction is dominated by a single effective outer-wall transmission channel. The tube-resolved decomposition in Table~\ref{tau2} shows that the inner-wall contribution is nearly suppressed, whereas the outer walls carry almost all of the inter-DWCNT transmission. The reasonably good
agreement indicates that the DWCNT--DWCNT junction can still be viewed, to first approximation, as an effective coupled-waveguide system.

Adding the third wall changes the overlap-length dependence of the junction transmission (Figure~\ref{osc_mwcnt}(d--f)), so that the TWCNT--TWCNT junction no longer follows the trends observed for the SWCNT--SWCNT and DWCNT--DWCNT systems. The transmission is no longer organised into a clean broad envelope with well-defined large maxima and minima.  At $B=0$, it instead increases on average with overlap length, both at $E=0$ and at $E=-0.3$\,eV, and the near-zero transmission window observed for the DWCNT--DWCNT junction is absent. \color{black} A finite \color{black} magnetic field introduces broader modulations with $L$, but without a simple systematic shortening of the oscillation period or a monotonic quenching of the transmission maxima. In the language of the chain minimal model, the simple alternation between forward-current-like envelope maxima and standing-wave minima is therefore no longer cleanly separable, because several wall-derived pathways contribute simultaneously with different phases.

The energy-resolved spectra further confirm that the TWCNT--TWCNT junction cannot be reduced to a two-waveguide problem (Figure~\ref{osc_mwcnt}(h,i), \color{black} Figure~\ref{bench}\color{black}).  At $B=0$, both fast and slow energy-dependent modulations are present, and for the long overlap ($L$=88.54\,nm) the transmission exceeds 2 near $E=-0.5$\,eV, showing that several wall-derived channels contribute to transport. At fixed overlap length, a perpendicular magnetic field reshapes and damps these oscillations rather than producing the regular field-dependent response observed in simpler junctions. Accordingly, the TWCNT--TWCNT oscillations cannot be described, even qualitatively, by the coupled-wave expression in Eq.~\eqref{eq:yakobson}. The reason is that the additional wall opens extra coherent pathways. Electrons can tunnel between the two multi-walled nanotubes and, at the same time, redistribute among walls within each nanotube. This interpretation is supported by the tube-resolved decomposition in Table~\ref{tau2}, which shows that the wall contributions in the TWCNT--TWCNT junction are strongly unequal, unlike in the isolated TWCNT. The inner wall contributes least, whereas the outer wall carries the largest part of the transmission. Thus, the third wall does not simply add an independent conducting channel, but changes the way transmission is distributed across the whole junction.

 To test whether this behaviour is specific to the armchair (5,5)@(10,10)@(15,15) + 
 (5,5) @(10,10)@(15,15) junction, and whether it persists upon adding a fourth wall, we additionally calculated the zigzag (12,0)@(21,0)@(30,0) + (12,0)@(21,0)@(30,0) TWCNT--TWCNT junction and the four-walled (5,5)@(10,10)@(15,15)@(20,20) + (5,5)@(10,10)@(15,15)@(20,20) junction (Fig.~\ref{new_mwcnts} and Table~\ref{tau3}). In both systems, the wall-resolved contributions are strongly unequal and vary with electron energy and magnetic field. Our results, although based on a limited set of MWCNT architectures, suggest that additional walls do not provide independent channels that simply add to the total transmission. Instead, they modify the coupled channel structure and redistribute transmission among the walls. 

\color{black}{Although the main transport calculations were performed for idealised junction geometries, the SIESTA-optimised (PBE-D3-relaxed) structures exhibit measurable wall deformations, as quantified by the coefficient of variation of the nanotube radius (Table~\ref{tab:cv_radius}). In particular, the nanotube walls within the contact regions undergo structural relaxation to maximise the interfacial contact. These structural changes modify the ideal concentric-channel picture and the detailed transmission spectra. The main qualitative features, including the energy-dependent oscillations and the overall magnetic-field dependence, are preserved, while the relaxed transmission curves exhibit more pronounced
	irregularities (see Supplementary Information Fig.~\ref{trans-relaxed}).}
\color{black}

Taken together, these results show that the transition from SWCNT to multi-walled junctions is not governed by simple channel counting. The DWCNT--DWCNT junction remains close to an effective single-channel coupled system, because inter-junction transport is dominated by the outer walls. It therefore forms an intermediate case between SWCNTs and wider MWCNTs. \color{black} Junctions with three or more walls break \color{black}this simple picture: additional wall-resolved pathways redistribute transmission across the junction and make it more sensitive to magnetic-field-induced interference. This provides a microscopic explanation for the experimental trend in Figure~\ref{sem}(e): the SWCNT fibre shows only a weak conductance change up to 60\,T, whereas the MWCNT fibre, after a small low-field conductance enhancement, undergoes a much stronger high-field suppression.  
Thus, higher conductance is not achieved simply by packing together as many nominally conductive nanotubes as possible, but by preserving simple, direct and robust intertube transport pathways.

\FloatBarrier

\section{Conclusions}
To identify design principles for high-conductance CNT-based networks that remain robust under finite magnetic fields, we combined an analytically solvable minimal model, optical interference analogies, atomistic TB--NEGF calculations with the orbital effect of the field included through the Peierls phase, and ultrahigh-field experiments on CNT fibres. The minimal chain model shows that the key transport motifs later found in the CNT junctions can already arise from coherent interference in a coupled one-dimensional system. Guided by this insight, we interpret the atomistic CNT results using a coupled-waveguide language, which provides intuitive descriptions of how coherent interference appears in different junction geometries. For simple SWCNT--SWCNT junctions, the transmission is controlled by nanotube diameter, chirality, overlap length, electrochemical potential and magnetic field, making the local junction geometry decisive for conductance. Multi-junction and loop architectures show that additional overlaps do not simply add conducting paths but can instead restrict transmission to narrow resonant windows and introduce magnetic-field-induced transmission nodes. Similarly, additional walls do not simply add independent channels. DWCNT--DWCNT junctions remain close to an effective SWCNT-like coupled-channel system because inter-junction transmission is carried mainly by the outer walls. 
\color{black} In the three- and four-walled junctions studied here, additional walls redistribute \color{black}  transmission unevenly among interwall-coupled pathways and \color{black} create \color{black} additional destructive-interference conditions. This explains why wider MWCNT junctions can be less conductive and more field-sensitive despite  containing more \color{black} nominally \color{black} conducting walls. This microscopic picture is in accord with our ultrahigh-field measurements on directly spun CNT fibres produced by the same general FC-CVD route, with feedstock chemistry selected to favour either SWCNT or MWCNT formation. Under comparable cryogenic measurement conditions, the SWCNT fibre shows only a weak conductance change up to 60\,T, whereas the MWCNT fibre exhibits lower conductance and a much stronger high-field suppression.

Overall, the best design strategy to obtain high, field-stable conductance in CNT-based networks is not to pack together as many nominally conductive, undoped nanotubes as possible, nor to enforce a perfectly identical set of junction lengths. It is to build lightly doped SWCNT- or DWCNT-based networks with simple intertube contacts and a controlled spread of overlap lengths, so that some percolating paths remain close to high-transmission conditions even under magnetic field.

\section*{Acknowledgements}
T.K. and K.Z.M. gratefully acknowledge the Interdisciplinary Centre for Mathematical and Computational Modelling at the University of Warsaw, Poland (Grant No. G47-5) for providing computer facilities and technical support.  T.K., K.Z.M and I.V.L. also acknowledge the technical and human support provided by the DIPC Supercomputing Center, Spain. T.K. and K.Z.M. are grateful to the Agencia Estatal de Investigacion, Ministerio de Ciencia e Innovacion, Spain for funding this research under Proyectos de Generacion de Conocimiento 2022 program, PID2022-139776NB-C65. K.Z.M also would like to thank the European Commission (Marie Sklodowska-Curie Cofund Programme; grant no. H2020-MSCA-COFUND-2020-101034228-WOLFRAM2) for supporting this research. J.A.M. acknowledges the support from Centera2 project (FENG.02.02-IP.02.01-IP.05-T0004/23) funded with IRA FENG program of Foundation for Polish Science, and co-financed by the EU FENG Programme. I.V.L.  acknowledges support from the EuroHPC JU under the MAX (Materials design at the Exascale) project (grant no. 101093374), and from the Spanish MCIN/AEI/10.13039/501100011033 and the European Union NextGenerationEU/PRTR through grant no. PCI2022-134972-2. ICN2 is supported by the CERCA programme (Generalitat de Catalunya) and the Severo Ochoa Centres of Excellence programme (grant no. CEX2021-001214-S), funded by MCIN/AEI/10.13039/501100011033. A portion of this work was performed at the National High Magnetic Field Laboratory, which is supported by National Science Foundation Cooperative Agreement No. DMR-2128556*, the State of Florida, and the U.S. Department of Energy.

\section*{Author contributions}
The theoretical concept of the work was developed jointly by M.M., K.Z.M. and T.K. K.Z.M. led and coordinated the study and defined the integration of the experimental and theoretical parts. T.K. performed the numerical simulations under the supervision of K.Z.M. and contributed to the development of the sisl implementation required for magnetic-field simulations.  M.M. developed the toy-model calculations and wrote the corresponding parts of the manuscript. M.M., T.K., J.A.M.\color{black}, I.V.L. \color{black} and K.Z.M. contributed to the interpretation of the theoretical results. I.V.L. contributed to the development of the SIESTA implementation required for magnetic-field simulations and performed the \color{black} DFT\color{black} structural optimisations \color{black}in SIESTA\color{black}. A.E.L.-R. performed the high-field measurements with J.S.B., analysed the experimental data, and wrote the experimental sections of the manuscript. J.S.B. also carried out the SEM characterisation. F.F.B. supported the high-field experiments at Los Alamos. K.K., A.E.L.-R. and K.Z.M. secured funding. T.K.  prepared the first version of the manuscript, which was subsequently revised by J.A.M, M.M.\color{black}, I.V.L. \color{black} and K.Z.M. All authors discussed the results and contributed to the final manuscript.

\section*{Competing Interests}

The authors declare no competing interests.

\section*{Supplementary Material}

The online version contains supplementary material available below.

\FloatBarrier
\clearpage
\thispagestyle{plain} 

\begin{center}
    \LARGE \textbf{Supplementary Material} \\[1.5em]
    
    \large
    for \\[1.5em]
    
\LARGE{Taming quantum interference: a route to high electrical conductance in carbon nanotube assemblies}
\end{center}

\vspace{1.5cm}

\setcounter{section}{0}
\setcounter{subsection}{0}
\setcounter{figure}{0}
\setcounter{table}{0}
\setcounter{equation}{0}

\renewcommand{\thesection}{S\arabic{section}}
\renewcommand{\thesubsection}{S\arabic{section}.\arabic{subsection}}

\renewcommand{\thefigure}{S\arabic{figure}}
\renewcommand{\thetable}{S\arabic{table}}
\renewcommand{\theequation}{S\arabic{equation}}

\section{\color{black} Additional experimental characterisation  of the MWCNT fibre}

\begin{figure*}[th!]
	\centering
	\includegraphics[width=0.8\linewidth]{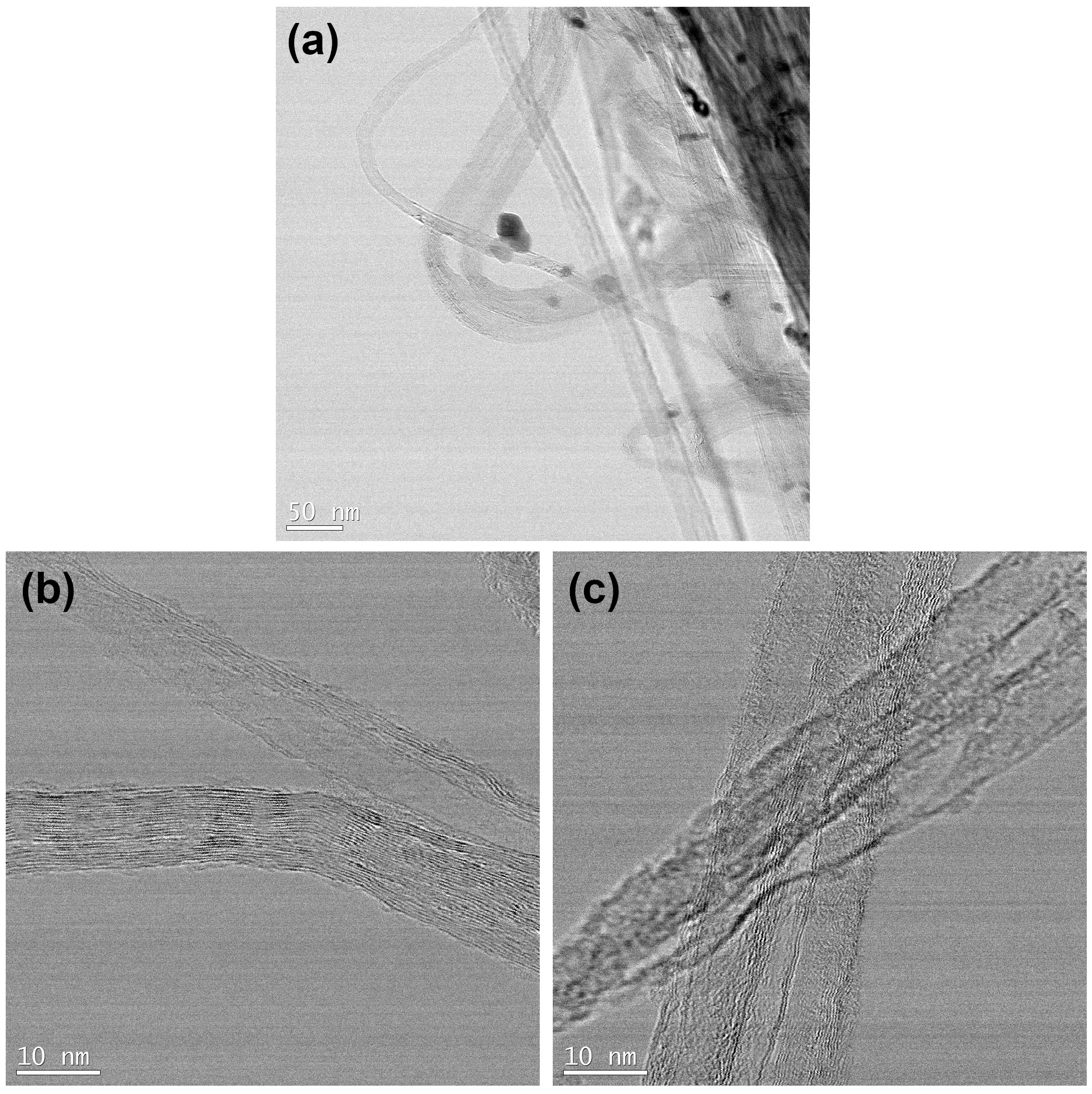}
	\caption{\color{black} High resolution TEM images of the MWCNT fibre used in the present study.	}
	\label{HRTEM}
\end{figure*}
\color{black}

\clearpage
\section{Calculation of the transmission in an atomic chain junction}
\label{supp:chain-analytics}

\begin{figure}[h]
	\begin{center}
		\includegraphics[width=0.9\textwidth]{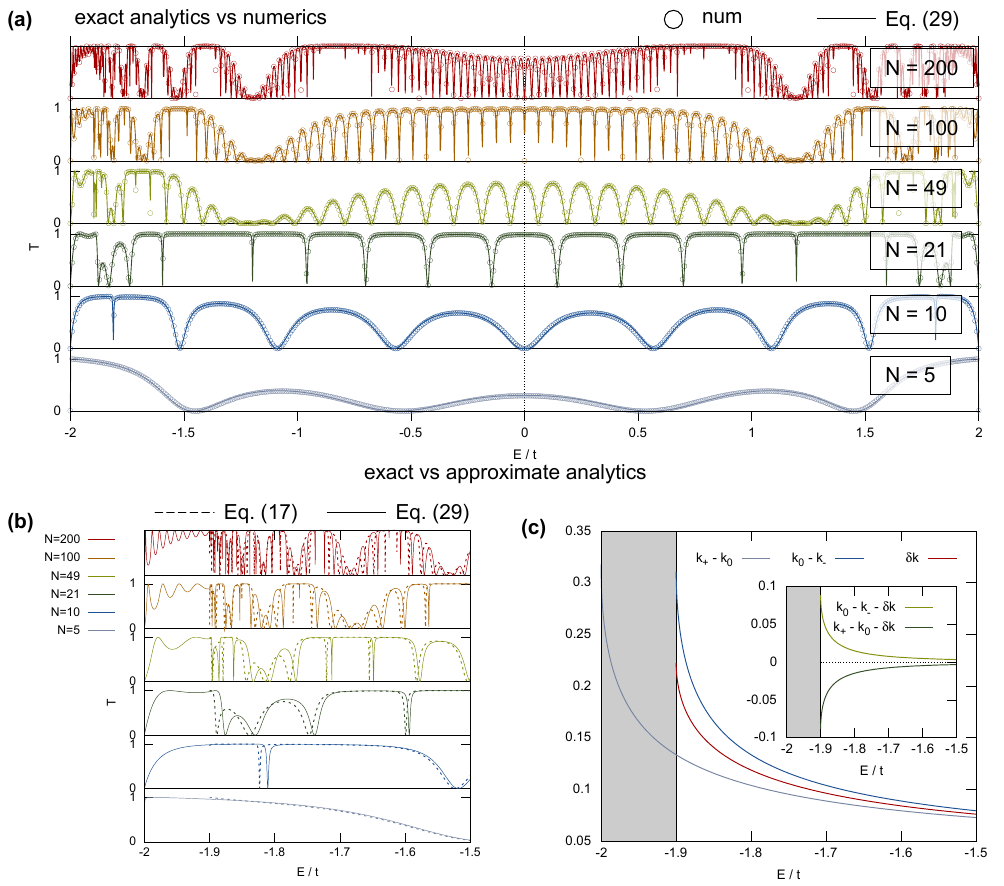}
		\caption{ \label{fig:junction-analytics-numerics}
			Comparison between numerical and the two analytical results. (a) 
			Analytics vs numerics. (b) Difference between the exact Eq.~\eqref{eq:chain-transmission-exact} and the approximate Eq.~\eqref{eq:chain-transmission-formula} in the energy range of maximum discrepancy. (c) Mismatch between $k_+,k_-$ and $K_0\pm \delta k$ as a function of energy for the valence band.}
	\end{center}
\end{figure}

The tight\color{black}-\color{black}binding Hamiltonian of a single monoatomic chain in real space is given by
\begin{equation}
	\label{eq:hamiltonian-chain-real-2quant}
	\hat{H}_{p=L,R} = t \sum_{j} \left( \hat{c}^\dag_{jp}\hat{c}_{j+1,p} + \hat{c}^\dag_{j+1,p}\hat{c}_{j,p}\right) .
\end{equation}
In the region of the junction\color{black}, \color{black} the two chains form a ladder, described by a Hamiltonian with an additional degree of freedom, denoting the bottom ($\alpha$) or top ($\beta$) chain.
\begin{equation}
	\label{eq:hamiltonian-ladder-real-2quant}
	\hat{H}_C = \sum_{j=1}^{N-1} \sum_{p=\alpha,\beta} t \bigl(\hat{c}^\dag_{jp}\hat{c}_{j+1,p} + \hat{c}^\dag_{j+1,p}\hat{c}_{j,p}\bigr) +
	\sum_{j=1}^{N} t' \left(\hat{c}^\dag_{j\alpha}\hat{c}_{j\beta} + \hat{c}^\dag_{j\beta}\hat{c}_{j\alpha}\right) .
\end{equation}
Using the Bloch ansatz for the wave function and setting the lattice constant $a=1$ to simplify the notation, we find
\begin{equation}
	\label{eq:bloch-wave-2quant}
	\twovector{\hat{c}_{k\alpha}^\dag}{\hat{c}^\dag_{k\beta}} = \frac{1}{\sqrt{N}} \sum_j e^{ikj} \twovector{\hat{c}^\dag_{j\alpha}}{\hat{c}^\dag_{j\beta}},
	\quad\textnormal{and conversely, }\quad
	\twovector{\hat{c}_{j\alpha}^\dag}{\hat{c}^\dag_{j\beta}} = \frac{1}{\sqrt{N}} \sum_k e^{-ikj} \twovector{\hat{c}^\dag_{k\alpha}}{\hat{c}^\dag_{k\beta}}.
\end{equation}
The Hamiltonian for single chains in reciprocal space is then
\begin{equation}
	\label{eq:hamiltonian-chain-reciprocal}
	\hat{H}_{L/R} = \sum_{k\in 1BZ} H_{L/R}(k)\;\hat{c}_{k,L/R}^\dag \hat{c}_{k,L/R},\quad
	H_{L/R}(k) = 2t\,\cos(k).
\end{equation}
In spinor notation we have for the ladder
\begin{equation}
	\hat{H}_C = \sum_{k\in 1BZ} H(k),
	\quad
	H_C(k) = \left(\begin{array}{cc}
		2t\cos k & t' \\
		t' & 2t\cos k
	\end{array}
	\right) = 2t\,\cos\,k\, \mathbbm{1} + t'\sigma_x.
\end{equation}
Its eigenstates are symmetric and antisymmetric combinations of single chain states, with equal weight:
\begin{equation}
	E_\pm(k) = 2t\,\cos k \pm t',\quad \ket{\psi_\pm(k)} =
	\frac{1}{\sqrt{N}} \sum_j e^{ikj} \twovector{1}{\pm1}.
\end{equation}
The wave function for an electron at energy $E$, incoming from the left on chain $\alpha$ and continuing across the junction to leave on the chain $\beta$\color{black}, \color{black} is given by
\begin{subequations}
	\label{eq:wavefunctions}
	\begin{equation}
		\psi_L(j)  =  e^{ikj} + r e^{-ikj},\quad k = \arccos\left(\frac{E}{2t}\right)
	\end{equation}
	\begin{equation}
		\psi_{C\alpha}(j)  =  c_{+,f} e^{ik_+j} + c_{+,b}e^{-ik_+j} + c_{-,f} e^{ik_-j} + c_{-,b}e^{-ik_-j},\quad
		k_\pm = \arccos\left(\frac{E\mp t'}{2t}\right),
	\end{equation}
	\begin{equation}
		\psi_{C\beta}(j)  =  c_{+,f} e^{ik_+j} + c_{+,b}e^{-ik_+j} - c_{-,f} e^{ik_-j} - c_{-,b}e^{-ik_-j} 
	\end{equation}
	\begin{equation}
		\psi_R(j)  =  t e^{ikj}.
	\end{equation}
\end{subequations}
In the central region\color{black}, \color{black}  it is a linear combination of the four wavevectors $\pm k_\pm$ corresponding to the energy $E$, as shown in Fig. 3(b) in the main text. The subscript $f/b$ stands for forward/backward propagating wave ( with $(k>0)/(k<0)$ ). The sign on the ``-'' components in the junction is due to the form of the eigenstate spinors.
The appropriate boundary conditions to apply \color{black} to \color{black} the wave function are those for the continuity of the wave function at  \color{black}the \color{black} interfaces ($x=0$ and $x=L:=N+1$):
\begin{equation}
	\label{eq:bc-wave}
	\psi_L(0) = \psi_{C\alpha}(0),\quad \psi_{C\alpha}(L)=0,
	\quad \psi_{C\beta}(0) = 0,\quad \psi_{C\beta}(L)=\psi_R(L),
\end{equation}
and \color{black} for \color{black} the continuity of the probability current along the unbroken chain, which means the continuity of the derivative:
\begin{equation}
	\label{eq:bc-derivative}
	\frac{\partial}{\partial x} \psi_L(x)|_{x=0} = \frac{\partial}{\partial x} \psi_{C\alpha}(x)|_{x=0},\quad
	\frac{\partial}{\partial x} \psi_{C\beta}(x)|_{x=L} = \frac{\partial}{\partial x} \psi_R(x)|_{x=L}.
\end{equation}
This set of linear equations for the six coefficients $r,c_{\pm f}, c_{\pm b},t$ can be solved analytically, yielding the transmission coefficients
\begin{subequations}
	\label{eq:chain-transmission-exact}
	\begin{equation}
		T(E) = |t(E)|^2 = \left| 16\,i e^{-i(k_0-k_--k_+)L} k_0\,\frac{(k_-\sin(k_+L) - k_+\sin(k_-L))}{\det\mathcal{M}} \right|^2,                          \end{equation}
	\begin{eqnarray}
		\det\mathcal{M} & = &(4k_0^2 + k_-^2 + k_+^2)\left(e^{2ik_-L}-1\right)\left(e^{2ik_+L}-1\right) - 4k_0k_-\left(e^{2ik_-L}+1\right)\left(e^{2ik_+L}-1\right) \\
		& & -4k_0k_+ \left(e^{2ik_-L}-1\right)\left(e^{2ik_+L}+1\right) + 2k_-k_+\left[\left(e^{2ik_-L}+1\right)\left(e^{2ik_+L} +1\right)+ 4e^{i(k_-+k_+)L}\right].\nonumber
	\end{eqnarray}
\end{subequations}
In the simplified approach where we assumed that $k_+-k_0 = k_0 - k_- =: \delta k$, we find the formula from the main text,
\begin{equation}
	\label{eq:transmission-chains-nofield}
	T(E,L) = 64\,k_0^2\frac{\left[k_0\,\cos(k_0L)\,\sin(\delta k\,L)-\delta k\,\cos(\delta k\,L)\,\sin(k_0 L)\right]^2}{\left|\left(k_0+\delta k +e^{-2i\delta k L}(k_0-\delta k)\right)^2 -4 e^{-i2(k_0-\delta k)L} k_0^2\right|^2}.
\end{equation}
The exact formula Eq.~\eqref{eq:chain-transmission-exact} agrees perfectly with the numerical results, as can be seen in Fig.~\ref{fig:junction-analytics-numerics}(a). The numerical results were obtained with \color{black}our  \color{black} own C++ code written using the Armadillo library~\cite{sanderson2025corr} and the Green's function technique described in the Methods.   

\begin{figure}[h]
	\begin{center}
		\includegraphics[width=0.9\textwidth]{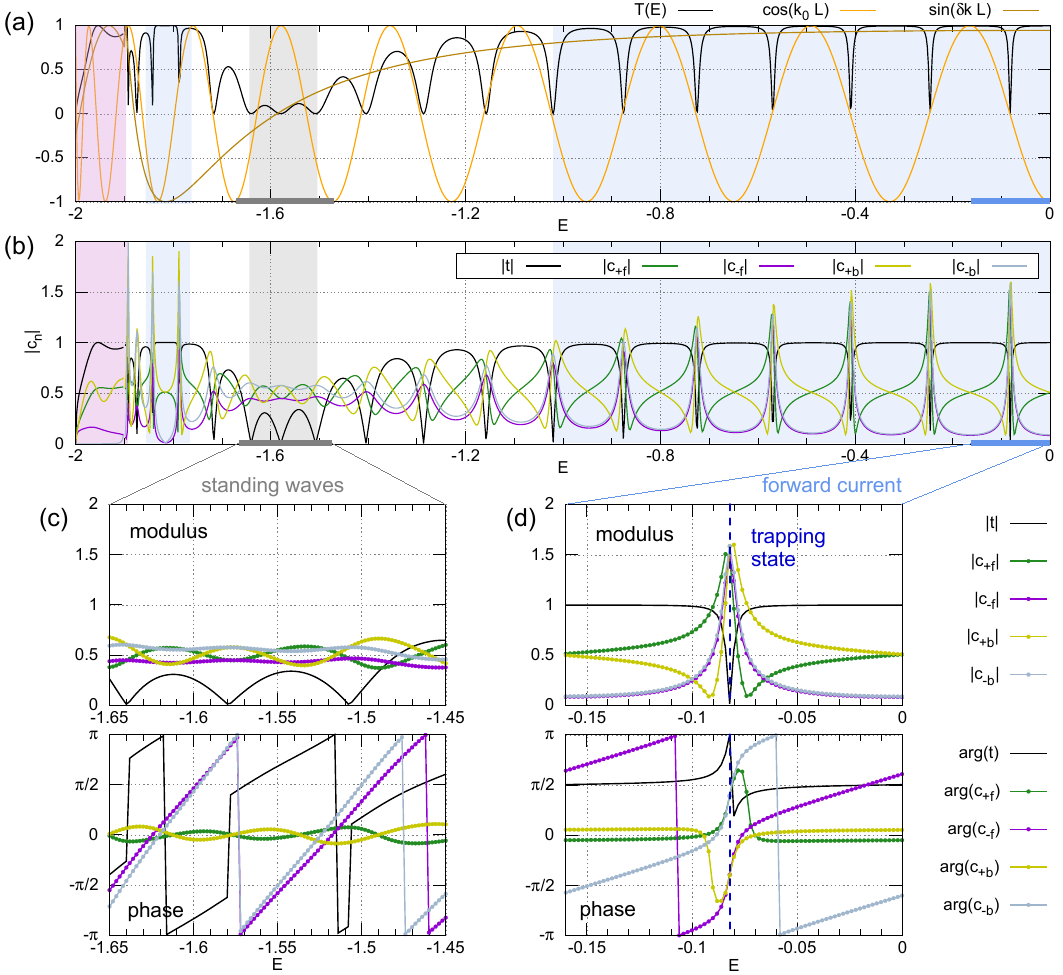}
		\caption{ \label{fig:junction-coefficients-extended}
			Composition of the wave functions in the junction with $N=37$ sites, $t=-1$ and $t'=-0.1$. (a) Transmission and the functions of the incoming momentum $k_0$ and momentum difference $\delta k$, which determine the transmission regime: $\cos(k_0L)$ and $\sin(\delta k L)$. (b) Absolute value of the coefficients $c_{\pm,f/b}$ from Eqs.~\ref{eq:wavefunctions}. In (a) and (b)\color{black}, \color{black} the coloured areas correspond to the transport regimes from the main text. (c) Absolute value and phase of the $c_{\pm,f/b}$ in the range of the standing waves regime. (d) The same for the forward current regime, with a trapping state in the \color{black} centre \color{black} of the range.
		}
	\end{center}
\end{figure}

\begin{figure}[h]
	\begin{center}
		\includegraphics[width=0.9\textwidth]{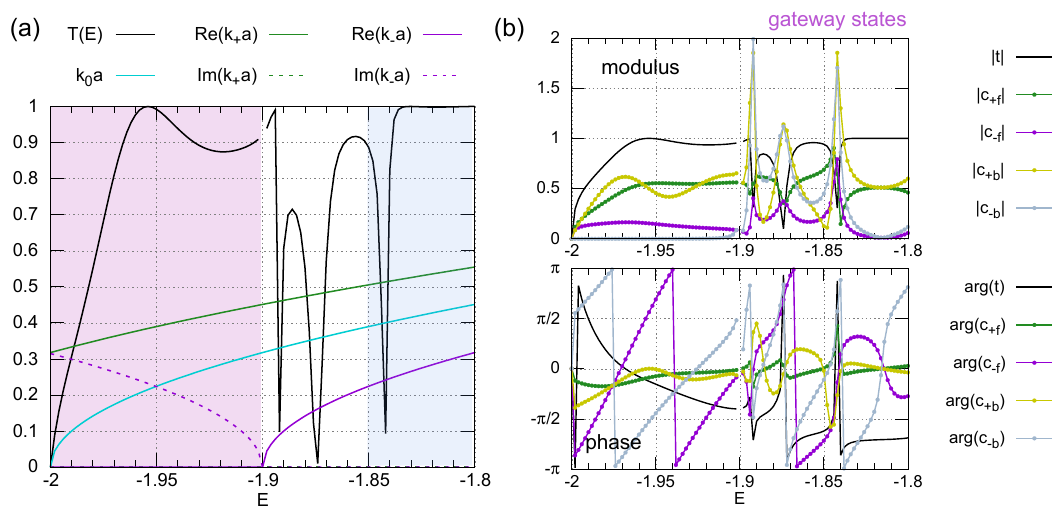}
		\caption{ \label{fig:junction-coefficients-gateway}
			Transmission and wave function composition in the gateway states regime. (a) $T(E)$ and the solutions for $k_\pm(E)$. For $E/t<-1.9$\color{black}, \color{black} $k_-(E)$ is purely imaginary. (b) modulus and (c) phase of the coefficients of the linear combination, $c_{\pm,f/b}$.
		}
	\end{center}
\end{figure}

The two analytical formulae for transmission yield very similar results, as can be seen in Fig.~\ref{fig:junction-analytics-numerics}(b). The source of \color{black}the \color{black} discrepancy is the assumption $k_+-k_0 = k_0 - k_- =\delta k$, which holds only away from the band edges, as shown in Fig.~\ref{fig:junction-analytics-numerics}(c).
Depending on $E$ and $L$, the linear combinations carrying the current in the junction vary, as shown in Fig.~\ref{fig:junction-coefficients-extended} and Fig\color{black}.\color{black}~\ref{fig:junction-coefficients-gateway}. They are different in each of the four regimes discussed in the main text.

{\em Forward currents}. When $\sin(\delta k L)\simeq 1$, the linear combination contains only the $k>0$ (forward) states with almost the same phase (see Fig.~\ref{fig:junction-coefficients-extended}(d)). It is then given by
\begin{equation}
	\psi_\alpha(x) \propto e^{ik_0x}\cos(\delta k x), \quad
	\psi_\beta(x) \propto e^{ik_0x}\sin(\delta k x).
\end{equation}
The amplitude of the wave function is maximal both at the entry to \color{black}  ($\psi_\alpha(0)$) and at the exit from the junction ($\psi_\beta(L)$)\color{black}.

{\em Trapping states}. Throughout the whole energy range we see sharp dips in transmission, occurring whenever $\cos(k_0L)=0$. At these energies the linear combination contains all four states with equal magnitudes, but different phases: $\arg(c_{+f})=\arg(c_{-b})\simeq \pi/2$, and $\arg(c_{+b})=\arg(c_{-f})\simeq -\pi/2$. The linear combination \color{black} then yields\color{black}
\begin{equation}
	\psi_\alpha(x) \propto \sin(\delta k x)\,\cos(k_0 x),\quad
	\psi_\beta(x) \propto \sin(k_0x)\,\cos(\delta k x).
\end{equation}
The wave function on chain $\alpha$ is forced to vanish at the entry into the junction by $\sin(\delta k x)$. This suppresses the transmission for any $\delta k$, but the sharp dip is especially noticeable in the regime of forward currents, $\sin(\delta k L)\simeq 1$\color{black}; \color{black} then $\cos(\delta k L) = 0$ and the wave function has a node also at the exit.

{\em Standing waves}. When $\sin(\delta k L) \simeq 0$, transmission is very low even if $\cos(k_0 L) \simeq \pm 1$. The magnitudes of all four contributions to the electronic wave function in the junction are again very similar (see Fig.~\ref{fig:junction-coefficients-extended}(c)). Their phases are different, $\arg(c_{+f})=\arg(c_{+b})\simeq 0$, and $\arg(c_{-b})=\arg(c_{-f})\simeq \pi$. The resulting wave function is
\begin{equation}
	\psi_\alpha(x) \propto \sin(k_0 x)\,\cos(\delta k x),
	\quad \psi_\beta(x) \propto \sin(\delta k x)\,\cos(k_0 x).
\end{equation}
The amplitude of the wave function is then suppressed at the entry by $\sin(k_0 x)$, and at the exit by the condition $\sin(\delta k L)\simeq 0$. Even if $\cos(k_0L)=1$ \color{black}, so that \color{black} we would expect a peak in transmission, this peak turns into a dip to 0.

{\em Gateway states}. In this regime, where one of the wave vectors is imaginary and describes a decaying wave, we find that the current can again be carried mostly by the forward propagating wave from the other branch of the dispersion. From the mathematical point of view the only role of the evanescent wave is to ensure the \color{black} fulfilment \color{black} of the boundary conditions, as a side effect facilitating the transport through the junction.

\section{Atomic chain in perpendicular magnetic field}
\label{supp:chain-magnetic}

With constant $\mathbf{B}=B\,\mathbf{e}_z$\color{black}, \color{black} we choose for $\mathbf{A}(\mathbf{r})$ the gauge in which translational symmetry along $x$ is preserved, $\mathbf{A}(\mathbf{r})=-yB\,\mathbf{e}_x$. The hopping integrals on each chain are then modified by the appropriate Peierls phase,
\begin{equation}
	\frac{ie}{h} \int_{ja}^{(j+1)a} A_x dx = -i\frac{2\pi}{\Phi_0} Ba\,y \quad\rightarrow\quad
	t_\alpha \rightarrow t\,e^{i\varphi},\quad t_\beta \rightarrow t\,e^{-i\varphi}, \quad \varphi := -\pi Ba\,d/\Phi_0,
\end{equation}
where $\Phi_0=h/e$ is the (nonsuperconducting) magnetic flux quantum, $a$ the lattice constant on each chain (we bring it back to use the concept of magnetic flux per unit cell), $\mp d$ the position of the chain $\alpha/\beta$ on the $y$ axis and $\varphi$ the flux per unit cell in the units of $\Phi_0$. The hopping $t'$, in the direction perpendicular to $\mathbf{A}$, remains unchanged. Because we keep the translational symmetry, we can still use Bloch's theorem\color{black}, \color{black} and $k$ is a good quantum number.

\begin{figure}[h]
	\includegraphics[width=\textwidth]{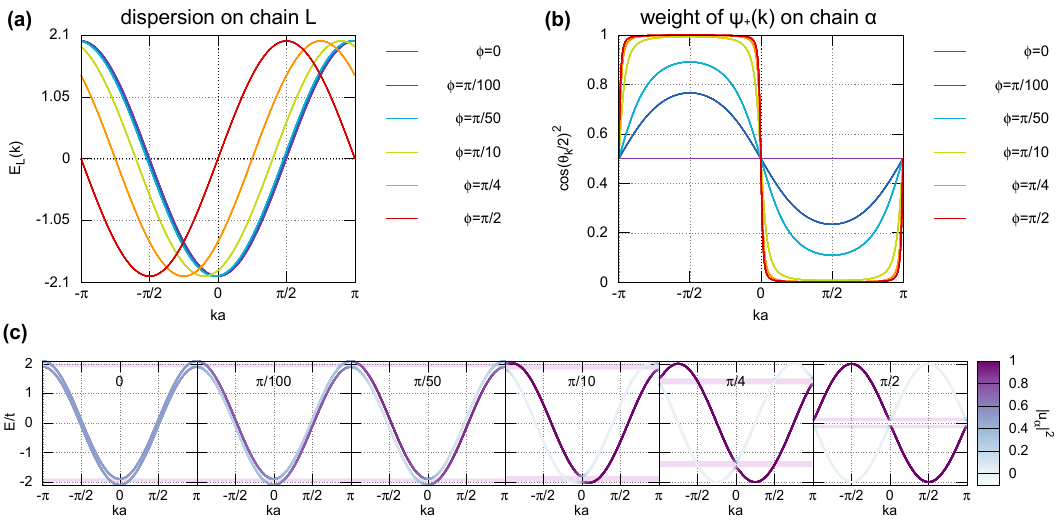}
	\caption{\label{fig:chains-magnetic-supp}
		Electronic energy levels and eigenstates in a chain junction in perpendicular magnetic field. (a) Dispersion in chain $L$ shifted by the Peierls phase $\varphi$. (b) Weight of the state $\ket{\psi_+(k)}$ on chain $\alpha$. (c) The two branches of electronic dispersion in the junction. Colour encodes the weight of the particular $\ket{\psi_\pm(k)}$ state on chain $\alpha$\color{black}; \color{black} violet background marks the range of gateway states.}
\end{figure}

The dispersion on individual side chains is now phase-shifted,
\begin{equation}
	E_{L/R}(k) =  2t\,\cos(ka \pm\varphi).
\end{equation}
The Hamiltonian of the central ladder is modified more strongly.
\begin{equation}
	H_C(k) = \twoarray{2t\cos(ka+\varphi)}{t'}{t'}{2t\cos(ka-\varphi)}
	= 2t\cos ka\,\cos\varphi \mathbbm{1} + 2t\sin ka\,\sin\varphi\; \sigma_z + t'\sigma_x.
\end{equation}
The dispersion of the ladder and its eigenstates are now
\begin{subequations}
	\begin{equation}
		E_\pm(k) = 2t\,\cos ka\,\cos\varphi \pm \sqrt{(t')^2 + (2t\,\sin ka\, \sin\varphi)^2},
	\end{equation}
	\begin{equation}
		\ket{\psi_+(k)} = \frac{1}{N} \sum_j e^{ikj}\twovector{\cos(\theta_k/2)}{\sin(\theta_k/2)},\quad
		\ket{\psi_-(k)} = \frac{1}{N} \sum_j e^{ikj} \twovector{\sin(\theta_k/2)}{-\cos(\theta_k/2)},
	\end{equation}
	\begin{equation}
		\theta_k := \arctan\left( \frac{t'}{2t\,\sin ka\,\sin\varphi}\right).
	\end{equation}
\end{subequations}
The dispersion itself isn't very sensitive to $\varphi$ (cf. Fig.~\ref{fig:chains-magnetic-supp}(a), where up to $\varphi=\pi/50$ the $E(k)$ curves are almost indistinguishable). In contrast, the effective angle $\theta_k$ depends on $\varphi$ and on $k$, \color{black} and this dependence is strong for $t' \ll t$, \color{black} as can be seen in Fig.~\ref{fig:chains-magnetic-supp}(b).
Since the spinorial part of the eigenstates encodes the distribution of a given Bloch state between the two chains, for $\varphi\not\approx \pi n$ the only regions of the Brillouin zone where \color{black} hybridisation \color{black} is possible ($\cos(\theta_k/2)\approx 1/\sqrt{2}$) are the close \color{black} neighbourhood \color{black} of $k=0$ and $k=\pm\pi/a$. Transmission is quickly suppressed with increasing $\varphi$ in the \color{black} two-band \color{black} energy range by the extremely skewed occupation of the two chains. In contrast, the gaps opened by the \color{black} hybridisation \color{black} (cf. Fig.~\ref{fig:chains-magnetic-supp}(c)) still host evanescent waves, which act as gateway states maintaining the transmission close to unity in these narrow energy ranges. 

\color{black}
\subsection{Including the electron spin}
The complete Hamiltonian of an electron in a magnetic field should also include a Zeeman term. Without spin-orbit coupling, the natural spin quantisation axis for the electrons is given by the external magnetic field, and the Zeeman Hamiltonian becomes $H_Z = g \mu_B B/2\sum_{i,\sigma} \sigma \hat{c}_{i\sigma}^\dag \hat{c}_{i\sigma}$, where $g$ is the gyromagnetic factor (in CNTs $g\simeq 2$) and $\mu_B$ the Bohr magneton. In our minimal model with all parameters arbitrary, it is more convenient to express it as $H_Z = g_{\mathrm{eff}} \varphi_B  \sum_{i,\sigma} \sigma \hat{c}_{i\sigma}^\dag \hat{c}_{i\sigma}$, where $g_{\mathrm{eff}} = -g\mu_B\Phi_0/(2\pi ad)$, $\varphi_B$ is the Peierls phase in the left chain,  and we choose the unit of $g_\mathrm{eff}$ to be $t/(2\pi)$.

\begin{figure}[h]
	\begin{center}
		\includegraphics[width=0.8\textwidth]{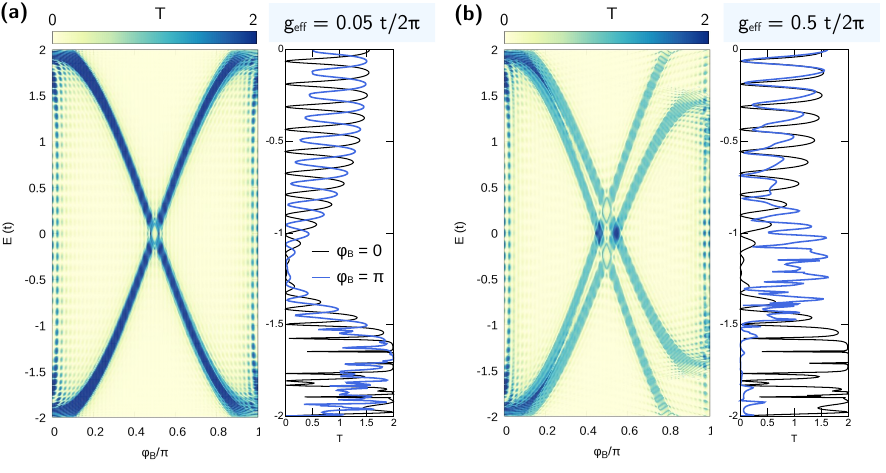}
	\end{center}
	\caption{\label{fig: response-zeeman}\color{black} Transmission in the spinful atomic chain model with the junction containing 49 unit cells.}
\end{figure}

Without spin-orbit coupling or magnetic impurities, the transmission of a spinful system $T_{(\sigma)}$ is a simple sum of two independent spin contributions shifted by the Zeeman effect, $T_{(\sigma)}(E) = T(E+g_{\mathrm{eff}}\varphi_B) + T(E-g_{\mathrm{eff}}\varphi_B)$, as shown in Fig.~\ref{fig: response-zeeman}.  The strength of the Zeeman shift depends on the value of $g_{\mathrm{eff}}$, yielding patterns where the two contributions can be distinguished ess clearly (panel (a)) or more clearly (panel (b)).
\color{black}

\section{Transmission through two atomic chain junctions in series}
\label{supp:chain-multijunction}

\begin{figure}[h]
	\begin{center}
		\includegraphics[width=\textwidth]{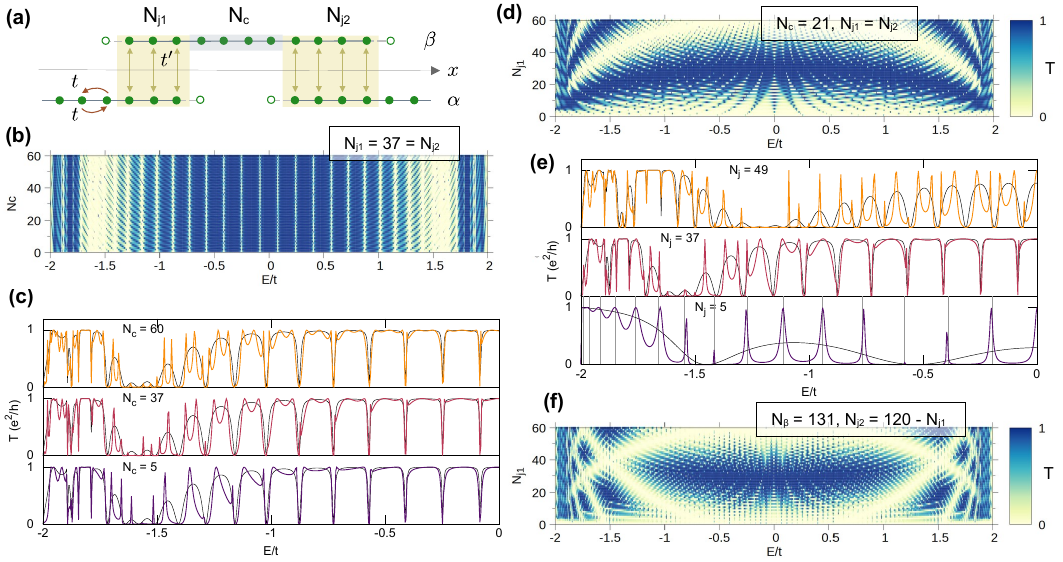}
	\end{center}
	\caption{\label{fig:chain-multijunction}
		(a) Two atomic junctions in series, with the length of the bridge chain $N_\beta = N_c + N_{j1}+N_{j2}$. (b) Transmission through a double junction with increasing bridge length and fixed junction lengths. (c) Line cuts of (b) at $N_c=5,37,60$. The black lines are $T(E)$ of a single junction with $N_j=37$. (d) Transmission through a symmetric ($N_{j1}=N_{j2}$) double junction with fixed contact length and increasing length of the junctions. (e) Line cuts across (d) for $N_j=5,37,49$ with black lines showing $T(E)$ for a single junction with the same length. Grey lines mark the energy levels in an isolated $\beta$ chain with $L=N_c+2N_j=31$. (f) Transmission with the length of the bridge chain fixed, and the total length of the two junctions $N_{j1}+N_{j2}=120$.}
\end{figure}

\begin{figure*}[h!tb]
	\centering
	\includegraphics[width=1.0\linewidth]{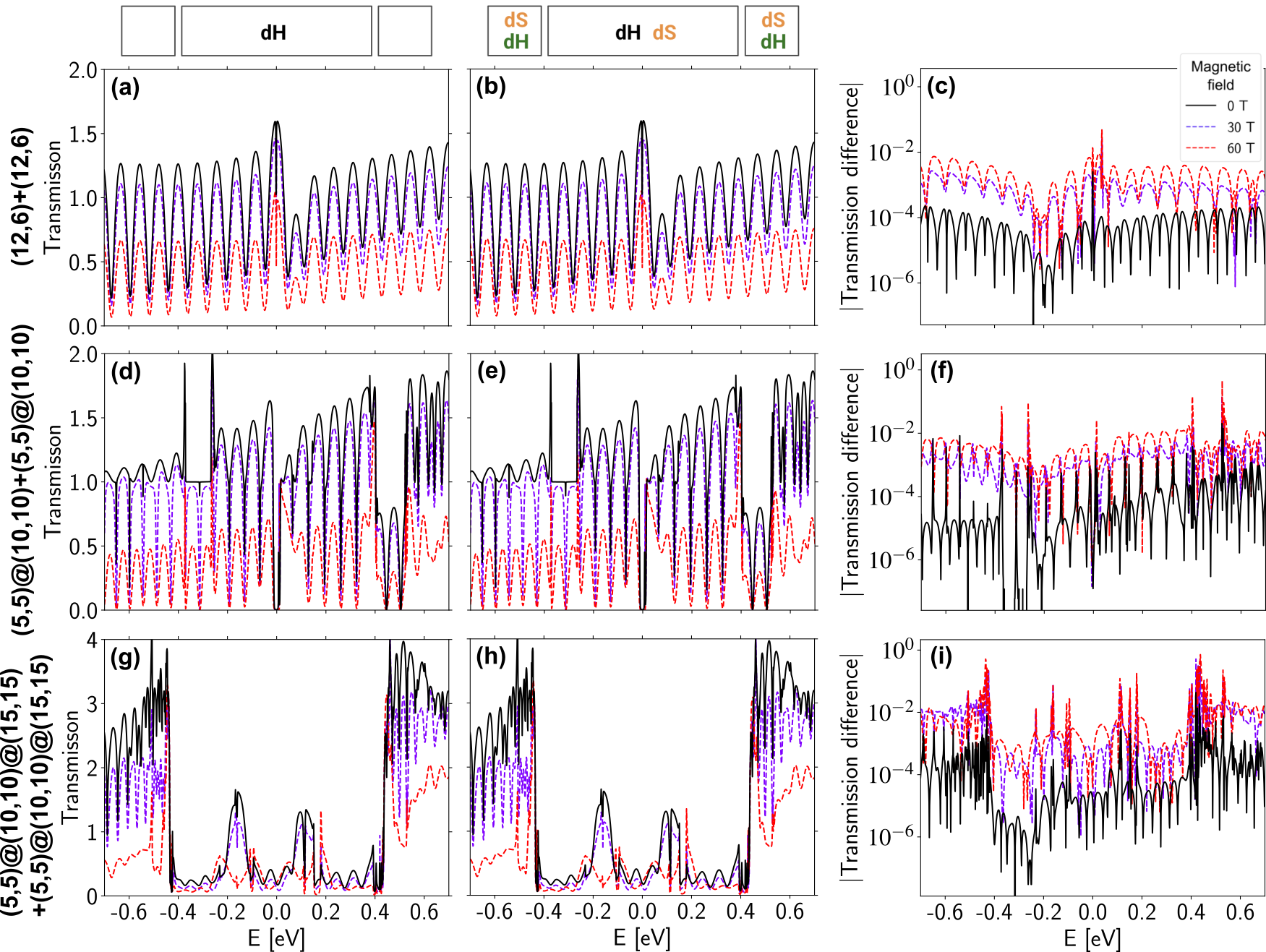}
	\caption{\color{black} Benchmarks of the transmission for the (a-c) (12,6)+(12,6) SWCNT junction with an overlap length of 20.29 nm (18 unit cells), (d-f) (5,5)@(10,10)+(5,5)@(10,10) DWCNT junction with an overlap length of 24.60 nm (100 unit cells), (g-i) (5,5)@(10,10)@(15,15)+(5,5)@(10,10)@(15,15) TWCNT junction with an overlap length of 24.60 nm (100 unit cells), in a perpendicular magnetic field, obtained using (a,d,g) the unmodified and (b,e,h) the modified versions of the \texttt{sisl} and \texttt{TBtrans} codes. (c,f,i) Absolute differences between the corresponding transmissions. }
	\label{bench}
\end{figure*}

\begin{figure*}[h!tb]
	\centering
	\includegraphics[width=1.0\linewidth]{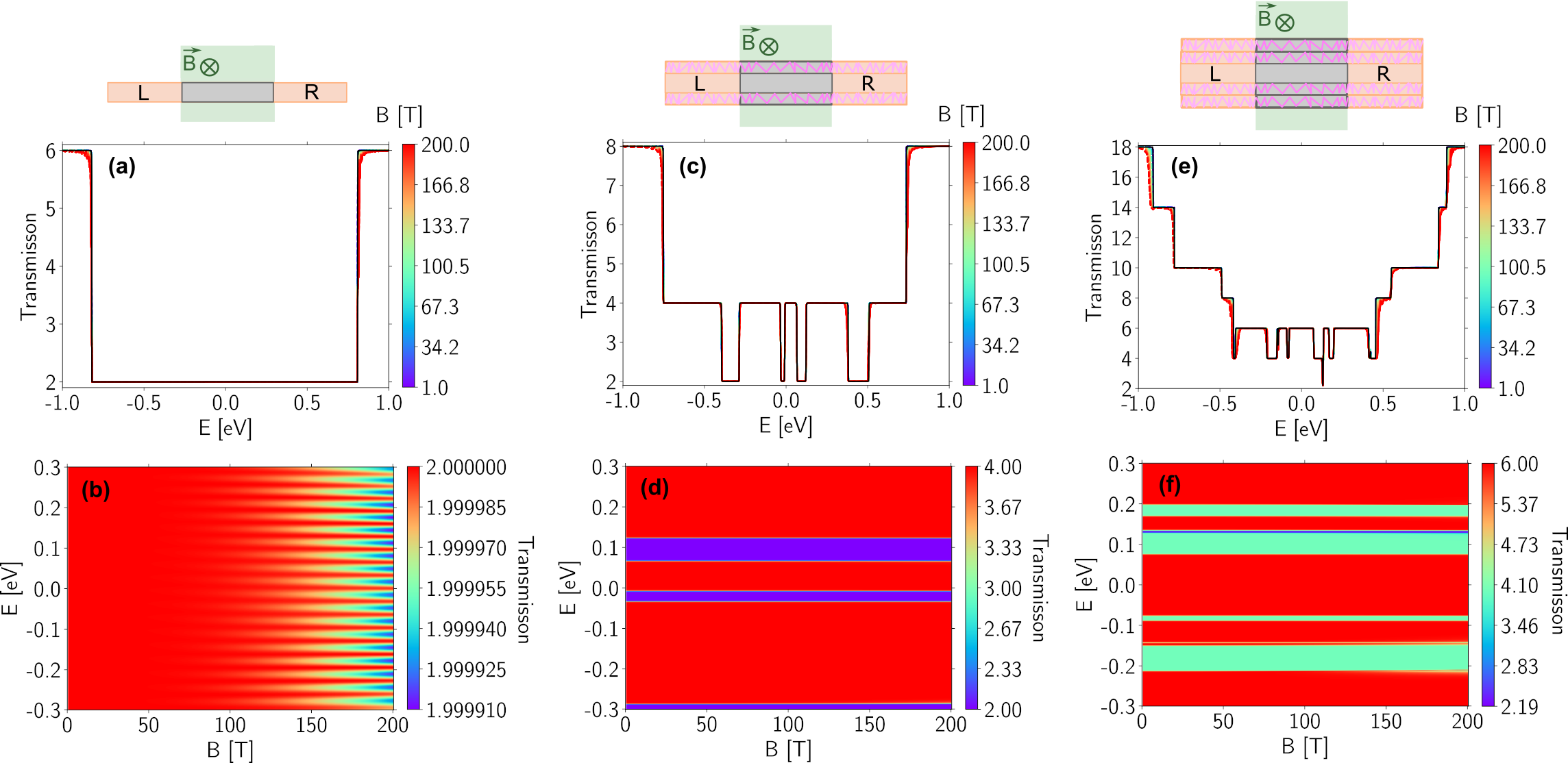}
	\caption{Computed zero-bias transmission spectra for (a) (9,9) SWCNT, (c) (5,5)@(10,10) DWCNT, and (e) (5,5)@(10,10)@(15,15) MWCNT as a function of electron energy $E$ and magnetic field $B$ up to 200\,T. (b,d,f) Corresponding transmission maps for the systems shown in panels (a), (c), and (e), respectively. For the SWCNT, the magnetic-field-induced change in transmission is of the order of 10$^{-5}$, whereas for the DWCNT and MWCNT it is of the order of 10$^{-1}$, but only within a narrow energy range.
	}
	\label{99_single}
\end{figure*}

\begin{figure*}[h!tb]
	\centering
	\includegraphics[width=0.70\linewidth]{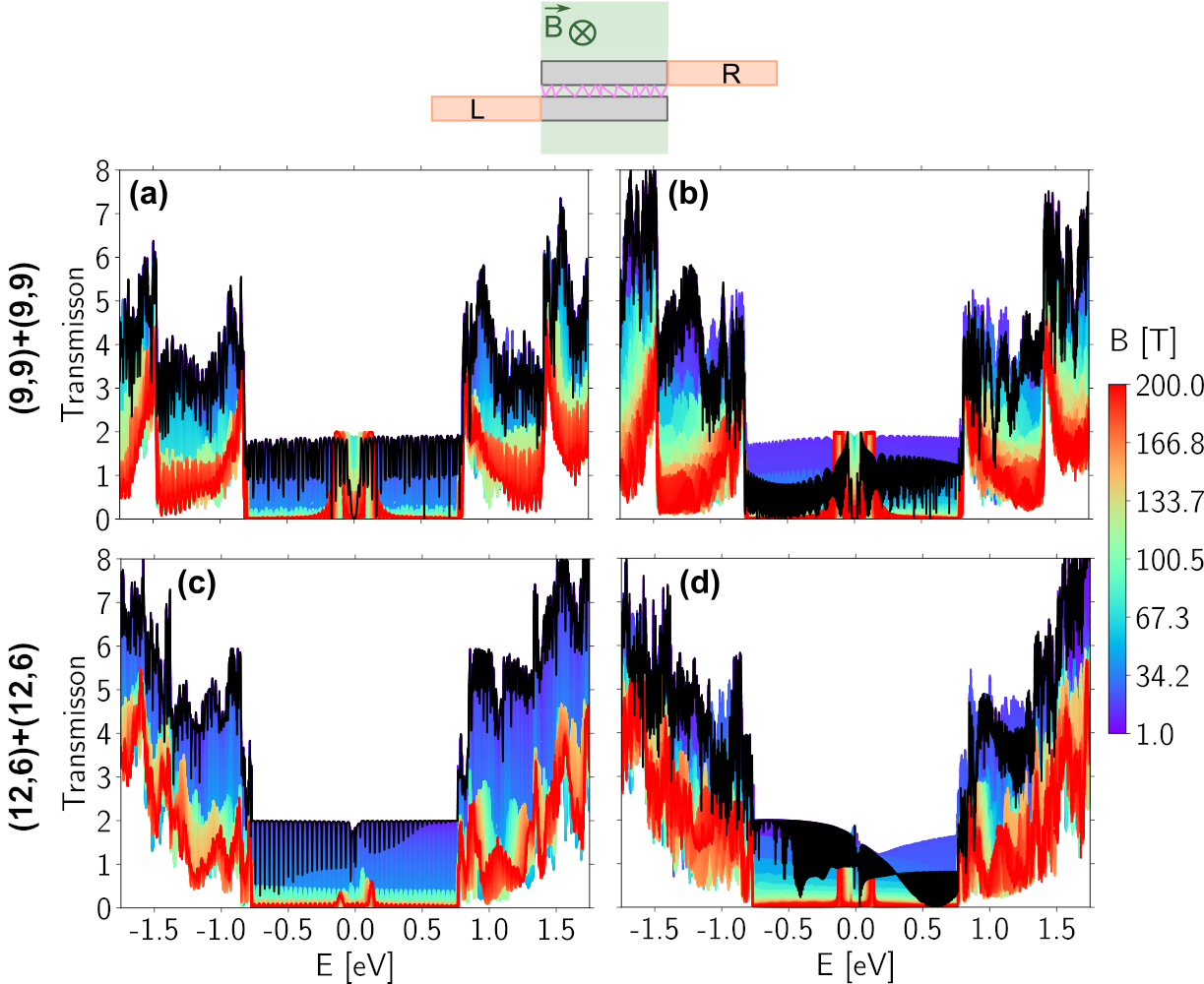}
	\caption{
		Computed zero-bias transmission spectra over an extended energy range for (a,b) (9,9)+(9,9) and (c,d) (12,6)+(12,6) metallic--metallic CNT junctions with overlap lengths of (a) 49.19\,nm, (b) 196.76\,nm, (c) 45.08\,nm, and (d) 180.33\,nm, as a function of electron energy $E$ and magnetic field $B$.
	}
	\label{transmission}
\end{figure*}

\begin{figure*}[h!tb]
	\centering
	\includegraphics[width=0.75\linewidth]{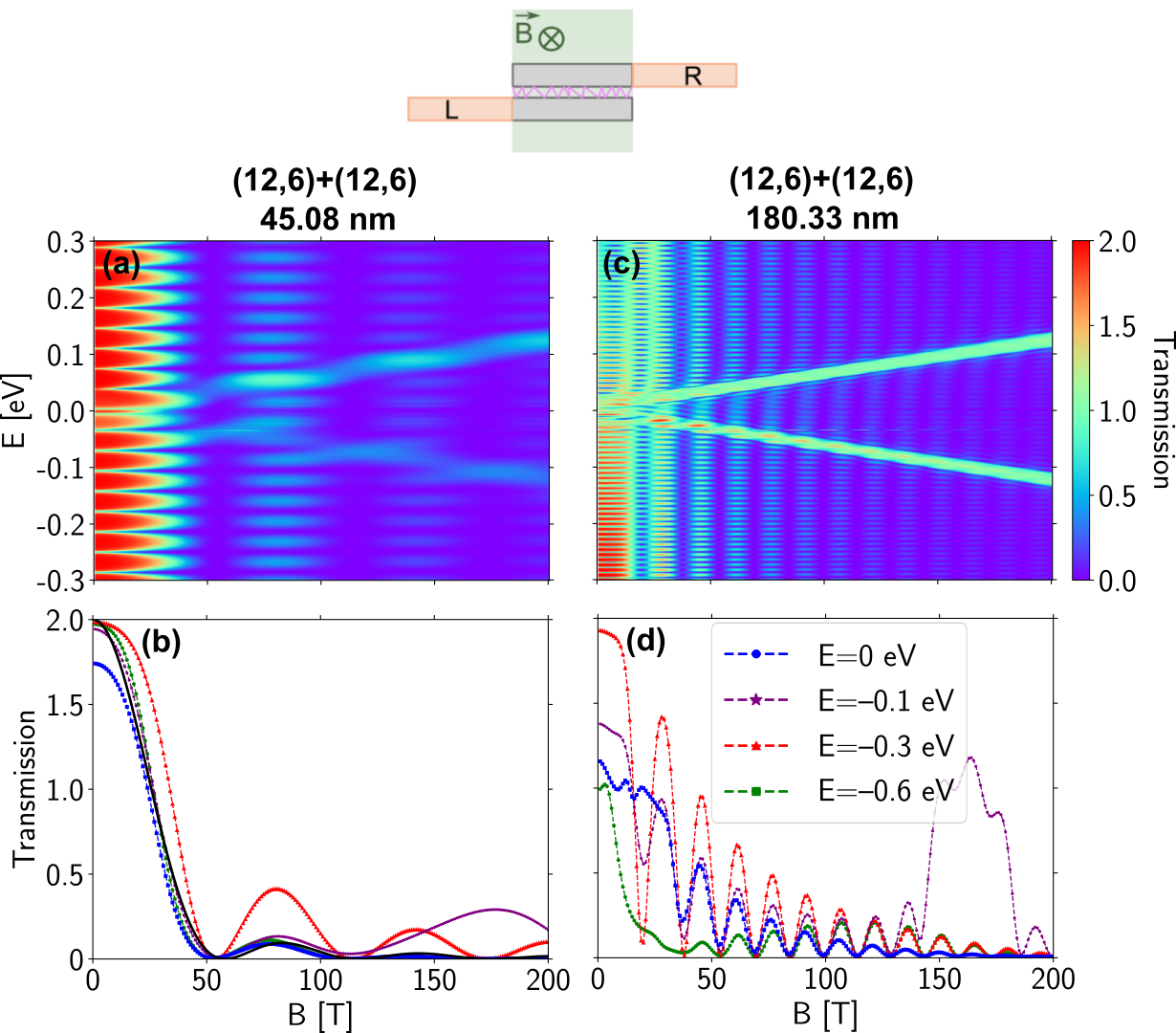}
	\caption{Computed zero-bias transmission maps for (12,6)+(12,6) CNT junctions as a function of electron energy $E$ and external perpendicular magnetic field $B$: (a) overlap length of 45.08\,nm (40 unit cells) and (c) overlap length of 180.33\,nm (160 unit cells). (b,d) Cross-sections of the corresponding transmission maps, plotted at $E=0$, -0.1, -0.3\,eV, and -0.6\,eV, showing transmission oscillations as a function of magnetic field $B$. In panel (b), the solid black line is the fitted Fraunhofer diffraction formula.
	}
	\label{55_map_SI}
\end{figure*}

\begin{figure*}[h!tb]
	\centering
	\includegraphics[width=0.7\linewidth]{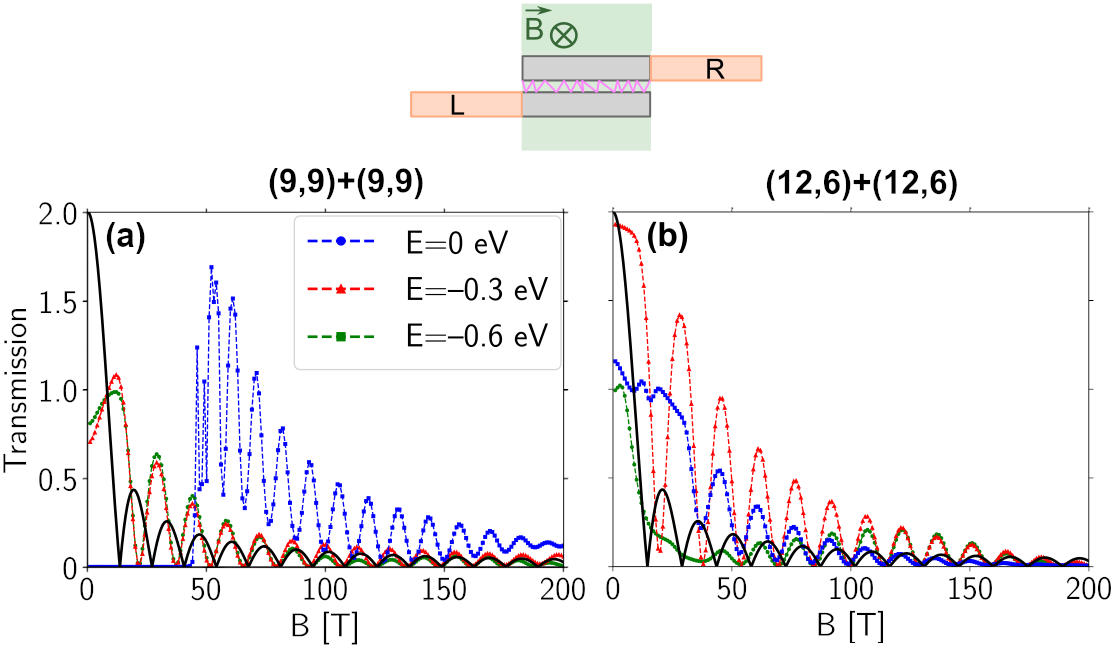}
	\caption{Comparison of the computed transmission $T(B)$ with the phenomenological long-junction envelope $T_{\mathrm{max}}\left| \sin(\pi\Phi/\Phi_0)/(\pi\Phi/\Phi_0) \right|$ (solid black line), where $\Phi=B(2r+d)L$, for long metallic--metallic SWCNT--SWCNT junctions: (a) (9,9)+(9,9) with an overlap length of 196.76\,nm (800 unit cells) and (b) (12,6)+(12,6) with an overlap length of 180.33\,nm (160 unit cells). The computed line cuts are shown for $E=0$, $-0.3$, and $-0.6$\,eV. For these long overlaps, repeated
		intertube hopping and internal reflections make the squared Fraunhofer transmission probability insufficient; the $\left|\sin x/x\right|$ form therefore serves only as a
		phenomenological envelope of the oscillatory $T(B)$ response.
	}
	\label{fraunhofer}
\end{figure*}

\begin{figure*}[h!tb]
	\centering
	\includegraphics[width=0.7\linewidth]{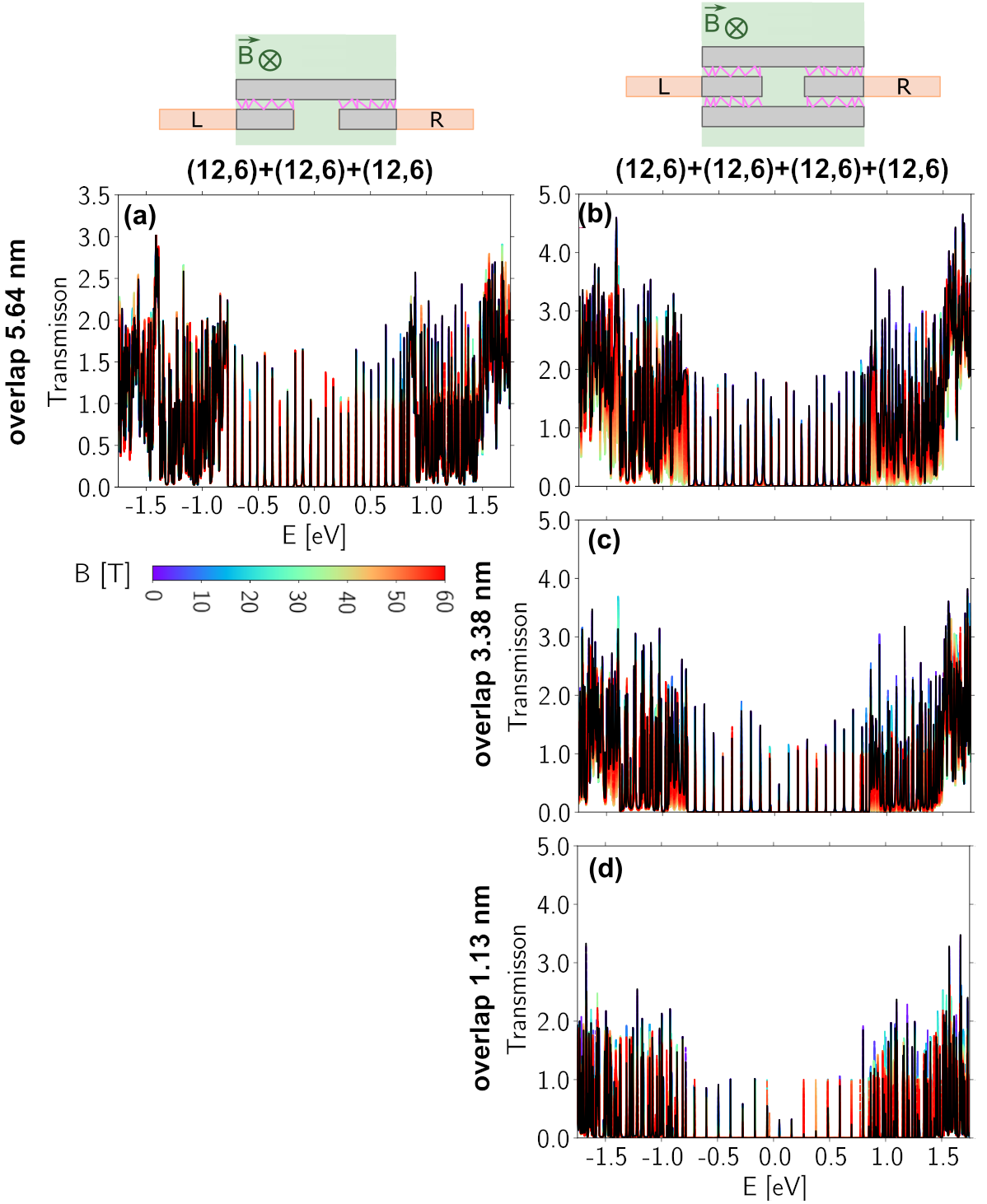}
	\caption{Computed zero-bias transmission spectra over an extended energy range for multi-junction and loop architectures made of metallic (12,6) CNTs without magnetic field (black line) and under an external perpendicular magnetic field up to 60\,T. Panel (a) corresponds to the multi-junction with overlap length 5.64\,nm (five unit cells), whereas panels (b–d) show loops with overlap lengths of 5.64, 3.38 (three unit cells), and 1.13\,nm (one unit), respectively.
	}
	\label{loops2}
\end{figure*}

\begin{figure}[h!]
	\centering
	\includegraphics[width=0.6\textwidth]{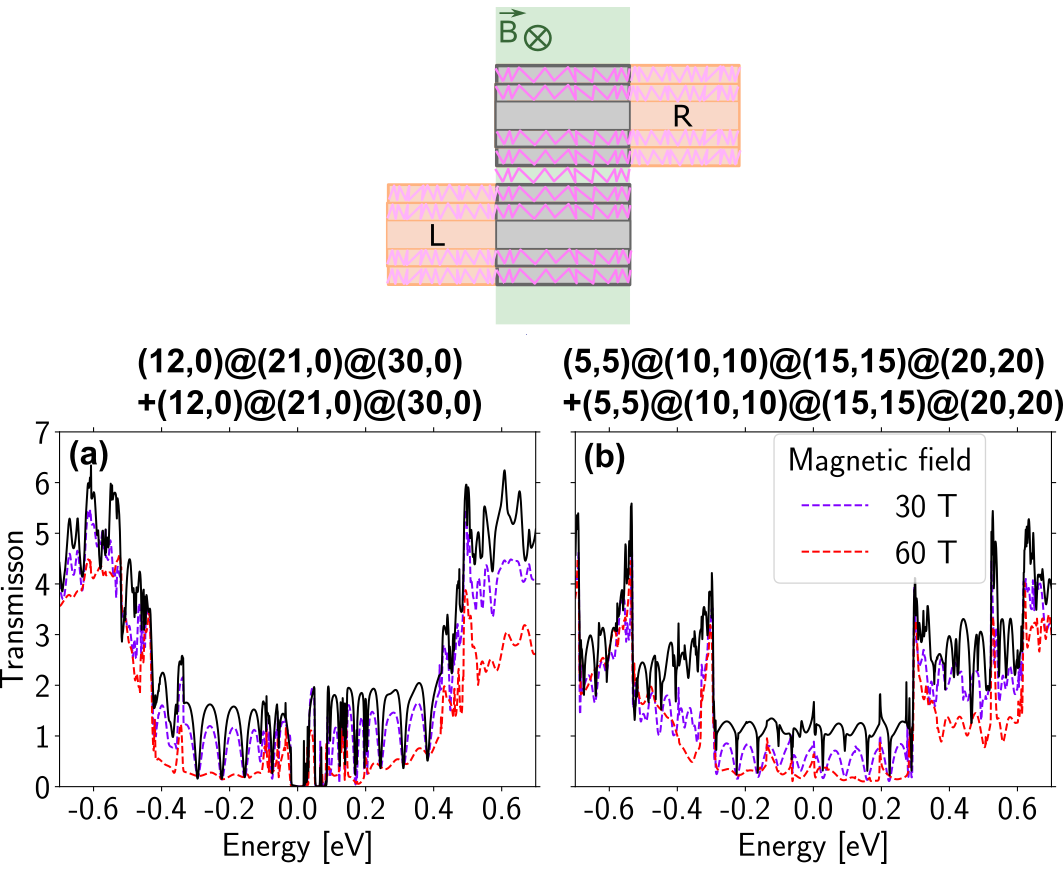}
	\caption{\color{black}  Transmission at B=0 (black line), 30 T (purple line), 60 T (red line) of (a) (12,0)@(21,0)@(30,0)+(12,0)@(21,0)@(30,0) and (b) (5,5)@(10,10)@(15,15)@(20,20)+(5,5)@(10,10)@(15,15)@(20,20) CNT junctions. }
	\label{new_mwcnts}
\end{figure}

When the two junctions are connected in series as sketched in Fig.~\ref{fig:chain-multijunction}, we might expect the electron transfer probability, i.e. the transmission, to be roughly a product of the individual transmissions of each junction. Nevertheless, since our transport is fully coherent, the fact that the chain $\beta$ bridging the two \color{black} $\alpha$ chains \color{black} is finite forces the wave function to have nodes at both ends of the bridge chain $\beta$. The length of the bridge might therefore play some role.

As we see from the numerical results shown in Fig.~\ref{fig:chain-multijunction}(b) and the line cuts in Fig.~\ref{fig:chain-multijunction}(c), in a symmetric set-up similar to that in Section 3.4 of the main text, once the junction lengths are fixed the overall character of transmission almost doesn't depend on the length of the single chain part of the bridge ($N_c=N_\beta-2N_{j}$). Its main effect is to superpose a finer oscillation on the transmission of a single junction. This oscillation is negligible in the forward current regime due to the very low reflection coefficient, so that transmission is indeed a product of transmissions of individual junctions. When the reflection coefficient increases, in the standing wave and intermediate regimes, the second junction splits the transmission into distinct sub-peaks. In Fig.~\ref{fig:chain-multijunction}(d)\color{black} and \color{black} (e)\color{black}, \color{black} a complementary result is shown, for a system with fixed $N_c$ but increasing length of the junctions. Here again, the result is very similar to that in Fig.~\ref{fig:chain-junction-basic}(c) of the main text -- the large scale transmission features in a symmetric double junction are the same as in a single junction of the same length. The line cuts shown in Fig.~\ref{fig:chain-multijunction}(e) show more clearly the pattern of finer oscillations. The grey lines for $N_j=5$ mark the position of electronic energy levels in an isolated $\beta$ chain, with the overall $T(E)$ profile reminiscent of Fig.~\ref{loop} (a) in the main text. When the double junction is not symmetric, transmission is roughly that of a product of the two individual transmissions. This is illustrated in Fig.~\ref{fig:chain-multijunction}(f), where we show the numerical results for the double junction with fixed bridge length $N_\beta=131$ and $N_{j2}=120-N_{j1}$.

\section{Coupled-mode derivation of the CNT-CNT junction transmission formula
}\label{sec:kappa_derivation}

Let us consider two identical nanotubes forming a junction with an overlap region of length $L$, extending from $z=0$ to $z=L$ (the scheme above the panels in Figure~\ref{transmission_zoom}). For an electron with fixed energy $E$, the uncoupled modes in the two nanotubes have the same longitudinal wave vector, $k_1=k_2=k$, corresponding to the phase-matched limit of the coupled-mode description. Within this description, the electronic wavefunction in the overlap region can be written as
\begin{equation}
	\Psi(z)=a(z)U_1e^{ikz}+b(z)U_2e^{ikz},
	\label{eq:S1_bloch}
\end{equation}
where $U_1e^{ikz}$ and $U_2e^{ikz}$ are the Bloch modes of the two nanotubes, while $a(z)$ and $b(z)$ are slowly varying amplitudes. The effective intertube coupling is described by the coefficient $\varkappa$, which has units of inverse length. The coupled-mode
equations then read
\begin{equation}
	\begin{aligned}
		\frac{da}{dz} &= i\varkappa b,\\
		\frac{db}{dz} &= i\varkappa a .
	\end{aligned}
	\label{eq:S1_coupled_modes}
\end{equation}
For an electron injected into nanotube 1, the boundary conditions are $a(0)=A_0$ and $b(0)=0$. The solution is
\begin{equation}
	\begin{aligned}
		a(z) &= A_0\cos(\varkappa z),\\
		b(z) &= iA_0\sin(\varkappa z).
	\end{aligned}
	\label{eq:S1_mode_solution}
\end{equation}
Thus, after propagation over a distance $z$ in the overlap region, the amplitude for remaining in nanotube 1 is $\cos(\varkappa z)$, whereas the amplitude for transfer to nanotube 2 is $i\sin(\varkappa z)$.

We now apply this result to the full overlap region of length $L$. Each passage across the overlap region gives the propagation phase $e^{ikL}$. Therefore, a passage that transfers the electron from nanotube 1 to nanotube 2 contributes $i\sin(\varkappa L)e^{ikL}$, whereas a passage that keeps the electron in the same nanotube contributes $\cos(\varkappa L)e^{ikL}$.

The total transmission amplitude is obtained by summing the coherent paths that connect the left and right electrodes. The shortest transmitted path contains one intertube transfer and therefore has the amplitude $i\sin(\varkappa L)e^{ikL}$. The next contribution contains one additional round trip in the overlap region before transmission. This round trip gives two factors of $\cos(\varkappa L)$ and two additional propagation phases, so that the total contribution is $\cos^2(\varkappa L)i\sin(\varkappa L)e^{3ikL}$. Keeping these two
contributions gives
\begin{equation}
	\begin{aligned}
		T &= \left| i\sin(\varkappa L)e^{ikL} +\cos^2(\varkappa L)  i\sin(\varkappa L)e^{3ikL}
		\right|^2 \\
		&= \sin^2(\varkappa L) \left[ 1+2\cos^2(\varkappa L)\cos(2kL) +\cos^4(\varkappa L)
		\right].
	\end{aligned}
	\label{eq:S1_two_paths}
\end{equation}
A more complete expression is obtained by allowing repeated transfers between the two nanotubes. The corresponding amplitudes form a geometric series,
\begin{equation}
	\begin{aligned}
		T &= \left| i\sin(\varkappa L)e^{ikL} +\cos^2(\varkappa L) i\sin(\varkappa L)e^{3ikL}
		+\cos^2(\varkappa L) \left[i\sin(\varkappa L)\right]^3 e^{5ikL} +\ldots \right|^2 \\
		&= \left| i\sin(\varkappa L)e^{ikL} +\cos^2(\varkappa L)e^{2ikL} \sum_{n=0}^{\infty}
		\left[i\sin(\varkappa L)e^{ikL}\right]^{2n+1} \right|^2 \\
		&= \left| \frac{i\sin(\varkappa L)e^{ikL} \left(1+e^{2ikL}\right)}
		{1+\sin^2(\varkappa L)e^{2ikL}} \right|^2 .
	\end{aligned}
	\label{eq:S1_geometric_series}
\end{equation}

The zero-field transmission of one coherent channel is therefore
\begin{equation}
	T(L)= \frac{ 2\sin^2(\varkappa L)\left[1+\cos(2kL)\right]} {1+2\sin^2(\varkappa L)\cos(2kL)
		+\sin^4(\varkappa L)} .
	\label{eq:S1_transmission}
\end{equation}

For junctions formed by different nanotubes, wave-vector mismatch would introduce a detuning term in the coupled-mode equations unless the dominant coupled modes are phase matched. The formula derived here therefore applies to that limit. It is equivalent to the formula introduced in Ref.~\cite{Xu2013corr}, but follows from the coupled-mode description and from summing repeated coherent paths in the overlap region.

\section{Fraunhofer-like interference from Aharonov--Bohm \\
	phases in CNT--CNT junctions}\label{sec:fraunhofer}

We consider a SWCNT--SWCNT junction of overlap length $L$ placed in a perpendicular magnetic field $B$. In the Fraunhofer-like approximation, the electron can tunnel between the two nanotubes at any position $z$ along the overlap region, with $0\leq z\leq L$. Each tunnelling path encloses a different magnetic flux and therefore acquires a different Aharonov--Bohm phase.

For a tunnelling event occurring at position $z$, the enclosed area is approximated as $(2r+d)z$, where $r$ is the nanotube radius and $d$ is the distance between the nanotubes. The corresponding magnetic flux is \begin{equation}
	\Phi(z)=B(2r+d)z .
\end{equation}
The phase accumulated by this path is therefore
\begin{equation}
	\varphi(z)=2\pi\frac{\Phi(z)}{\Phi_0} =2\pi\frac{B(2r+d)z}{\Phi_0},
\end{equation}
where $\Phi_0=h/e$ is the magnetic flux quantum.

The total transmission amplitude is obtained by summing coherently over all possible tunnelling positions along the overlap. Assuming a uniform tunnelling amplitude per unit length, the \color{black} normalised \color{black} transmission amplitude is
\begin{equation}
	t(B)=\frac{1}{L}\int_0^L \exp\left[ 2\pi i\frac{B(2r+d)z}{\Phi_0} \right] dz .
\end{equation}
Evaluating the integral gives
\begin{equation}
	t(B) =\frac{ \exp\left[ 2\pi i\frac{B(2r+d)L}{\Phi_0} \right]-1}
	{2\pi i L B(2r+d)/\Phi_0}  
	= \exp\left(i\pi\frac{\Phi}{\Phi_0}\right) \frac{
		\sin\left(\pi\frac{\Phi}{\Phi_0}\right)} {\pi\frac{\Phi}{\Phi_0}},
\end{equation}
where
\begin{equation}
	\Phi=B(2r+d)L
\end{equation}
is the total magnetic flux through the effective junction area.

The transmission probability is the squared modulus of the transmission
amplitude,
\begin{equation}
	\frac{T(B)}{T(0)} = |t(B)|^2 = \left[ \frac{ \sin\left(\pi\frac{\Phi}{\Phi_0}\right)} {\pi\frac{\Phi}{\Phi_0}} \right]^2 .
	\label{eq:Fraunhofer_transmission_SI}
\end{equation}
Equivalently,
\begin{equation}
	T(B)=T_{\mathrm{max}} \left[ \frac{ \sin\left(\pi\frac{B(2r+d)L}{\Phi_0}\right)}
	{\pi\frac{B(2r+d)L}{\Phi_0}} \right]^2 ,
	\label{eq:Fraunhofer_fit_SI}
\end{equation}
where $T_{\mathrm{max}}=T(0)$ in the ideal \color{black} normalised \color{black} case, or an effective amplitude when the expression is used as a fit to the computed transmission.

Equation~\eqref{eq:Fraunhofer_fit_SI} is the electronic analogue of the Fraunhofer diffraction envelope. In optical diffraction, the intensity is obtained by squaring the coherently summed field amplitudes emitted from different points of an aperture. Here, the corresponding quantity is the transmission probability, obtained by squaring the coherent sum of transmission amplitudes associated with different tunnelling positions in the CNT--CNT overlap region.

\section{Aharonov--Bohm phase condition for transmission nodes in loops
}\label{sec:loop}

When an electron propagates from one electrode to the other through the symmetric loop, it can travel through the upper or lower nanotube arm. These two partial waves enclose a region with non-zero magnetic flux $\Phi$ and therefore acquire an Aharonov--Bohm phase. In the notation used in Figure~\ref{loop}(e), the four overlap regions have length $L$, the finite upper and lower nanotube arms have length $z$, the separation between the two overlap regions is $\ell$, the nanotube radius is $r$, and the intertube distance is $d$. The limiting paths are shown schematically above Figure~\ref{loop}(b,g): the red path
encloses the maximal flux, whereas the yellow path encloses the minimal flux.

The Aharonov--Bohm phase difference between the two paths is
\begin{equation}
	\Delta \phi = 2\pi\frac{\Phi}{\Phi_0},
\end{equation}
where $\Phi_0=h/e$ is the magnetic flux quantum. For a symmetric loop, the phase shifts of the two partial waves can be written as
\begin{equation}
	\exp\left[i\left(kz+2\pi\frac{\Phi/2}{\Phi_0}\right)\right]
	\qquad \text{and} \qquad
	\exp\left[i\left(kz-2\pi\frac{\Phi/2}{\Phi_0}\right)\right],
\end{equation}
where $k$ is the electron wave vector along the nanotube arm. Thus, if the incoming wave amplitude is $\psi(0)$, the outgoing wave after recombination is
\begin{equation}
	\psi(z) = \psi(0) \left[ e^{i\left(kz-2\pi\frac{\Phi/2}{\Phi_0}\right)}
	+ e^{i\left(kz+2\pi\frac{\Phi/2}{\Phi_0}\right)} \right]
	= 2\psi(0)e^{ikz} \cos\left(\pi\frac{\Phi}{\Phi_0}\right).
\end{equation}
To obtain the transmission probability, we take the squared modulus of
the outgoing wave,
\begin{equation}
	|\psi(z)|^2  =    4|\psi(0)|^2
	\cos^2\left(\pi\frac{\Phi}{\Phi_0}\right).
\end{equation}
Thus, the magnetic-field dependence of the transmission is governed by
the factor $\cos^2(\pi\Phi/\Phi_0)$. The first transmission minimum occurs when
\begin{equation}
	\pi\frac{\Phi}{\Phi_0}=\frac{\pi}{2},
\end{equation}
so that
\begin{equation}
	\Phi_*=\frac{\Phi_0}{2}=B^*S.
\end{equation}

The two limiting areas are obtained from the paths shown above Figure~\ref{loop}(b,g). For the minimum-flux path, indicated by yellow arrows,
\begin{equation}
	S_{\min}=2\ell(r+d),
\end{equation}
whereas for the maximum-flux path, indicated by red arrows,
\begin{equation}
	S_{\max}=2(\ell+2L)(3r+d).
\end{equation}
There is also the area corresponding to the average electron path:
\begin{equation}
	S_{av}=2(\ell+L)(2r+d).
\end{equation}
This gives two limiting values, $B^*_{\min}$ and $B^*_{\max}$, and the average one $B^*_{av}$, listed in Table~\ref{loopB}. Comparing them with the numerical values of $B^*$ extracted from the transmission line cuts shows that the numerical minima take values of $B^*_{av}$. This indicates that the dominant electronic path encircles the average magnetic flux.

\newpage
\section{Additional Results}

\FloatBarrier

\begin{table}[h!]
	\centering
	\caption{Structural and magnetic parameters of the (9,9)+(9,9)+(9,9)+(9,9) (loop$_{\mathrm{(9,9)}}$) and (12,6)+(12,6)+(12,6)+(12,6) (loop$_{\mathrm{(12,6)}}$) loops. L is the overlap length shown in Fig.~\ref{loop}(e). $S_{\min}$, $S_{\max}$, and $S_{av}$ are the areas enclosed by the paths of minimum, maximum, and average magnetic flux, the first two indicated by yellow and red trajectories in the loop scheme above the panels. $B^*$ is the numerically determined field at which the transmission vanishes, while $B^*_{\min}=\phi_0/(2S_{\min})$, $B^*_{\max}=\phi_0/(2S_{\max})$, and $B^*_{av}=\phi_0/(2S_{av})$.}
	\begin{tabular}{|l|l|l|l|l|l|l|l|l|}
		\hline
		system & L [nm] & $B^*$ [T] & $S_{\min}$ [nm$^2$] & $B^*_{\min}$ [T] & $S_{\max}$ [nm$^2$] & $B^*_{\max}$ [T] & $S_{av}$ [nm$^2$] & $B^*_{av}$ [T] \\ 
		\hline
		\multirow{3}{*}{loop$_{\mathrm{(9,9)}}$}   & 4.92 & 39 & 23.26 & 88.91 & 95.89 & 21.57 & 53.57 & 38.60 \\
		& 2.46 & 45 & 23.26 & 88.91 & 74.58 & 27.73 & 45.92 & 45.03 \\
		& 0.49 & 52 & 23.26 & 88.91 & 57.54 & 35.94 & 39.80 & 51.96 \\
		\hline
		\multirow{3}{*}{loop$_{\mathrm{(12,6)}}$}  & 5.64 & 36 & 23.76 & 87.05 & 104.30 & 19.83 & 57.02 & 36.27 \\
		& 3.38 & 42 & 23.76 & 87.05 & 84.44 & 24.49 & 49.89 & 41.45 \\
		& 1.13 & 48 & 23.76 & 87.05 & 64.57 & 32.02 & 42.76 & 48.36 \\
		\hline
	\end{tabular}
	\label{loopB}
\end{table}

\begin{table}[h!]
	\centering
	\caption{Tube-resolved transmission contributions $\tau_{tube}$ at selected energies (0.0\,eV, -0.1\,eV, -0.3\,eV) for the (5,5) SWCNT, (5,5)@(10,10) DWCNT, and (5,5)@(10,10)@(15,15) TWCNT, each with a length of 24.60\,nm (100 unit cells), calculated with and without an applied magnetic field.}
	\setlength{\tabcolsep}{4pt}
	\renewcommand{\arraystretch}{1.15}
	\small
	\begin{tabular}{|l|l|l|l|l|}
		\hline
		\textbf{CNT} & \textbf{$\tau_{\mathrm{tube}}$} & \textbf{at $0.0\,eV$} & \textbf{at $-0.1\,eV$} & \textbf{at $-0.3\,eV$} \\ \hline
		\multirow{1}{*}{\makecell[l]{\textbf{SWCNT$_{B=0\,T}$}}}
		& $\tau_{\mathrm{total}}$ & 0.477 [100\,$\%$] & 1.274 [100\,$\%$] & 1.806 [100\,$\%$] \\ \hline
		\multirow{1}{*}{\makecell[l]{\textbf{SWCNT$_{B=60\,T}$}}}
		& $\tau_{\mathrm{total}}$ & 0.599 [100\,$\%$] & 1.366 [100\,$\%$] & 1.267 [100\,$\%$] \\ \hline
		\multirow{3}{*}{\makecell[l]{\textbf{DWCNT$_{B=0\,T}$}}}
		& $\tau_{\mathrm{inner}}$ & 1.982 [49.58\,$\%$] & 1.979 [49.47\,$\%$] & 0.931 [46.52\,$\%$] \\
		& $\tau_{\mathrm{outer}}$ & 2.016 [50.42\,$\%$] & 2.021 [50.53\,$\%$] & 1.070 [53.48\,$\%$] \\
		& $\tau_{\mathrm{total}}$ & 3.998 [100\,$\%$] & 4.000 [100\,$\%$] & 2.000 [100\,$\%$] \\ \hline
		\multirow{3}{*}{\makecell[l]{\textbf{DWCNT$_{B=60\,T}$}}}
		& $\tau_{\mathrm{inner}}$ & 1.982 [49.59\,$\%$] & 1.979 [49.47\,$\%$] & 0.930 [46.52\,$\%$] \\
		& $\tau_{\mathrm{outer}}$ & 2.015 [50.41\,$\%$] & 2.021 [50.53\,$\%$] & 1.0697 [53.48\,$\%$] \\
		& $\tau_{\mathrm{total}}$ & 3.997 [100\,$\%$] & 4.000 [100\,$\%$] & 2.000 [100\,$\%$] \\ \hline
		\multirow{4}{*}{\makecell[l]{\textbf{TWCNT$_{B=0\,T}$}}}
		& $\tau_{\mathrm{inner}}$  & 1.984 [33.06\,$\%$] & 1.980 [33.05\,$\%$] & 1.974 [32.90\,$\%$] \\
		& $\tau_{\mathrm{middle}}$ & 2.010 [33.51\,$\%$] & 2.003 [33.42\,$\%$] & 2.006 [32.44\,$\%$] \\
		& $\tau_{\mathrm{outer}}$  & 2.006 [33.43\,$\%$] & 2.009 [33.53\,$\%$] & 2.020 [32.66\,$\%$] \\
		& $\tau_{\mathrm{total}}$  & 6.000 [100\,$\%$] & 5.992 [100\,$\%$] & 6.000 [100\,$\%$] \\ \hline
		\multirow{4}{*}{\makecell[l]{\textbf{TWCNT$_{B=60\,T}$}}}
		& $\tau_{\mathrm{inner}}$  & 1.984 [33.06\,$\%$] & 1.980 [33.08\,$\%$] & 1.974 [32.90\,$\%$] \\
		& $\tau_{\mathrm{middle}}$ & 2.010 [33.51\,$\%$] & 1.998 [33.37\,$\%$] & 2.005 [33.42\,$\%$] \\
		& $\tau_{\mathrm{outer}}$  & 2.006 [33.43\,$\%$] & 2.008 [33.54\,$\%$] & 2.021 [33.68\,$\%$] \\
		& $\tau_{\mathrm{total}}$  & 6.000 [100\,$\%$] & 5.986 [100\,$\%$] & 6.000 [100\,$\%$] \\ \hline
	\end{tabular}
	\label{tau1}
\end{table}

\begin{table}[h!]
	\centering
	\caption{Maximum transmission at $E=-0.3\,eV$ for the (5,5)+(5,5) SWCNT--SWCNT,
		(5,5)@(10,10)+(5,5)@(10,10) DWCNT--DWCNT, and (5,5)@(10,10)@(15,15)+(5,5)@(10,10)@(15,15) TWCNT--TWCNT junctions, for all considered overlap lengths. Percentages given in brackets are relative to the corresponding zero-field value.}
	\begin{tabular}{|l|l|l|l|}
		\hline
		\textbf{junction} & \textbf{at $B=0\,T$} &
		\textbf{at $B=30\,T$} & \textbf{ at $B=60\,T$} \\
		\hline
		\textbf{SWCNT--SWCNT} & 2.000 [100\%] & 1.932 [96.6\%] & 1.624 [81.2\%] \\
		\hline
		\textbf{DWCNT--DWCNT} & 1.000 [100\%] & 0.969 [96.9\%] & 0.786 [78.6\%] \\
		\hline
		\textbf{TWCNT--TWCNT} & 2.258 [100\%] & 0.211 [9.3\%] & 0.300 [13.3\%] \\
		\hline
	\end{tabular}
	\label{tmax}
\end{table}

\begin{table}[h!]
	\centering
	\caption{Tube-resolved transmission contributions $\tau_{tube}$ at selected energies (0.0\,eV, -0.1\,eV, -0.3\,eV), for the (5,5)+(5,5) SWCNT--SWCNT, (5,5)@(10,10)+(5,5)@(10,10) DWCNT--DWCNT, and (5,5)@(10,10)@(15,15)+(5,5)@(10,10)@(15,15) TWCNT--TWCNT junctions, each with an overlap length of 24.60\,nm (100 \color{black} unit  \color{black} cells), calculated with and without an applied magnetic field. A negative transmission sign indicates only the opposite
		direction of electron flow and does not correspond to negative
		conductance.}
	\label{tab:junction_tau}
	\setlength{\tabcolsep}{4pt}
	\renewcommand{\arraystretch}{1.15}
	\small
	\begin{tabular}{|l|l|l|l|l|}
		\hline
		\textbf{junction} & \textbf{$\tau_{\mathrm{tube}}$} & \textbf{at $0.0\,eV$} & \textbf{at $-0.1\,eV$} & \textbf{at $-0.3\,eV$} 
		\\ \hline
		\multirow{2}{*}{\makecell[l]{\textbf{SWCNT--SWCNT$_{B=0\,T}$}\\}}
		& $\tau_{\mathrm{bottom}}$ & 0.439 [\,$100\%$] & 1.773 [\,$100\%$] & 1.769 [\,$100\%$] \\
		& $\tau_{\mathrm{top}}$    & -0.439 [\,$100\%$] & -1.773 [\,$100\%$] & -1.769 [\,$100\%$] \\ \hline
		\multirow{2}{*}{\makecell[l]{\textbf{SWCNT--SWCNT$_{B=60\,T}$}\\}}
		& $\tau_{\mathrm{bottom}}$ & 0.558 [\,$100\%$] & 1.496 [\,$100\%$] & 1.251 [\,$100\%$] \\
		& $\tau_{\mathrm{top}}$    & -0.558 [\,$100\%$] & -1.496 [\,$100\%$] & -1.251 [\,$100\%$] \\ \hline
		\multirow{6}{*}{\makecell[l]{\textbf{DWCNT--DWCNT$_{B=0\,T}$}\\}}
		& $\tau_{\mathrm{inner,bottom}}$ & 1.50$\cdot 10^{-4}$ [\,$4.25\%$] & 0.042 [\,$2.87\%$] & 0.013 [\,$1.31\%$] \\
		& $\tau_{\mathrm{outer,bottom}}$ & 3.33$\cdot 10^{-3}$ [\,$94.24\%$] & 1.431 [\,$97.54\%$] & 0.987 [\,$98.72\%$] \\
		& $\tau_{\mathrm{total,bottom}}$ & 3.53$\cdot 10^{-3}$ [100\,$\%$] & 1.467 [100\,$\%$] & 1.000 [100\,$\%$] \\
		& $\tau_{\mathrm{inner,top}}$    & -6.88$\cdot 10^{-5}$ [\,$1.95\%$] & -0.031 [\,$2.10\%$] & -0.007 [\,$0.66\%$] \\
		& $\tau_{\mathrm{outer,top}}$    & -3.41$\cdot 10^{-3}$ [\,$96.55\%$] & -1.442 [\,$98.30\%$] & -0.994 [\,$99.37\%$] \\
		& $\tau_{\mathrm{total,top}}$    & -3.53$\cdot 10^{-3}$ [100\,$\%$] & -1.467 [100\,$\%$] & -1.000 [100\,$\%$] \\ \hline
		\multirow{6}{*}{\makecell[l]{\textbf{DWCNT--DWCNT$_{B=60\,T}$}\\}}
		& $\tau_{\mathrm{inner,bottom}}$ & 3.38$\cdot 10^{-3}$ [\,$45.03\%$] & 0.006 [\,$0.97\%$] & 0.002 [\,$0.70\%$] \\
		& $\tau_{\mathrm{outer,bottom}}$ & 4.12$\cdot 10^{-3}$ [\,$54.97\%$] & 0.631 [\,$98.98\%$] & 0.283 [\,$99.30\%$] \\
		& $\tau_{\mathrm{total,bottom}}$ & 7.50$\cdot 10^{-3}$ [100\,$\%$] & 0.637 [100\,$\%$] & 0.285 [100\,$\%$] \\
		& $\tau_{\mathrm{inner,top}}$    & -3.09$\cdot 10^{-5}$ [\,$4.15\%$] & -0.010 [\,$1.57\%$] & -0.002 [\,$0.70\%$] \\
		& $\tau_{\mathrm{outer,top}}$    & -7.15$\cdot 10^{-4}$ [\,$95.85\%$] & -0.627 [\,$98.38\%$] & -0.283 [\,$99.30\%$] \\
		& $\tau_{\mathrm{total,top}}$    & -7.46$\cdot 10^{-4}$ [100\,$\%$] & -0.637 [100\,$\%$] & -0.285 [100\,$\%$] \\ \hline
		\multirow{8}{*}{\makecell[l]{\textbf{TWCNT--TWCNT$_{B=0\,T}$}\\}}
		& $\tau_{\mathrm{inner,bottom}}$  & 0.023 [\,$18.80\%$] & 0.010 [\,$6.73\%$] & 0.021 [\,$13.50\%$] \\
		& $\tau_{\mathrm{middle,bottom}}$ & 0.028 [\,$22.32\%$] & 0.035 [\,$24.22\%$] & 0.035 [\,$22.20\%$] \\
		& $\tau_{\mathrm{outer,bottom}}$  & 0.073 [\,$58.88\%$] & 0.099 [\,$69.06\%$] & 0.102 [\,$64.30\%$] \\
		& $\tau_{\mathrm{total,bottom}}$  & 0.123 [100\,$\%$] & 0.144 [100\,$\%$] & 0.158 [100\,$\%$] \\
		& $\tau_{\mathrm{inner,top}}$     & -0.023 [\,$19.00\%$] & -0.011 [\,$7.24\%$] & -0.021 [\,$12.83\%$] \\
		& $\tau_{\mathrm{middle,top}}$    & -0.027 [\,$22.21\%$] & -0.035 [\,$24.08\%$] & -0.034 [\,$21.43\%$] \\
		& $\tau_{\mathrm{outer,top}}$     & -0.072 [\,$58.79\%$] & -0.100 [\,$68.68\%$] & -0.106 [\,$65.74\%$] \\
		& $\tau_{\mathrm{total,top}}$     & -0.123 [100\,$\%$] & -0.145 [100\,$\%$] & -0.161 [100\,$\%$] \\ \hline
		\multirow{8}{*}{\makecell[l]{\textbf{TWCNT--TWCNT$_{B=60\,T}$}\\}}
		& $\tau_{\mathrm{inner,bottom}}$  & 0.017 [\,$10.03\%$] & 0.025 [\,$30.12\%$] & 0.093 [\,$49.84\%$] \\
		& $\tau_{\mathrm{middle,bottom}}$ & 0.037 [\,$22.25\%$] & 0.0144 [\,$17.36\%$] & 0.086 [\,$45.99\%$] \\
		& $\tau_{\mathrm{outer,bottom}}$  & 0.113 [\,$67.72\%$] & 0.044 [\,$52.53\%$] & 0.008 [\,$4.17\%$] \\
		& $\tau_{\mathrm{total,bottom}}$  & 0.166 [100\,$\%$] & 0.083 [100\,$\%$] & 0.186 [100\,$\%$] \\
		& $\tau_{\mathrm{inner,top}}$     & -0.038 [\,$27.63\%$] & -0.008 [\,$10.74\%$] & -0.011 [\,$7.16\%$] \\
		& $\tau_{\mathrm{middle,top}}$    & -0.030 [\,$21.92\%$] & -0.012 [\,$15.69\%$] & -0.018 [\,$11.12\%$] \\
		& $\tau_{\mathrm{outer,top}}$     & -0.070 [\,$50.45\%$] & -0.054 [\,$73.57\%$] & -0.129 [\,$81.72\%$] \\
		& $\tau_{\mathrm{total,top}}$     & -0.138 [100\,$\%$] & -0.074 [100\,$\%$] & -0.158 [100\,$\%$] \\ \hline
	\end{tabular}
	\label{tau2}
\end{table}

\begin{table}[h!]
	\centering
	\caption{\color{black} Tube-resolved transmission contributions $\tau_{tube}$ at selected energies (0.0\,eV, -0.1\,eV, -0.3\,eV) for the  (12,0)@(21,0)@(30,0)+(12,0)@(21,0)@(30,0) junction consisting of two three-walled CNTs and (5,5)@(10,10)@(15,15)@(20,20)+(5,5)@(10,10)@(15,15)@(20,20) junction consisting of two four-walled CNTs with overlap lengths of 21.30\,nm (50 \color{black} unit \color{black} cells) and 24.60\,nm (100 \color{black} unit \color{black} cells), respectively, calculated with and without an applied magnetic field. A negative transmission sign indicates only the opposite
		direction of electron flow and does not correspond to a negative
		conductance. The numbers describe the CNT walls with the innermost tube depicted as 1.}
	\setlength{\tabcolsep}{4pt}
	\renewcommand{\arraystretch}{1.15}
	\footnotesize
	\color{black}
	\begin{tabular}{|l|l|l|l|l|}
		\hline
		\textbf{junction} & \textbf{$\tau_{\mathrm{tube}}$} & \textbf{at $0.0\,eV$} & \textbf{at $-0.1\,eV$} & \textbf{at $-0.3\,eV$} 
		\\ \hline
\multirow{8}{*}{%
	\makecell[l]{%
		\textbf{(12,0)@(21,0)@(30,0)+}\\
		\textbf{(12,0)@(21,0)@(30,0)$_{B=0\,T}$}%
}}
		& $\tau_{\mathrm{1,bottom}}$ & 3.40$\cdot 10^{-5}$ [\,$0.65\%$] & 0.002 [\,$0.13\%$] & 0.022 [\,$3.40\%$] \\
		& $\tau_{\mathrm{2,bottom}}$ & 5.00$\cdot 10^{-3}$ [\,$95.19\%$] & 0.523 [\,$32.43\%$] & 0.068 [\,$10.36\%$] \\
		& $\tau_{\mathrm{3,bottom}}$ & 2.19$\cdot 10^{-4}$ [\,$4.16\%$] & 1.087 [\,$67.44\%$] & 0.563 [\,$86.24\%$] \\
		& $\tau_{\mathrm{total,bottom}}$ & 5.25$\cdot 10^{-3}$ [100\,$\%$] & 1.612 [100\,$\%$] & 0.653 [100\,$\%$] \\
		& $\tau_{\mathrm{1,top}}$ & -3.40$\cdot 10^{-5}$ [0.65\,$\%$] & -0.002 [0.13\,$\%$] & -0.022 [3.40\,$\%$] \\
		& $\tau_{\mathrm{2,top}}$    & -5.00$\cdot 10^{-3}$ [\,$95.19\%$] & -0.523 [\,$32.43\%$] & -0.068 [\,$10.36\%$] \\
		& $\tau_{\mathrm{3,top}}$    & -2.19$\cdot 10^{-4}$ [\,$4.16\%$] & -1.087 [\,$67.44\%$] & -0.563 [\,$86.24\%$] \\
		& $\tau_{\mathrm{total,top}}$    & -5.25$\cdot 10^{-3}$ [100\,$\%$] & -1.612 [100\,$\%$] & -0.653 [100\,$\%$] \\ \hline
\multirow{8}{*}{%
	\makecell[l]{%
		\textbf{(12,0)@(21,0)@(30,0)+}\\
		\textbf{(12,0)@(21,0)@(30,0)$_{B=60\,T}$}%
}}
		& $\tau_{\mathrm{1,bottom}}$ & 3.56$\cdot 10^{-5}$ [\,$0.86\%$] & 0.001 [\,$0.17\%$] & 0.031 [\,$13.97\%$] \\
		& $\tau_{\mathrm{2,bottom}}$ & 3.64$\cdot 10^{-3}$ [\,$88.09\%$] & 0.168 [\,$32.00\%$] & 0.064 [\,$29.00\%$] \\
		& $\tau_{\mathrm{3,bottom}}$ & 4.57$\cdot 10^{-4}$ [11.05\,$\%$] & 0.356 [67.83\,$\%$] & 0.125 [57.02\,$\%$] \\
		& $\tau_{\mathrm{total,bottom}}$ & 4.14$\cdot 10^{-3}$ [100\,$\%$] & 0.525 [100\,$\%$] & 0.219 [100\,$\%$] \\
		& $\tau_{\mathrm{1,top}}$ & -7.41$\cdot 10^{-6}$ [0.18\,$\%$] & -0.001 [0.12\,$\%$] & -0.029 [13.38\,$\%$] \\
		& $\tau_{\mathrm{2,top}}$    & -3.90$\cdot 10^{-3}$ [\,$94.29\%$] & -0.200 [\,$38.08\%$] & -0.065 [\,$29.44\%$] \\
		& $\tau_{\mathrm{3,top}}$    & -2.29$\cdot 10^{-4}$ [\,$5.53\%$] & -0.325 [\,$61.80\%$] & -0.125 [\,$57.18\%$] \\
		& $\tau_{\mathrm{total,top}}$    & -4.14$\cdot 10^{-3}$ [100\,$\%$] & -0.525 [100\,$\%$] & -0.219 [100\,$\%$] \\ \hline
\multirow{10}{*}{%
	\makecell[l]{%
		\textbf{(5,5)@(10,10)@(15,15)@(20,20)+}\\
		\textbf{(5,5)@(10,10)@(15,15)@(20,20)$_{B=0\,T}$}%
}}
		& $\tau_{\mathrm{1,bottom}}$  & 0.024 [\,$1.50\%$] & 0.011 [\,$0.90\%$] & 0.019 [\,$0.51\%$] \\
		& $\tau_{\mathrm{2,bottom}}$ & 0.327 [\,$20.43\%$] & 0.128 [\,$10.63\%$] & 0.091 [\,$2.38\%$] \\
		& $\tau_{\mathrm{3,bottom}}$  & 0.012 [\,$0.76\%$] & 0.015 [\,$1.26\%$] & 1.329 [\,$34.72\%$] \\
		& $\tau_{\mathrm{4,bottom}}$  & 1.247 [77.32\,$\%$] & 1.052 [87.20\,$\%$] & 2.388 [62.40\,$\%$] \\
		& $\tau_{\mathrm{total,bottom}}$  & 1.601 [100\,$\%$] & 1.206 [100\,$\%$] & 3.827 [100\,$\%$] \\
		& $\tau_{\mathrm{1,top}}$  & -0.018 [1.14\,$\%$] & -0.007 [0.58\,$\%$] & -0.019 [0.49\,$\%$] \\
		& $\tau_{\mathrm{2,top}}$     & -0.350 [\,$21.90\%$] & -0.123 [\,$10.17\%$] & -0.159 [\,$4.15\%$] \\
		& $\tau_{\mathrm{3,top}}$    & -0.007 [\,$0.45\%$] & -0.007 [\,$0.61\%$] & -1.236 [\,$32.29\%$] \\
		& $\tau_{\mathrm{4,top}}$     & -1.224 [\,$76.50\%$] & -1.069 [\,$88.64\%$] & -2.414 [\,$63.07\%$] \\
		& $\tau_{\mathrm{total,top}}$     & -1.601 [100\,$\%$] & -1.206 [100\,$\%$] & -3.827 [100\,$\%$] \\ \hline
\multirow{10}{*}{%
	\makecell[l]{%
		\textbf{(5,5)@(10,10)@(15,15)@(20,20)+}\\
		\textbf{(5,5)@(10,10)@(15,15)@(20,20)$_{B=60\,T}$}%
}}
		& $\tau_{\mathrm{1,bottom}}$  & 0.006 [\,$1.71\%$] & 0.001 [\,$0.51\%$] & 0.002 [\,$0.58\%$] \\
		& $\tau_{\mathrm{2,bottom}}$ & 0.057 [\,$16.89\%$] & 0.058 [\,$25.07\%$] & 0.099 [\,$29.38\%$] \\
		& $\tau_{\mathrm{3,bottom}}$  & 0.024 [\,$6.93\%$] & 0.011 [\,$4.98\%$] & 0.025 [\,$7.32\%$] \\
		& $\tau_{\mathrm{4,bottom}}$  & 0.253 [$74.47\%$] & 0.160 [69.44\,$\%$] & 0.211 [62.73\,$\%$] \\
		& $\tau_{\mathrm{total,bottom}}$  & 0.339 [\,$100\%$] & 0.230 [\,$100\%$] & 0.336 [\,$100\%$] \\
		& $\tau_{\mathrm{1,top}}$    & -0.008 [\,$2.36\%$] & -0.0001 [\,$0.05\%$] & -0.003 [\,$0.75\%$] \\
		& $\tau_{\mathrm{2,top}}$     & -0.084 [\,$24.73\%$] & -0.048 [\,$20.90\%$] & -0.087 [\,$25.87\%$] \\
		& $\tau_{\mathrm{3,top}}$     & -0.037 [\,$10.85\%$] & -0.0001 [\,$0.03\%$] & -0.001 [\,$0.43\%$] \\
		& $\tau_{\mathrm{4,top}}$     & -0.211 [\,$62.06\%$] & -0.182 [\,$79.01\%$] & -0.248 [\,$73.81\%$] \\
		& $\tau_{\mathrm{total,top}}$     & -0.339 [100\,$\%$] & -0.230 [100\,$\%$] & -0.336 [100\,$\%$] \\ \hline
	\end{tabular}
	\label{tau3}
\end{table}

\begin{table}[h!]
	\centering
	\caption{Coefficient of variation of the nanotube radius
		(CV)~\cite{milowska2015}, defined as the ratio of the standard deviation to the mean of the nanotube radius, for the constituent nanotubes in the optimised (5,5)@(10,10) + (5,5)@(10,10) DWCNT--DWCNT and (5,5)@(10,10)@(15,15)+(5,5)@(10,10)@(15,15) TWCNT--TWCNT junctions, compared with the corresponding unrelaxed structures.}
	\label{tab:cv_radius}
	\setlength{\tabcolsep}{4pt}
	\renewcommand{\arraystretch}{1.15}
	\small
	\begin{tabular}{|l|l|l|l|}
		\hline
		\textbf{junction} & \textbf{nanotube} &
		\textbf{CV ideal} & \textbf{CV relaxed} \\
		\hline
		\multirow{4}{*}{DWCNT--DWCNT}
		& inner, bottom & 4.6$\cdot$10$^{-7}$ & \color{black} 3.0$\cdot$10$^{-3}$   \\
		& outer, bottom & 2.0$\cdot$10$^{-7}$ & \color{black} 4.9$\cdot$10$^{-3}$    \\
		& inner, top    & 4.6$\cdot$10$^{-7}$ & \color{black} 2.9$\cdot$10$^{-3}$  \\
		& outer, top    & 2.0$\cdot$10$^{-7}$ & \color{black} 5.2$\cdot$10$^{-3}$   \\
		\hline
		\multirow{6}{*}{TWCNT--TWCNT}
		& inner, bottom  & 4.6$\cdot$10$^{-7}$ & \color{black} 3.7$\cdot$10$^{-3}$   \\
		& middle, bottom & 2.0$\cdot$10$^{-7}$ & \color{black} 4.6$\cdot$10$^{-3}$   \\
		& outer, bottom  & 2.3$\cdot$10$^{-7}$ & \color{black} 7.4$\cdot$10$^{-3}$   \\
		& inner, top     & 4.6$\cdot$10$^{-7}$ & \color{black} 3.9$\cdot$10$^{-3}$   \\
		& middle, top    & 2.0$\cdot$10$^{-7}$ & \color{black} 5.1$\cdot$10$^{-3}$   \\
		& outer, top     & 2.3$\cdot$10$^{-7}$ & \color{black} 7.6$\cdot$10$^{-3}$  \\
		\hline
	\end{tabular}
\end{table}

\begin{figure}[h!]
	\centering
	\includegraphics[width=0.7\textwidth]{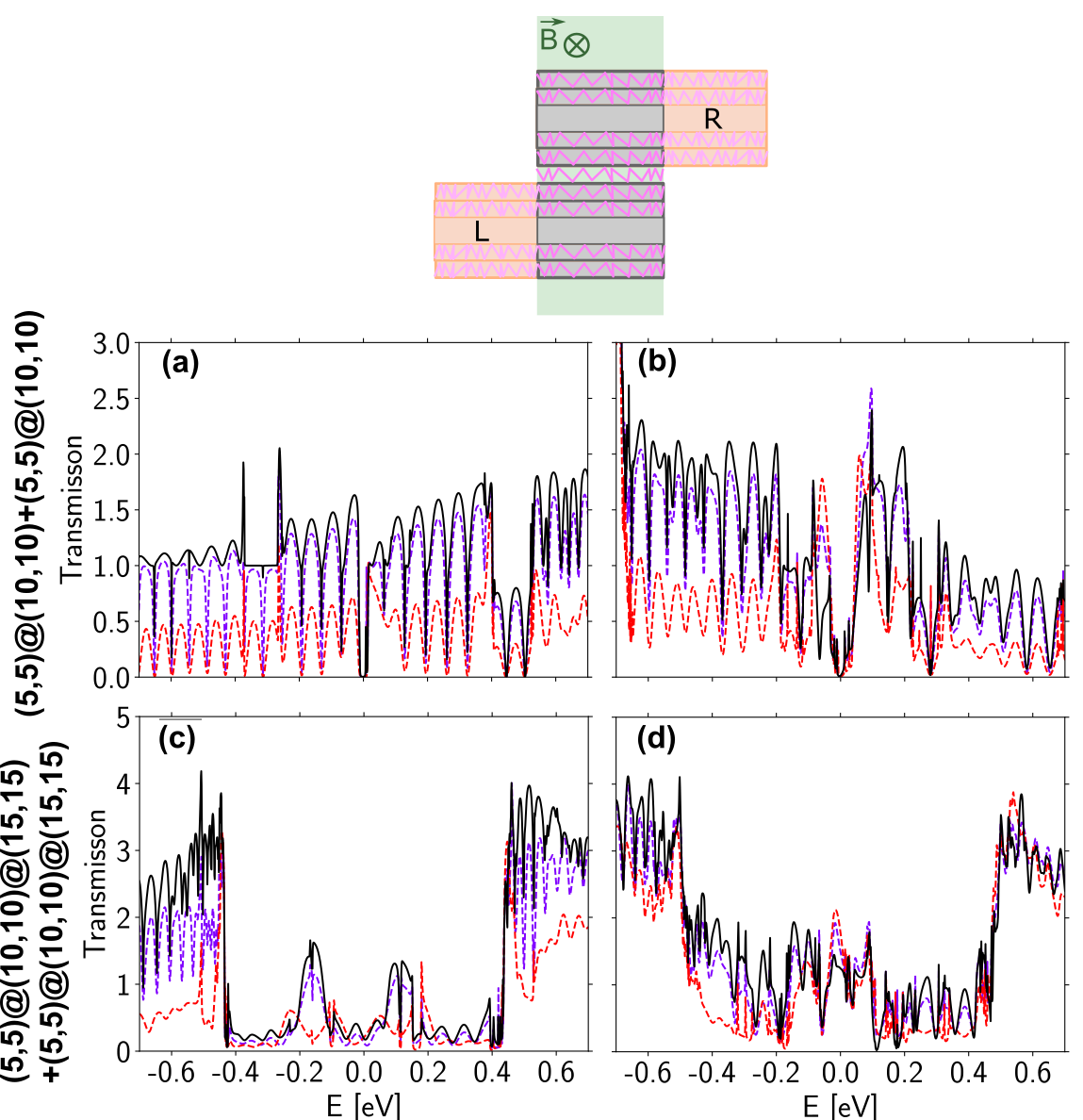}
	\caption{\color{black}Transmission spectra through the DWCNT junction (5,5)@(10,10)+(5,5)@(10,10) in the (a) ideal and (b) PBE-D3 relaxed geometries, and through the TWCNT junction (5,5)@(10,10)@(15,15)+(5,5)@(10,10)@(15,15) in the (c) ideal and (d) PBE-D3 relaxed geometries. All junctions have an overlap length of 24.60 nm (100 unit cells) and are subjected to a perpendicular magnetic field. The black, purple and red lines correspond to $B=0$, 30 T and 60\,T, respectively.}
	\label{trans-relaxed}
\end{figure}

\FloatBarrier
\clearpage
\bibliographystyle{unsrt}
\bibliography{biblio}

\end{document}